\documentclass{article}

\usepackage{arxiv}

\usepackage{amsmath}
\usepackage[utf8]{inputenc} 
\usepackage[T1]{fontenc}    
\usepackage{hyperref}       
\usepackage{url}            
\usepackage{booktabs}       
\usepackage{amsfonts}       
\usepackage{nicefrac}       
\usepackage{microtype}      
\usepackage{cleveref}       
\usepackage{lipsum}         
\usepackage{graphicx}
\usepackage{natbib}
\usepackage{doi}
\usepackage{bbm}
\usepackage{multirow}

\usepackage{amsfonts}
\usepackage{mathrsfs}
\usepackage{graphicx}
\usepackage{caption}
\usepackage{subcaption}
\usepackage[skip=1pt]{subcaption}
\usepackage{changepage}
\usepackage{booktabs}
\usepackage[shortlabels]{enumitem}
\usepackage{caption}
\newtheorem{definition}{Definition}
\newtheorem{theorem}{Theorem}[section]

\newtheorem{prop}[theorem]{Proposition}
\usepackage{algorithm}
\usepackage{algpseudocode}

\usepackage{color, colortbl}	
\definecolor{Gray}{gray}{0.9}

\title{A Stochastic Process Model for Time Warping Functions}

\author{ Yijia Ma  \qquad\qquad Xinyu Zhou \qquad \qquad Wei Wu\\ 
	Department of Statistics, 
	Florida State University\\
	Tallahassee, FL 32306 \\
	\texttt{ym19f@my.fsu.edu }\quad\texttt{xz19c@my.fsu.edu}\quad\texttt{wwu@stat.fsu.edu}
}

\date{}

\renewcommand{\shorttitle}{\textit{A Stochastic Process Model for Time Warping Functions} }

\hypersetup{
pdftitle={A Stochastic Process Model for Time Warping Functions},
}

\begin{document}
\maketitle

\begin{abstract}
	Time warping function provides a mathematical representation to measure phase variability in functional data. Recent studies have developed various approaches to estimate optimal warping between functions and provide non-Euclidean models.  However, a principled, linear, generative model on time warping functions is still under-explored. This is a highly challenging problem because the space of warping functions is non-linear with the conventional Euclidean metric. To address this problem, we propose a stochastic process model for time warping functions, where the key is to define a linear, inner-product structure on the time warping space and then transform the warping functions into a sub-space of the $\mathbb L^2$ Euclidean space. With certain constraints on the warping functions, this transformation is an isometric isomorphism. In the transformed space, we adopt the $\mathbb L^2$ basis in the Hilbert space for representation.  This new framework can easily build generative model on time warping by using different types of stochastic process. It can also be used to conduct statistical inferences such as functional PCA, functional ANOVA, and functional regressions.  Furthermore, we demonstrate the effectiveness of this new framework by using it as a new prior in the Bayesian registration, and propose an efficient gradient method to address the important maximum a posteriori estimation. We illustrate the new Bayesian method using simulations which properly characterize nonuniform and correlated constraints in the time domain. Finally, we apply the new framework to the famous Berkeley growth data and obtain reasonable results on modeling, resampling, group comparison, and classification analysis. 
\end{abstract}



\section{Introduction}
Temporal phase variability has been a central topic in the field of functional data analysis. In function registration or alignment, the goal is often to separate phase and amplitude variabilities, where the phase variation is represented using a time warping function.  
In many studies, finding the aligned functions is the main goal because warping is considered as a nuisance variable in the measurement process and its variability needs to be removed \citep{ramsay2006functional}.
However, in other cases, phase is considered an essential and critical feature in the data \citep{marron2015functional}. For either purpose, one needs to estimate optimal time warpings to align functional observations properly. A common space of warping functions is defined as $\Gamma=\{\gamma:[0,1]\rightarrow [0,1]|\gamma (0)=0, \gamma(1)=1,0<\dot\gamma(t)<\infty\}$ ($\dot \gamma$ denotes the derivative of $\gamma$) \citep{srivastava2011registration}, which is obviously a nonlinear space under the conventional Euclidean metric. 
Over the past two-to-three decades, this has been an active topic, and various approaches have been developed for robust and efficient estimation.   
Early approaches formulated a least-square problem by representing warping function with a linear combination of B-spline basis functions, and the warping can be obtained by estimating the corresponding coefficients \citep{ramsay1998curve, gervini2004self, james2007curve, eilers2004parametric}. Recent approaches conducted registration by minimizing the Fisher-Rao metric \citep{srivastava2011registration,wu2014analysis}, or resampling with Bayesian registration \citep{cheng2014analysis,cheng2016bayesian,kurtek2017geometric,lu2017bayesian}. 
In particular, there have been attempts to analyze the phase variations for functional regression \citep{hadjipantelis2014analysis,gervini2015warped}, classification \citep{tucker2013generative} and functional PCA \citep{lee2016combined,happ2019general}. 

We point out that most of these studies focused on estimating optimal warping function for alignment, but not building a probabilistic model on it.  Studies in Bayesian registration have examined statistical models on time warping, whereas the methods mainly focus on the simplified Dirichlet distribution \citep{cheng2016bayesian} or Gaussian process in the tangent space of the nonlinear Hilbert unit sphere \citep{kurtek2017geometric,lu2017bayesian}. Because of this non-linearity, the modeling of time warping is still very challenging in the field.   To explore the benefits in a linear, inner product space, we will at first define linear and inner-product operations in the warping functions.  Our goal is to build an isometric isomorphism to transform the warping space to a subspace in the Euclidean $\mathbb L^2$ space and then adopt stochastic process such as Gaussian process to model the transformed functions.   
 
To achieve this goal, we at first use the fact that a warping function in $\Gamma$ can be equivalently represented as a probability density function by simply taking its derivative. In addition, it is well known that there exists an isometric isomorphism between the density space and a Euclidean sub-space in $\mathbb L^2$ under the Centered Logratio transformation \citep{egozcue2006hilbert}. Based on these results, we propose to model the time warping function in three steps: 1) transform warping function to density function, 2) transform density function to Euclidean sub-space, and 3) develop a stochastic process model in the Euclidean sub-space. A similar idea was explored in \citep{happ2019general}, where the focus was on joint-modeling phase and amplitude components. 
Our proposed framework is partially motivated by this study, whereas our goal is to build a principled model on time warping functions.   

To the best of our knowledge, our new framework is the first linear inner-product model on the time warping functions.  The framework has the following four apparent features:  Firstly, the model has a principled theoretical framework; it is given in an explicitly generative form, and various time warping functions can be easily sampled from this model. Secondly, the model can be effectively learned from given observations. We will evaluate the effectiveness of this model by conducting resampling and comparing the outcome with original data. Thirdly, the proposed model can be adopted as a new prior in the Bayesian registration framework. Fourthly, we can conduct various statistical inferences on the original data by building one-to-one mapping from the time warping space to a Euclidean space.  We will illustrate a two-sample test and a logistic regression in a real-world dataset in the transformed space. 

The rest of this manuscript is organized as follows: In Section \ref{sec: method}, we at first define a space that contains only warping functions with bounded derivative and show it is isometrically isomorphic to a Euclidean inner-product space by applying the Centered Logratio transformation to the derivative of the warping function. We then extend the inner-product space to a Hilbert space so that we can build a stochastic-process-based model with bounded Hilbert basis functions. In addition, we describe how to estimate a model from observations and provide two simulation experiments. In Section \ref{Sec: BayesReg}, we present a new Bayesian framework for registration with a Gaussian process prior by using our model, and illustrate the method with three simulations. A real-world dataset is given in Section \ref{sec: app} to illustrate the statistical inferences under the proposed framework. Finally, we summarize our work in Section \ref{sec6}.   All mathematical details are given in the appendices.

\section{Methods}
\label{sec: method}

\subsection{Warping functions with bounded derivatives}
\label{sec: bounded}
\begin{figure}[h]
	\centering
	\begin{subfigure}[h]{0.3\textwidth}
		\includegraphics[width=\textwidth]{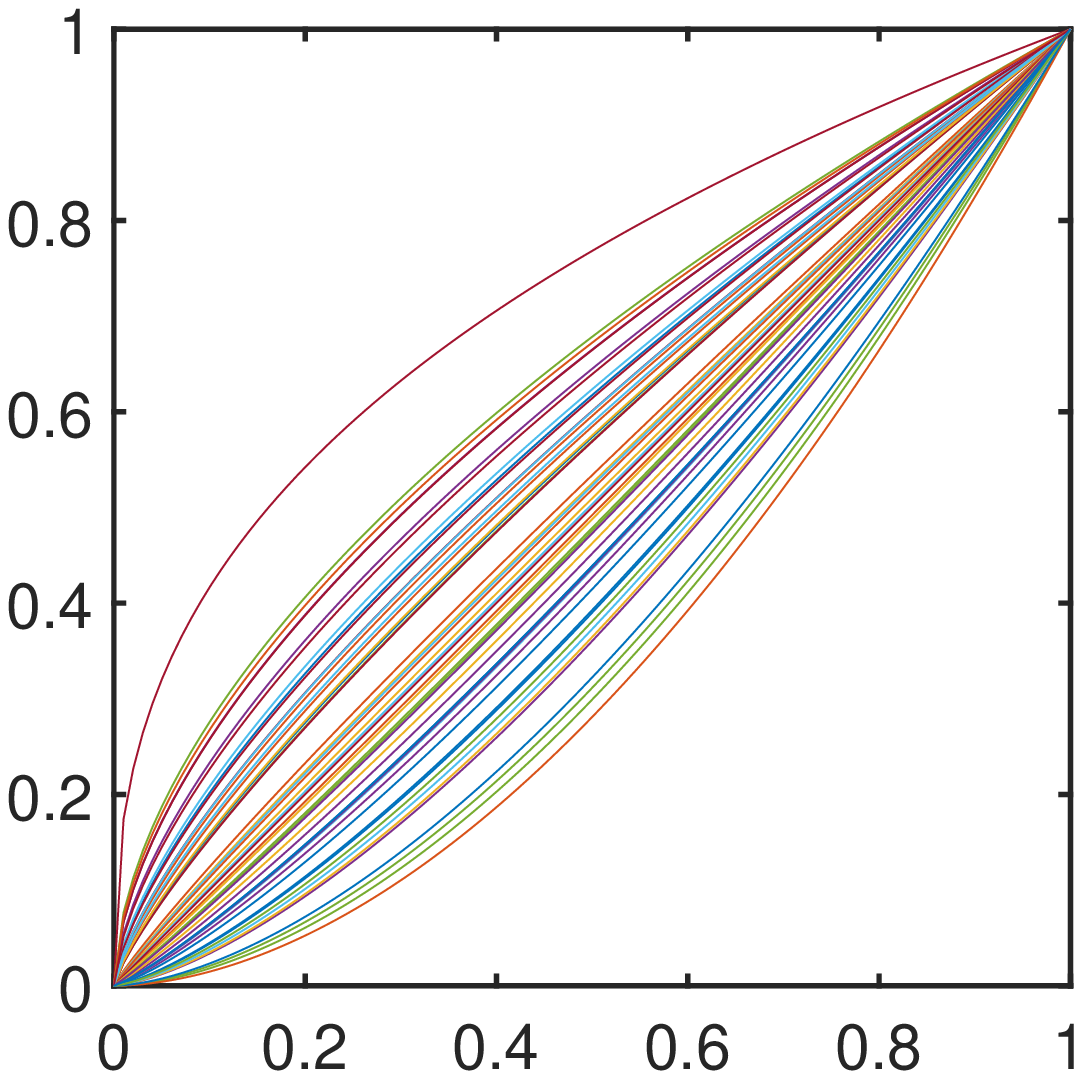}
		\caption{$\gamma_i(t) = t^{\alpha_i}$}
	\end{subfigure}\hspace{-0.2cm}
	\begin{subfigure}[h]{0.3\textwidth}
		\includegraphics[width=\textwidth]{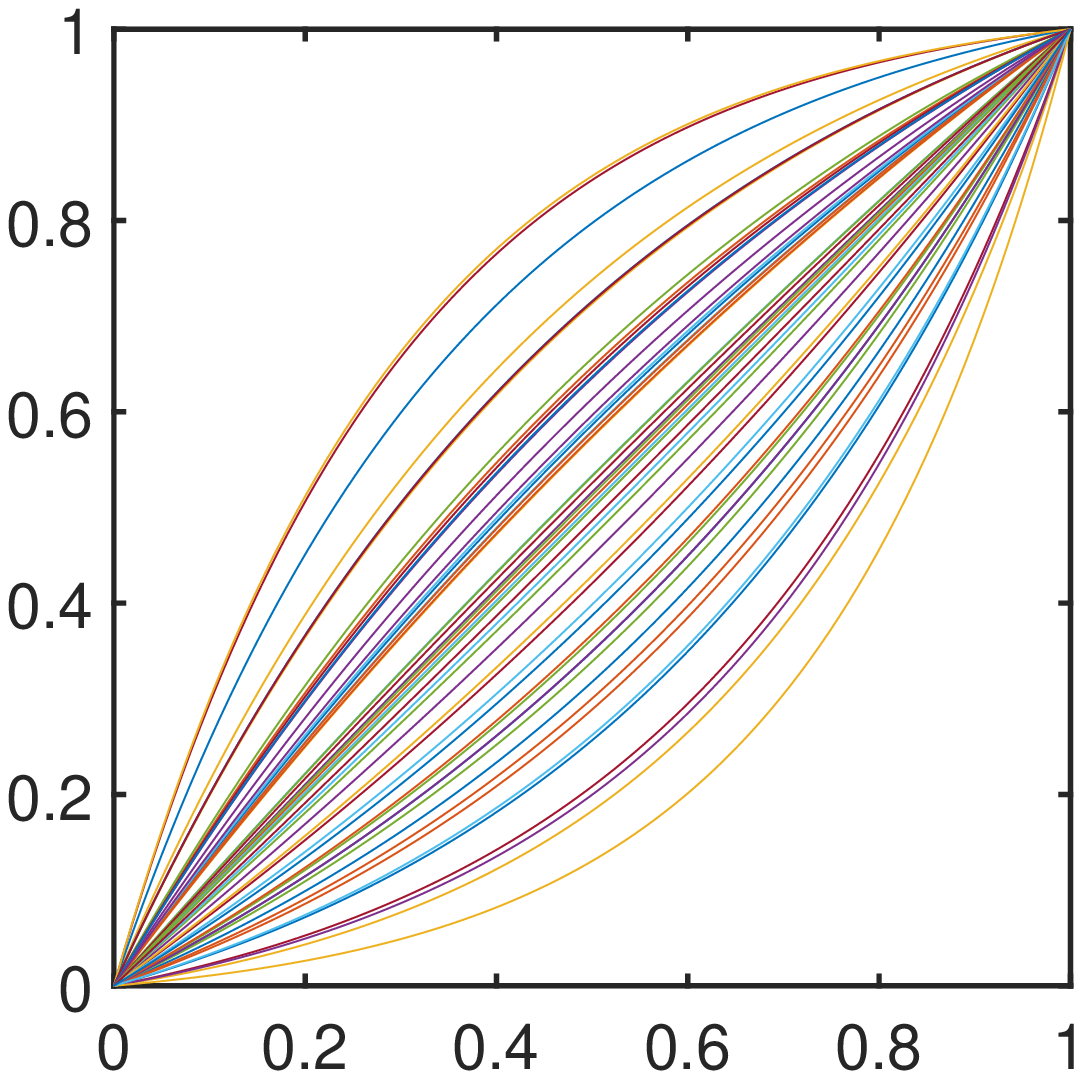}
		\caption{$\gamma_i(t) = \frac{e^{a_it}-1}{e^{a_i}-1}$}
	\end{subfigure}
	\caption{Simulations of 50 warping functions using two typical methods, respectively. (a) $\gamma_i(t) = t^{\alpha_i}$ with $\alpha_i\sim G(5,0.2)$, i.e., Gamma distribution with mean 1 and variance 0.2. (b) $\gamma_i(t) = \frac{e^{a_it}-1}{e^{a_i}-1}$  with $a_i\sim N(0,4)$, i.e. normal distribution with mean 0 and variance 4.}
	\label{fig.simpleGamma}
\end{figure}

Time warping functions have been studied extensively in the literature, and a common space for all warpings in a finite domain $[0, 1]$ is defined as 
\begin{equation}
	\Gamma=\{\gamma:[0,1]\rightarrow [0,1]|\gamma (0)=0, \gamma(1)=1,0<\dot\gamma(t)<\infty\}.
	\label{eq:gamma}
\end{equation}
A simple function $\gamma(t) = t^\alpha, \,\alpha>0$, is often used as a time warping example.  See Figure \ref{fig.simpleGamma}(a) for a few example curves.  
This paper aims to provide a stochastic-process-based model on those warping functions. To make the mathematical representation feasible, we need to provide certain basic assumptions on the process. For example, we often assume the process is second-order (such as a Gaussian process) so that the standard covariance-based methods can be adopted. However, $\dot \gamma$ in Equation \eqref{eq:gamma} is simply positive without any other constraints, which makes it challenging to develop an appropriate model.   

One simple and effective solution is to provide a lower and an upper bound on the derivative function, and this can lead to finite integrations such as the $\mathbb L^p$ norms for most models. That is, we should have various frameworks to model time warping functions in the following domain:
\begin{equation}
	\Gamma_1=\{\gamma:[0,1]\rightarrow [0,1]|\gamma (0)=0, \gamma(1)=1, 0<m_\gamma< \dot\gamma(t)<M_\gamma<\infty\}.
	\label{eq:gamma1}
\end{equation}
In this new domain the two bounds $m_\gamma$ and $M_\gamma$ vary with respect to the function $\gamma$.  One typical example in $\Gamma_1$ is $\gamma(t) = \frac{e^{at}-1}{e^{a}-1}$, with $a \neq 0$.  A few example curves of this type of warping is given in Figure  \ref{fig.simpleGamma}(b).  
We point out that  $\gamma(t) = t^\alpha$ is not in $\Gamma_1$.  Note that the derivative of warping function $\dot \gamma$ is essentially a probability density function on $[0, 1]$, and $\Gamma_1$ is a {\em group} with function composition. Motivated by the Centered Logratio (CLR)  transformation between a density space and a Euclidean space \citep{egozcue2006hilbert}, we aim to build an isometric isomorphism to transform $\Gamma_1$ to a proper Euclidean space.  At first, we need to build an inner-product structure on $\Gamma_1$. 

It is apparent that $\Gamma_1$ has constraints and is not even a vector space under the conventional $\mathbb L^2$ metric.  In this paper, we propose to define perturbation, power, and inner-product operations to make $\Gamma_1$ an inner-product space.  The mathematical proofs in these definitions are straightforward and omitted in this paper.  At first, our definition of linear operators is given as follows:

\begin{definition} (Linear Operations) \ 
	For $f, g \in \Gamma_1$ and $\alpha \in \mathbb{R}$, the perturbation with operator $\oplus_\Gamma: \Gamma_1 \times \Gamma_1 \rightarrow \Gamma_1$ is given by
	$$[f\oplus_\Gamma g](t) = \frac{\int_{0}^{t} \dot{f}(s)\dot{g}(s)\,ds}{\int_{0}^{1} \dot{f}(\tau)\dot{g}(\tau)\,d\tau}.$$
	The power operation with operator  $\odot_\Gamma:\mathbb{R} \times \Gamma_1 \rightarrow   \Gamma_1 $ is given by:
	$$[\alpha\odot_\Gamma f](t) = \frac{\int_{0}^{t} \dot{f}^\alpha(s)\,ds}{\int_{0}^{1} \dot{f}^\alpha(\tau)\,d\tau}. $$
	\label{def:linear}
\end{definition}

\noindent With these operations, $\Gamma_1$ is a vector space. In addition, we can define Euclidean geometry, the inner product, on $\Gamma_1$ to make it an inner-product space. 

\begin{definition} (Inner-Product)
	For $f,g \in \Gamma_1$, the inner product is defined as the functional $\langle\cdot,\cdot\rangle_\Gamma: \Gamma_1 \times \Gamma_1 \rightarrow \mathbb{R}$ in the following form:
	\[\langle f,g\rangle_\Gamma = \int_0^1 \log(\dot{f}(t))\log(\dot{g}(t))\,dt-\int_{0}^{1}\log(\dot{f}(s))\,ds\int_{0}^{1}\log(\dot{g}(t))\,dt.\]
	\label{def:ip}	
\end{definition}
\noindent With the inner-product in Definition \ref{def:ip}, the associated norm and metric distance can be easily defined. 
Based on the CLR transformation result in \citep{egozcue2006hilbert}, we select a Euclidean space under the conventional $\mathbb L^2$ norm in the following form: 
\begin{equation}
	H(0,1)=\Big\{h\in \mathbb L^2([0,1])|\int_{0}^{1} h(t)\,dt =0, -\infty<m_h< h(t)<M_h<\infty\Big\}.
	\label{eq:H}
\end{equation}

\noindent It is easy to see that $H(0,1)$ is a subspace of $\mathbb L^2([0,1])$.  The main result between $\Gamma_1$ and  $H(0,1)$ is given in the following theorem:

\begin{theorem}
	Given the mapping $\psi_B: \Gamma_1\rightarrow H(0,1)$:
	\begin{equation}
		h(t)=\psi_B(\gamma)(t)=\log(\dot{\gamma}(t))-\int_{0}^{1} \log(\dot{\gamma}(s))\,ds,
		\label{eq: twclt}
	\end{equation}
	the space $H(0,1)$ and $\Gamma_1$ are isometric isomorphism (under the linear and inner-product operations).  In particular, the inverse mapping $\psi_B^{-1}: H(0,1) \rightarrow \Gamma_1$ is given by:
	$$\gamma(t)=\psi_B^{-1}(h)(t)=\frac{\int_{0}^{t}\exp(h(s))\,ds}{\int_{0}^{1}\exp(h(\tau))\,d\tau}.$$
	\label{thm:iso1}
\end{theorem}

\subsection{Extension to a Hilbert space}
\label{sec:hilbert}
Hilbert space, a.k.a. complete inner-product space, is a natural extension of finite Euclidean spaces to the infinite-dimensional case.   Because the space is complete, all limiting operations are closed, and techniques in calculus can be directly used. 
In this paper, we focus on using a stochastic process to model time warping functions, where a key step is to transform the warping function into a space with the conventional $\mathbb L^2$ metric, and then the orthonormal basis representations can be fully exploited.  

However, we can see that the space $H(0,1)$ defined in Equation \eqref{eq:H} contains only bounded functions, and is therefore not a Hilbert space.  In this section, we aim to extend it to a Hilbert space in the following form: 
\begin{equation}
	E(0,1)=\Big\{h\in \mathbb L^2([0,1])|\int_{0}^{1} h(t)\,dt =0 \Big\}.
	\label{eq:e01}
\end{equation}
It is easy to verify that $E(0,1)$ is indeed the smallest Hilbert space containing the space $H(0,1)$. Basically, we just remove the lower bound $m_\gamma$ and upper bound $M_\gamma$ in Equation \eqref{eq:H}. 

As $E(0,1)$ is also a subspace of $\mathbb L^2([0,1])$, we can at first find a complete orthonormal system for $\mathbb L^2([0,1])$: 
$$\Big\{h_0(t)=1, h_{2j-1}(t)=\sqrt{2}\sin(2j\pi t), h_{2j}(t)=\sqrt{2}\cos(2j\pi t), j\geq 1,\,t\in[0,1]\Big\}.$$
In Equation \eqref{eq:e01}, the only constraint is that $\int_{0}^{1} h(t)\,dt = 0$.  Therefore, by removing the constant term $h_0(t)=1$, we obtain the complete orthonormal system in the space $E(0,1)$ as follows:
\begin{equation}
	B=\Big\{\phi_{2j-1}(t)=\sqrt{2}\sin(2j\pi t), \phi_{2j}(t)=\sqrt{2}\cos(2j\pi t), j\geq 1, t\in [0,1]\Big\}.
	\label{eq:cons}
\end{equation}

We also extend warping space $\Gamma_1$ in Equation \eqref{eq:gamma1} to the following form:
\begin{equation}
	\Gamma_2 = \Big\{\gamma:[0,1]\rightarrow [0,1]|\dot{\gamma}\in \mathcal F\Big\},
	\label{eq: Gamma2}
\end{equation}
where the $\mathcal F$ space is an extended probability density function space in \citep{egozcue2006hilbert} and given as 
\begin{equation}
	\mathcal {F}=\Big\{f:[0,1]\rightarrow \mathbb{R}|f>0,\, \log f\in \mathbb L^2([0,1])\Big\}.
	\label{eq:pdf space}
\end{equation}
In particular, two functions $f,g\in \mathcal{F}$ are equivalent, if $f=\alpha g$ a.e. in $(0,1)$ with $\alpha>0$. Moreover, if $\int_0^1{f(t)}dt<\infty$, then we take the representative to be the one satisfying $\int_0^1{f(t)}dt=1$, whereas if $\int_0^1{f(t)}dt = \infty$, then we take the representative to be the one satisfying $\int_0^1{\log( f(t))}dt=0$.  

The perturbation and power operations in $\mathcal{F}$ are defined as:
\[[f\oplus g](t) = f(t)g(t),\quad [\alpha \odot f](t) = [f(t)^\alpha], \quad f,g\in \mathcal{F},\, \alpha\in\mathbb{R} \]
Also, the inner product is given by:
\[\langle f,g\rangle = \int_0^1 \log (f(t)) \log (g(t))\,dt-\int_{0}^{1}\log (f(s))\,ds\int_{0}^{1}\log (g(t))\,dt.\]
It was shown in \citep{van2014bayes} that under the CLR transformation from $\mathcal F$ to $E(0,1): clr(f)=\log (f) - \int_{0}^{1}\log (f(s))\,ds$, $\mathcal{F}$ and $E(0,1)$ are isometric isomorphism and the inverse transformation is given by $$clr^{-1}(h)=\begin{cases}
	\frac{\exp (h)}{\int_{0}^{1}\exp (h(s))\,ds}, & \text {if $\int_{0}^{1}\exp (h(s))\,ds<\infty$} \\
	\exp (h),& \text {o.w.}
\end{cases}$$
Thus, if $\{\phi_j\}_{j\geq 1}$ is a set of bounded complete orthonormal basis functions in $E(0,1)$ (e.g. $B$ in Equation \eqref{eq:cons}), then $\{\psi_j\}_{j\geq 1}$, with $\psi_j=\frac{\exp[\phi_j]}{\int_{0}^{1}\exp[\phi_j(s)]\,ds}$, is also a complete orthonormal basis for $\mathcal{F}$.

It is easy to see that $\Gamma_2$ contains $\Gamma_1$.  However, we point out that a derivative operation is not a bijective mapping between $\Gamma_2$ and $\mathcal F$. For any $\gamma \in \Gamma_2$, we have $\dot \gamma \in \mathcal F$.  However, for any $f \in \mathcal F$, if $\int_0^1 f(t)dt = \infty$, then there will not be a $\gamma \in \Gamma_2$ such that $f = \dot \gamma$.  Moreover, the linear operations in $\mathcal F$ cannot be directly used in $\Gamma_2$.  This is obvious because the product of two density functions may not have a finite integration value on $[0, 1]$.

\subsection{Stochastic process model for time warpings}
\label{sec: MBF}
In Section \ref{sec:hilbert}, we have extended the bounded inner-product space $H(0,1)$ to a Hilbert space $E(0,1)$.  We also show that using the CLR transformation, the density function space $\mathcal {F}$ is isometrically isomorphic to the $E(0,1)$ space. This implies we can generate a model on space $E(0,1)$, and then use inverse CLR transformation to project it back into space $\mathcal{F}$. In addition, if the transformed function is integrable, we can project it back to the time warping space $\Gamma_2$. 
In this section, we will provide a second-order stochastic process model for CLR-transformed time warping functions. At first, we will review important background materials to introduce notation and make the content self-contained.

\subsubsection{Review of Mercer's theorem and Karhunen-Lo\`eve expansion}
Mercer's theorem is analogous to the multivariate singular value decomposition to functional variables and is commonly used in the Hilbert space of stochastic processes.  In the function domain of $[0,1]$, this theorem can be described as follows:  \\
\noindent \textbf{Mercer's Theorem:} 
Let the continuous kernel $K: [0, 1] \times [0, 1] \rightarrow \mathbb{R}$ be symmetric and nonnegative definite. Then, there is a complete orthonormal basis, called eigenfunctions, $\{e_i\}\in \mathbb L^2([0,1])$ and associated nonnegative eigenvalues $\{\lambda_i\}$ in decreasing order (i.e., $\lambda_1 \ge \lambda_2 \ge \cdots$) such that
\[K(s,t)=\sum_{i=1}^{\infty} \lambda_i e_i(s)e_i(t)\] 
for all $s,t\in [0,1]$.  Furthermore, the convergence is absolute and uniform.

Based on Mercer's theorem, Karhunen-Lo\`eve expansion represents a stochastic process as an infinite linear combination of orthogonal functions. This method is a covariance-based method, which generalizes the conventional principal component analysis.  With the time domain $[0,1]$, the expansion is given below: \\  
\noindent \textbf{Karhunen-Lo\`eve Expansion:}
Let $X=\{X(t): t\in [0,1]\}$ be a mean-square continuous stochastic process with mean zero and covariance functions $K(s,t)$, and let  $\{e_i\}$ and $\{\lambda_i\}$ be the eigenfunctions and eigenvalues using the kernel $K(s,t)$ in Mercer's theorem. Then $X(t)$ admits the following representation:
\[X(t)=\sum_{i=1}^{\infty}Z_i e_i(t),\]
where the convergence is uniform in $\mathbb L^2$ and coefficients $Z_i = \int_{0}^1 X(t) e_i(t)dt$.  Furthermore, random variables $Z_i$ have zero-mean, are uncorrelated and have variance $\lambda_i, i = 1, 2, \cdots$. 
In particular, if the process is a Gaussian process, then $Z_i$ are independent normal random variables with mean 0 and variance $\lambda_i$. 

\subsubsection{Model time warping via second-order stochastic process}
\label{model:MBF}
In this section, we will develop a new procedure to model time warping functions in $\Gamma_1$.  By the isometric isomorphism, we only need to model functions in $H(0,1)$.  The modeling process is based on second-order stochastic process representation in the Hilbert space $E(0,1)$ (smallest extension of $H(0,1)$).     

According to the Karhunen-Lo\`eve expansion, any mean-square continuous stochastic process can be represented as an infinite linear combination of a set orthonormal basis in $\mathbb L^2([0,1])$, where the basis functions are eigenfunctions for the covariance kernel $K(s,t)$. Thus, to model the process, we focus on building an appropriate kernel function. In principle, any complete orthonormal basis can be used. To characterize bounded functions in $H(0,1)$, we adopt the bounded Fourier set $B$ in Equation \eqref{eq:cons}.  The construction of the kernel is given in the following proposition, where the detailed proof is given in Appendix A. 
\begin{prop}
	For any non-negative sequence $\{\mu_i\}_{i=1}^{\infty}$ such that $\sum_{i=1}^{\infty}\mu_i<\infty$, let 
	$$K(s,t)=\sum_{i=1}^{\infty}\mu_i \phi_i(s) \phi_i(t), \ \ \  \text{ for all } s,t \in [0,1],$$ 
	where $\{\phi_i\}_{i=1}^\infty$, given in Equation \eqref{eq:cons}, is the complete orthonormal system in $E(0,1)$.  
	Then 1) $K$ converges absolutely and uniformly, and 2) $K$ is a continuous, symmetric, non-negative definite function.
	\label{prop}
\end{prop}

\noindent {\bf Remark:} By Proposition  \ref{prop}, we can easily construct a kernel function using convergent sequence $\{\mu_i\}$ and basis $\{\phi_i\}$.  Note that $\{\mu_i\}$ do not necessarily follow the decreasing order.  By the uniqueness of eigenvalues $\{\lambda_i\}$ in Mercer's theorem, $\{\lambda_i\}$ are in fact the ordered sequence (from large to small) of  $\{\mu_i\}$. 
In practical use, common choices of $\{\mu_i\}$ are $\mu_i = \frac{1}{i^s}$, or $\frac{1}{i (\log(i))^s}$, with $s \ge 2$.

Based on the above result, we can simulate a random process in $E(0,1)$ as follows:  Given the orthonormal basis $B = \{\phi_i(t)\}$ and non-negative sequence $\{\mu_i\}_{i=1}^{\infty}$ with convergent sum, we can generate a mean-centered second-order process $X$ in the following form: 
\begin{equation}
	X(t)=\sum_{i=1}^{\infty} G_i \phi_i(t)
	\label{eq:model}
\end{equation}
where $G_i$ are uncorrelated random variables with mean 0 and variance $\mu_i$.  Note that there is no constraint for the type of distribution on $G_i$, which fully characterizes the randomness in $X(t)$.  We can choose any distribution to explore all possible variabilities.  In particular, to generate a Gaussian process, we only need to set $G_i \sim N(0, \mu_i)$.

In practice, we can only simulate a second-order stochastic process $X(t)$ in Equation \eqref{eq:model} with finite $m$ terms in the sum.  That is, $\mu_i = 0$ and $G_i = 0$ when $i > m$.  In this case, it is always true that $\sum_{i=1}^\infty \mu_i < \infty$ and the corresponding covariance kernel $K$ is well-defined. This truncated version can be written as:
\begin{equation}
	X_{m}(t)=\sum_{i=1}^{m} G_i \phi_i(t)
	\label{eq:truKL}
\end{equation} 
As $\phi_i$ are all bounded functions, $X_m(t)$ is also bounded and therefore in $H(0,1)$.  By the isometric isomorphism between $H(0,1)$ and $\Gamma_1$, we can transform $X_m(t)$ to build a warping function.  In summary, the generative model of time warping function in $\Gamma_1$ is given in Algorithm \ref*{alg:MBF}. 

\begin{algorithm}[h]
	\caption{Generative model for warping function in $\Gamma_1$}
	\begin{algorithmic} 
		\Require Basis functions $\{\phi_i\}_{i=1}^m$ in Equation \eqref{eq:cons}. 
		\State Generate coefficient sequence $G_i$ with any probability distribution with mean 0 and variance $\mu_i, i = 1, \cdots, m$. 
		\State $X_{m}(t) = \sum_{i=1}^{m} G_i \phi_i(t)$
		\State $\gamma_m(t) =\frac{\int_{0}^{t}\exp(X_m(s))\,ds}{\int_{0}^{1}\exp(X_m(\tau))\,d\tau}$
		\State Output $\gamma_m$
	\end{algorithmic}
	\label{alg:MBF}
\end{algorithm}

\noindent {\bf Remark:} If we allow $m=\infty$ in Algorithm \ref{alg:MBF}, then we will need to add two conditions to simulate warping function: 1) $\sum_{i=1}^\infty \mu_i < \infty$, and 2) $\int_0^1 \exp \big(X_m(s)\big)ds  < \infty$.  In this case, the simulated warping function may not be in $\Gamma_1$ (i.e., bounded), but will be in $\Gamma_2$ as defined in Equation \eqref{eq: Gamma2}.

\subsubsection{Illustration}
\begin{figure}[h]
	\centering
	\begin{subfigure}[h]{0.22\textwidth}
		\includegraphics[width=\textwidth]{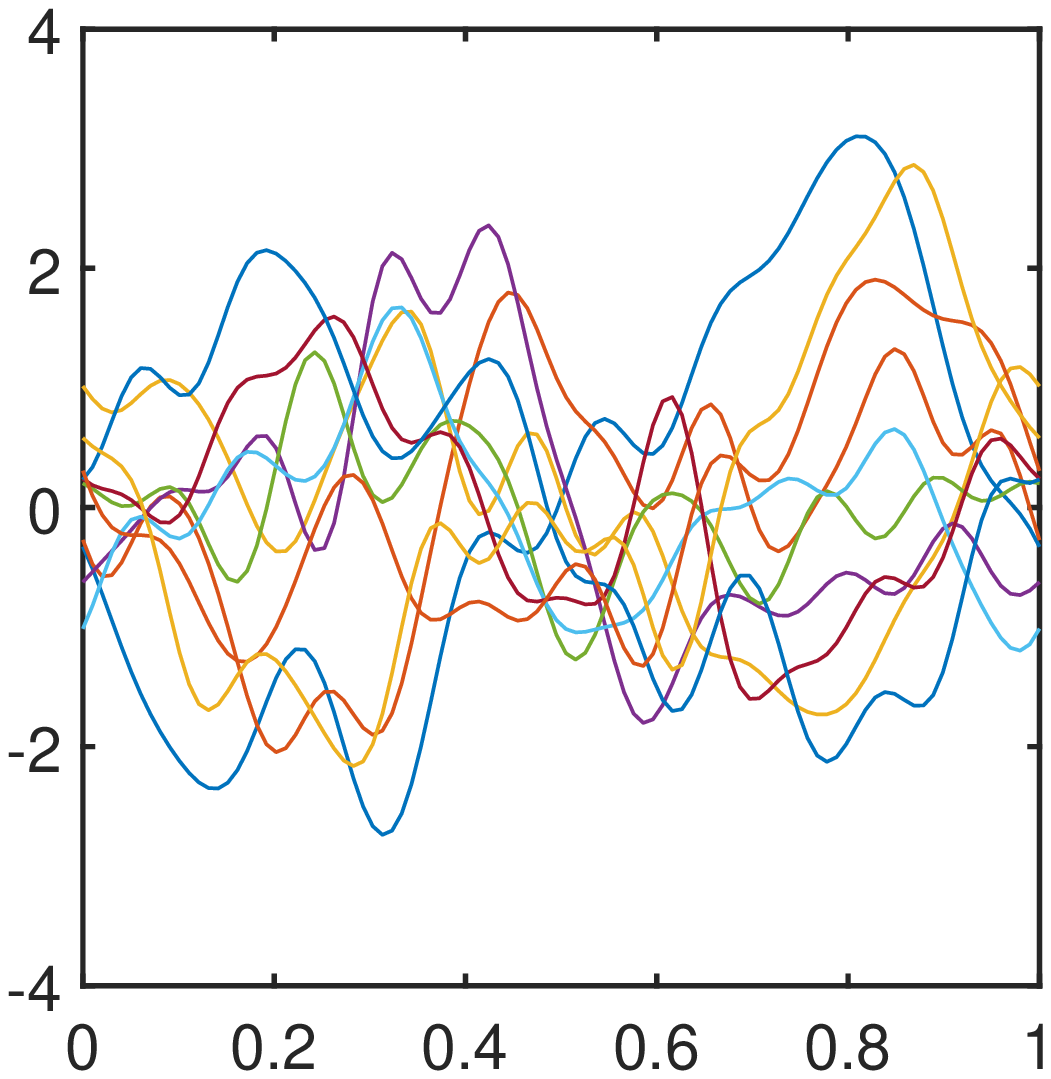}
		\vfill
		\includegraphics[width=\textwidth]{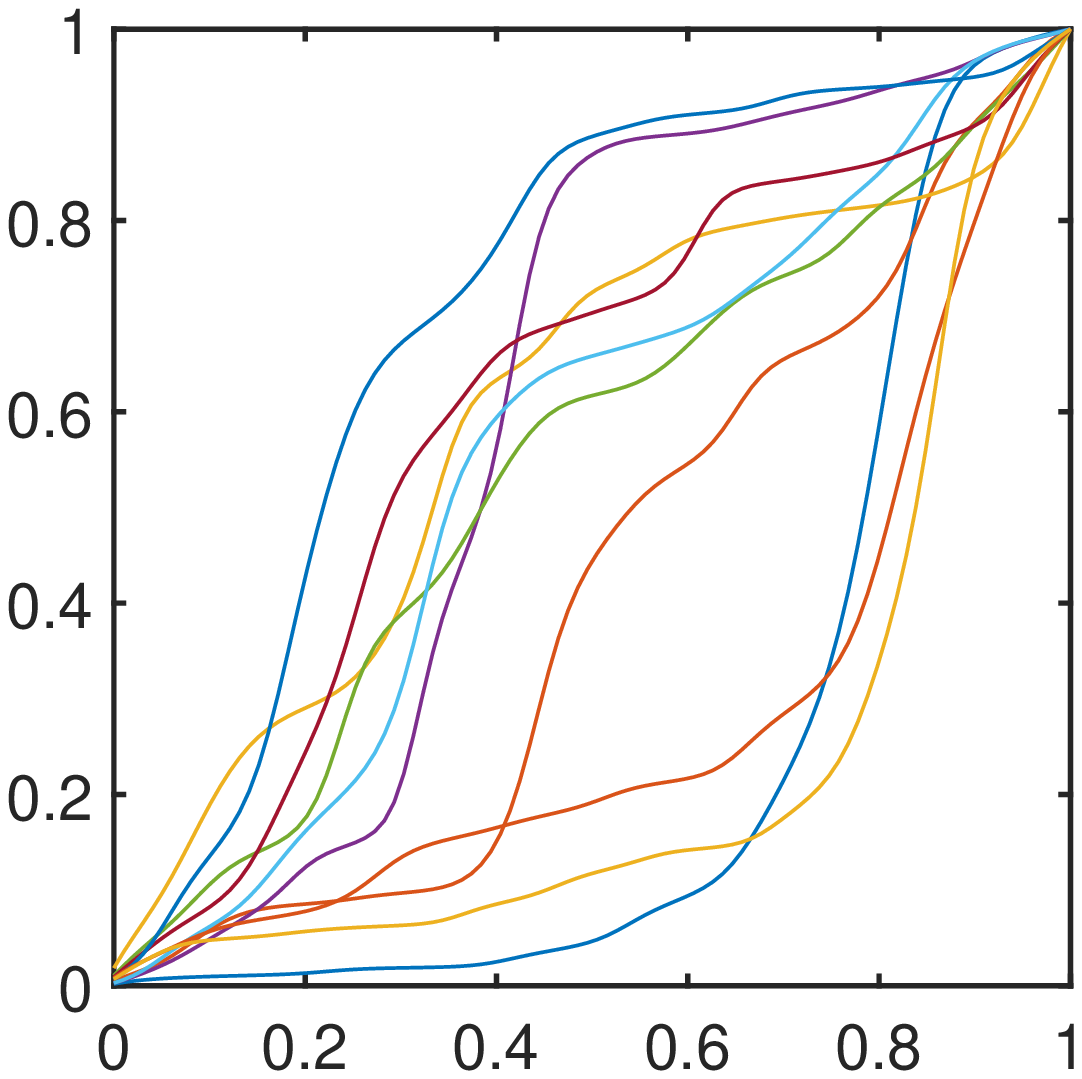}
		\caption{}
	\end{subfigure}
	\hspace{-0.7cm}
	\begin{subfigure}[h]{0.22\textwidth}
		\includegraphics[width=\textwidth]{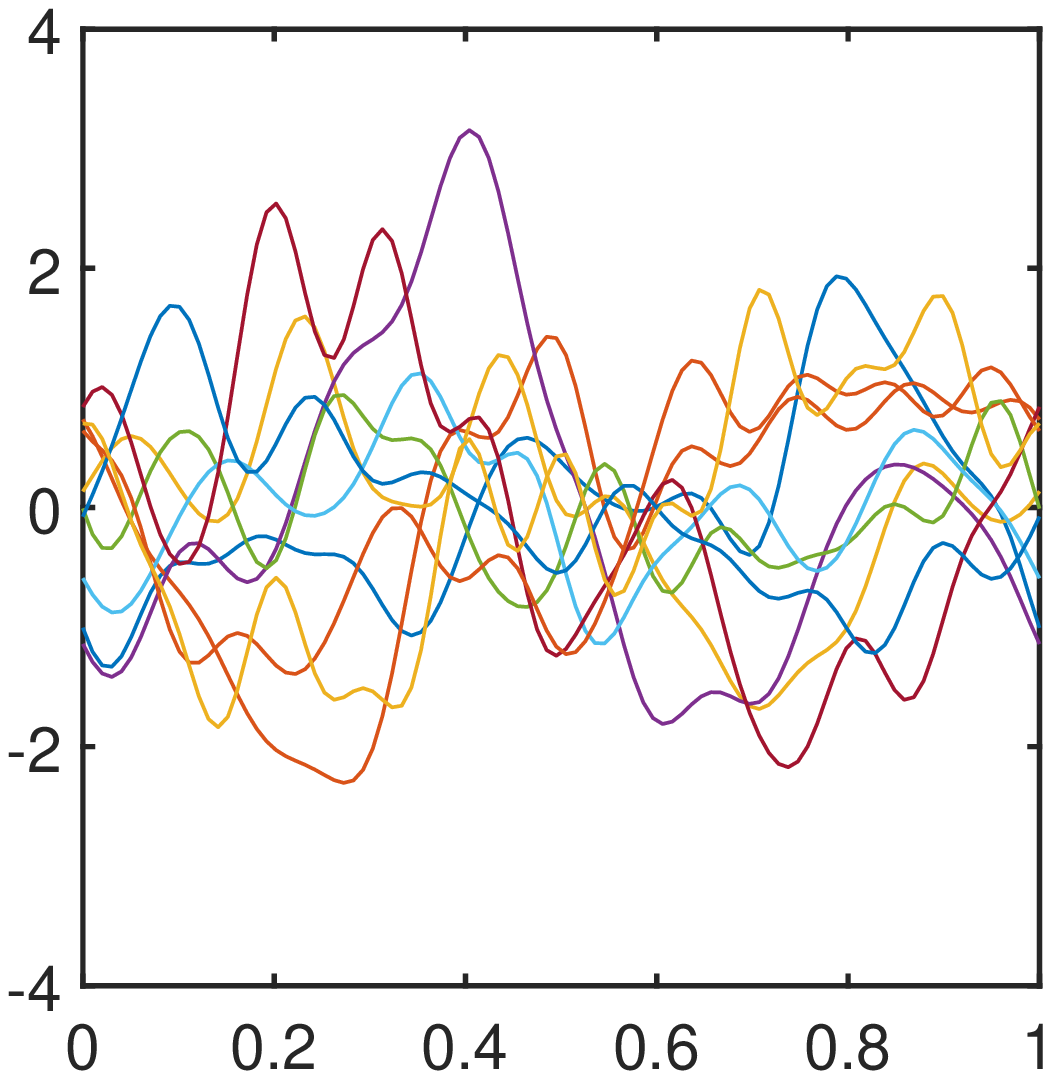}
		\vfill
		\includegraphics[width=\textwidth]{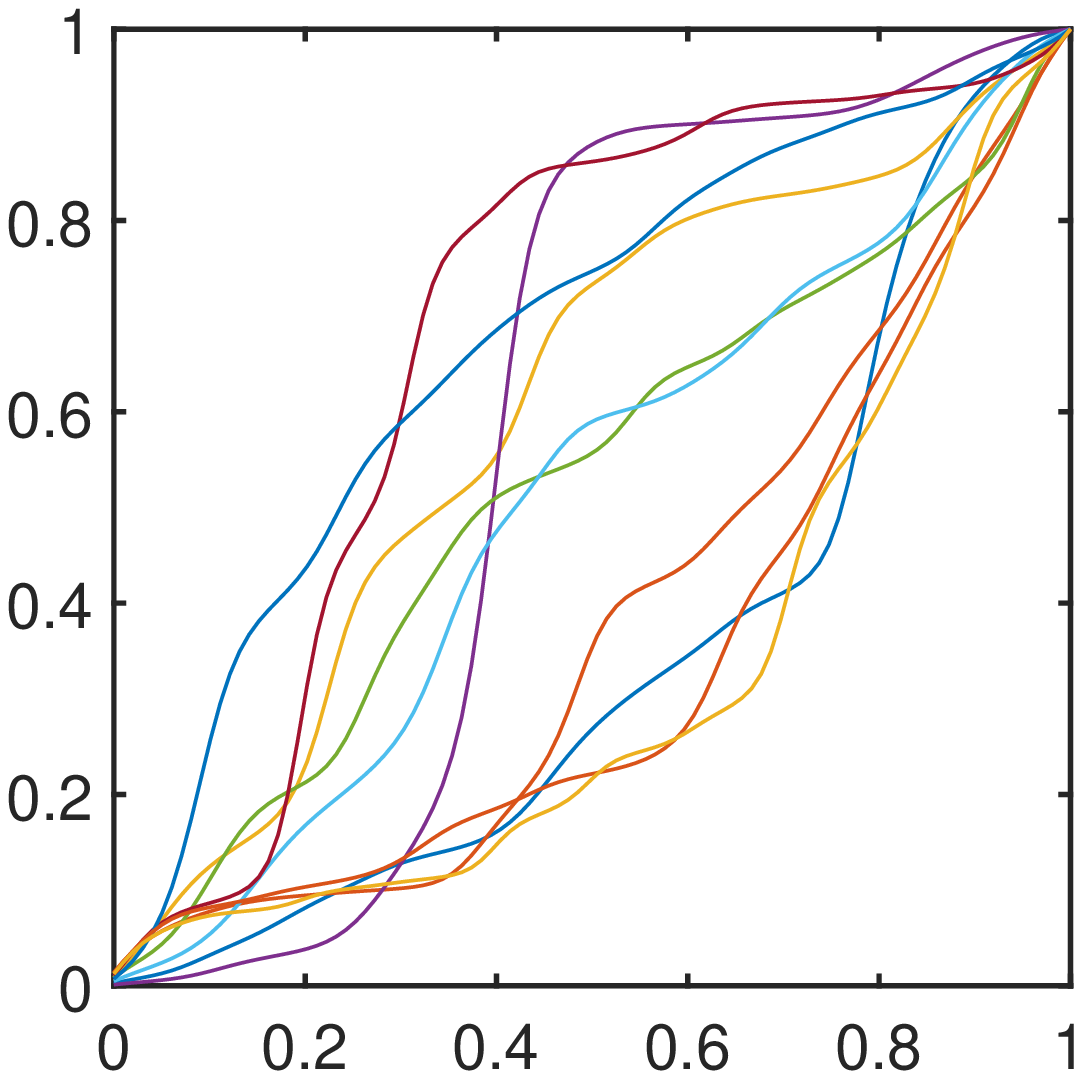}
		\caption{}
	\end{subfigure}	
	\hspace{-0.7cm}
	\begin{subfigure}[h]{0.22\textwidth}
		\includegraphics[width=\textwidth]{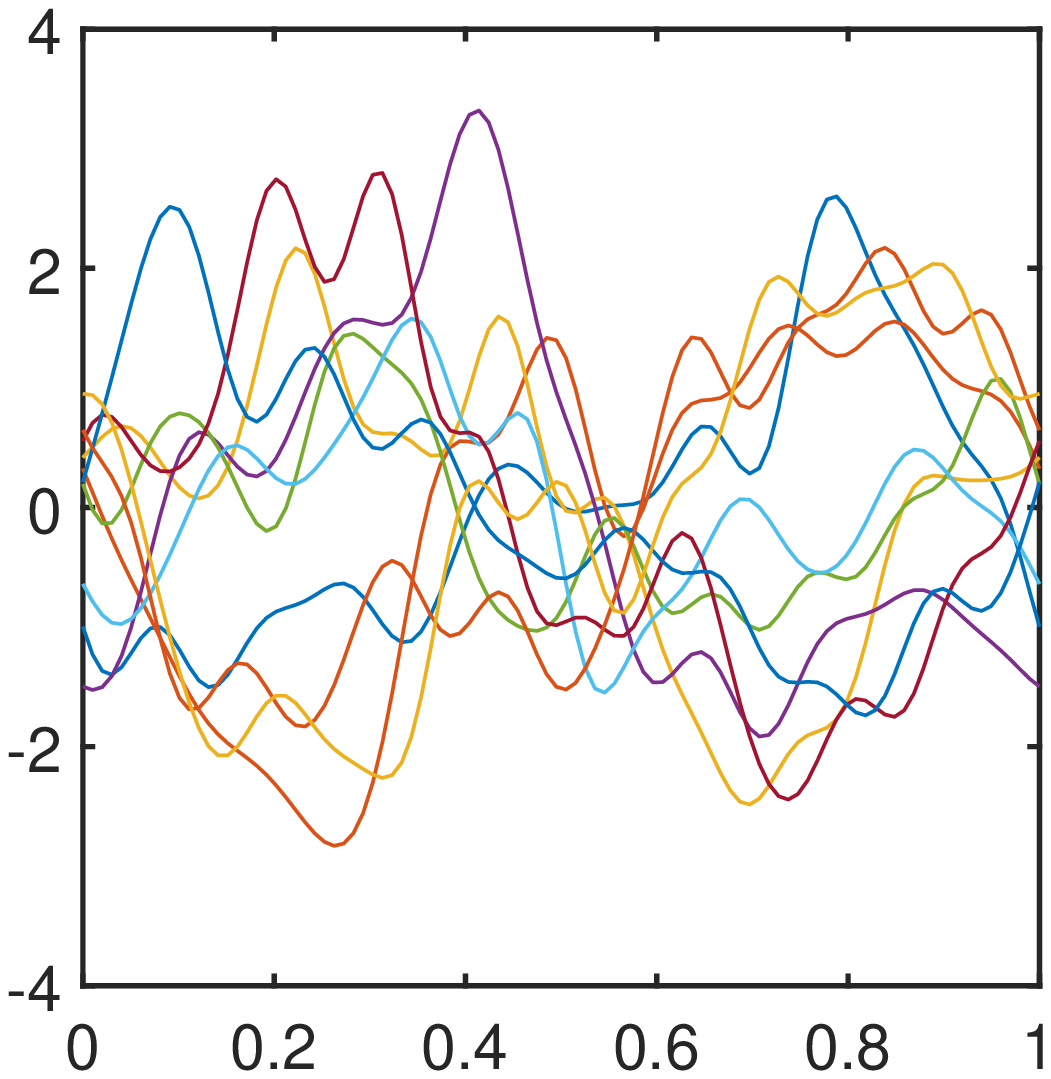}
		\vfill
		\includegraphics[width=\textwidth]{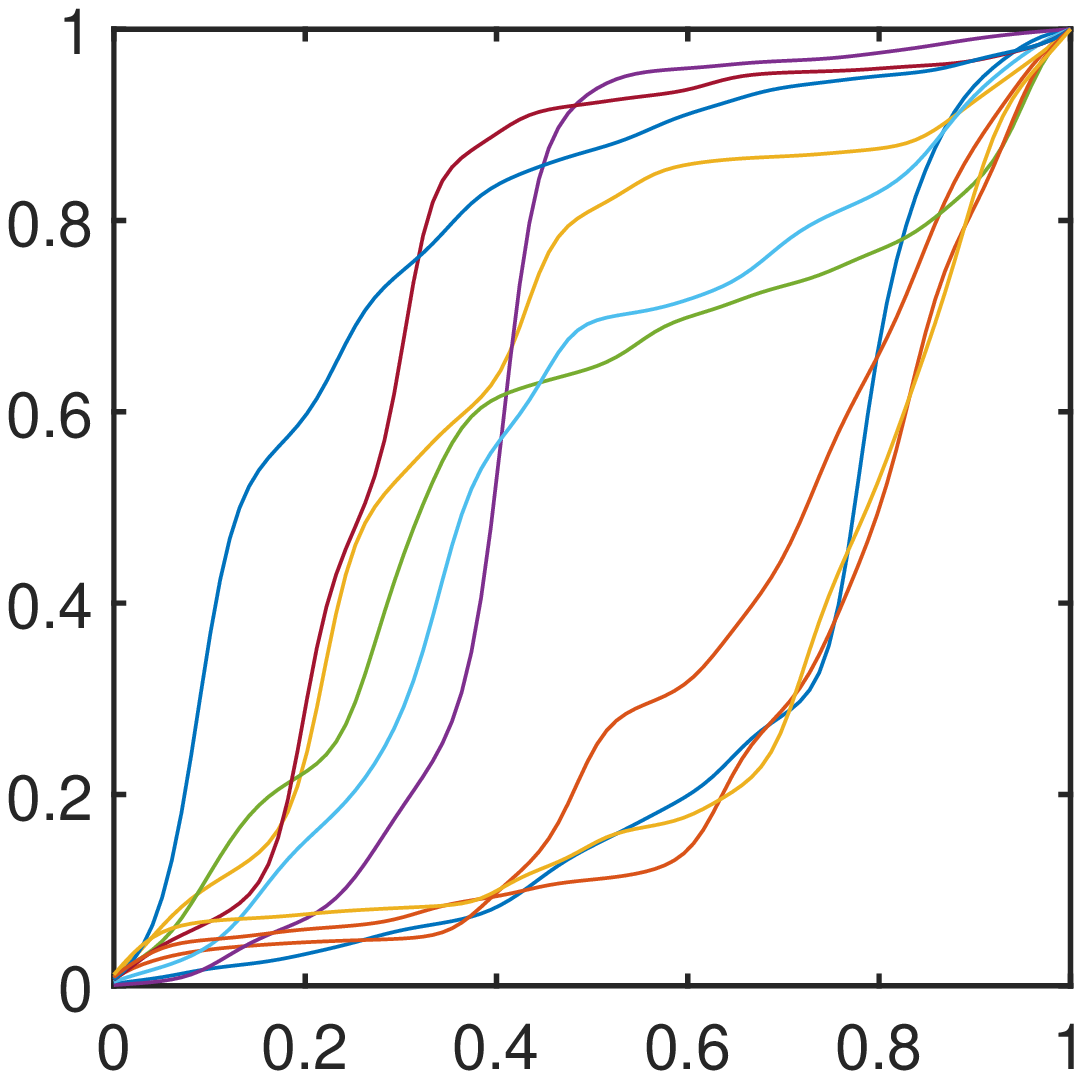}
		\caption{}
	\end{subfigure}
	\hspace{-0.7cm}
	\begin{subfigure}[h]{0.22\textwidth}
		\includegraphics[width=\textwidth]{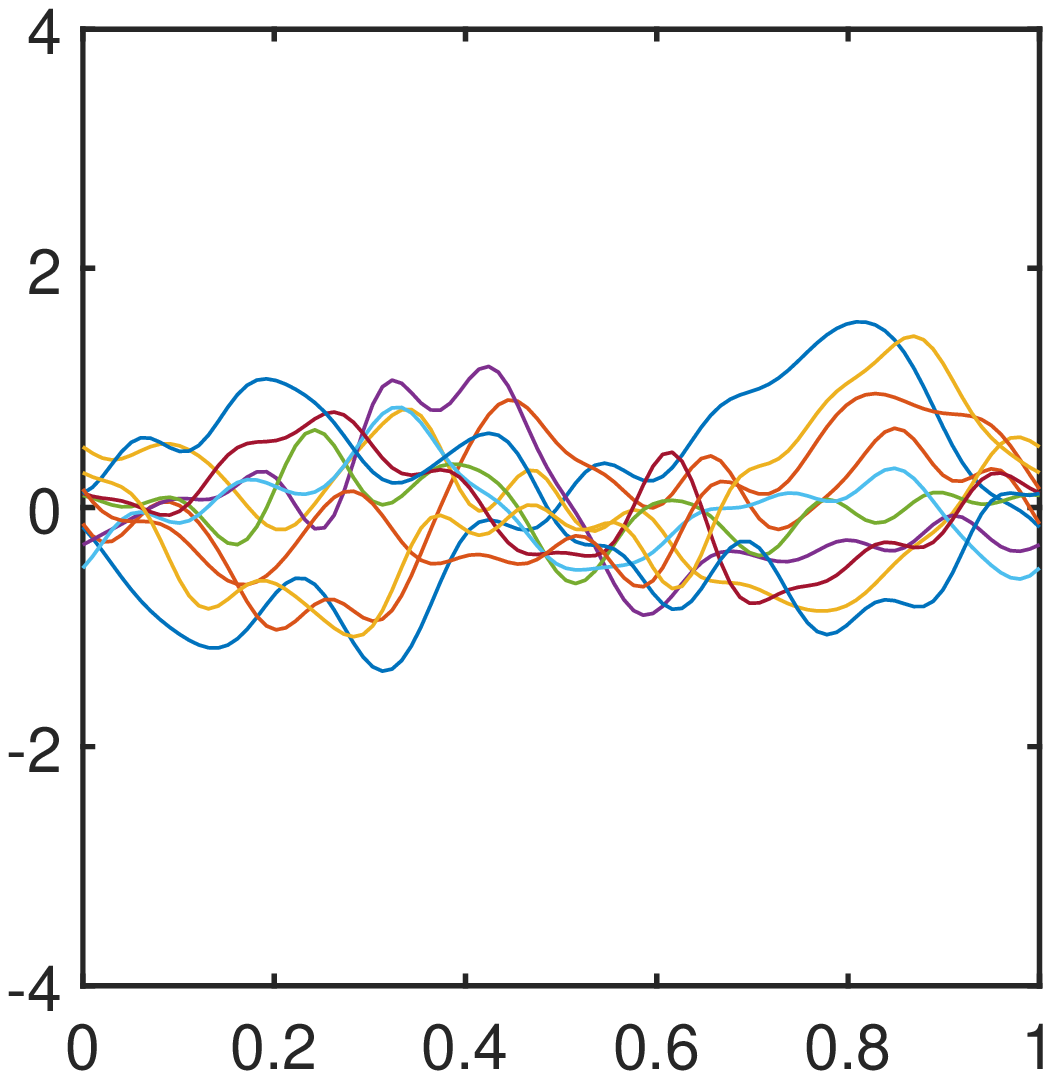}
		\vfill
		\includegraphics[width=\textwidth]{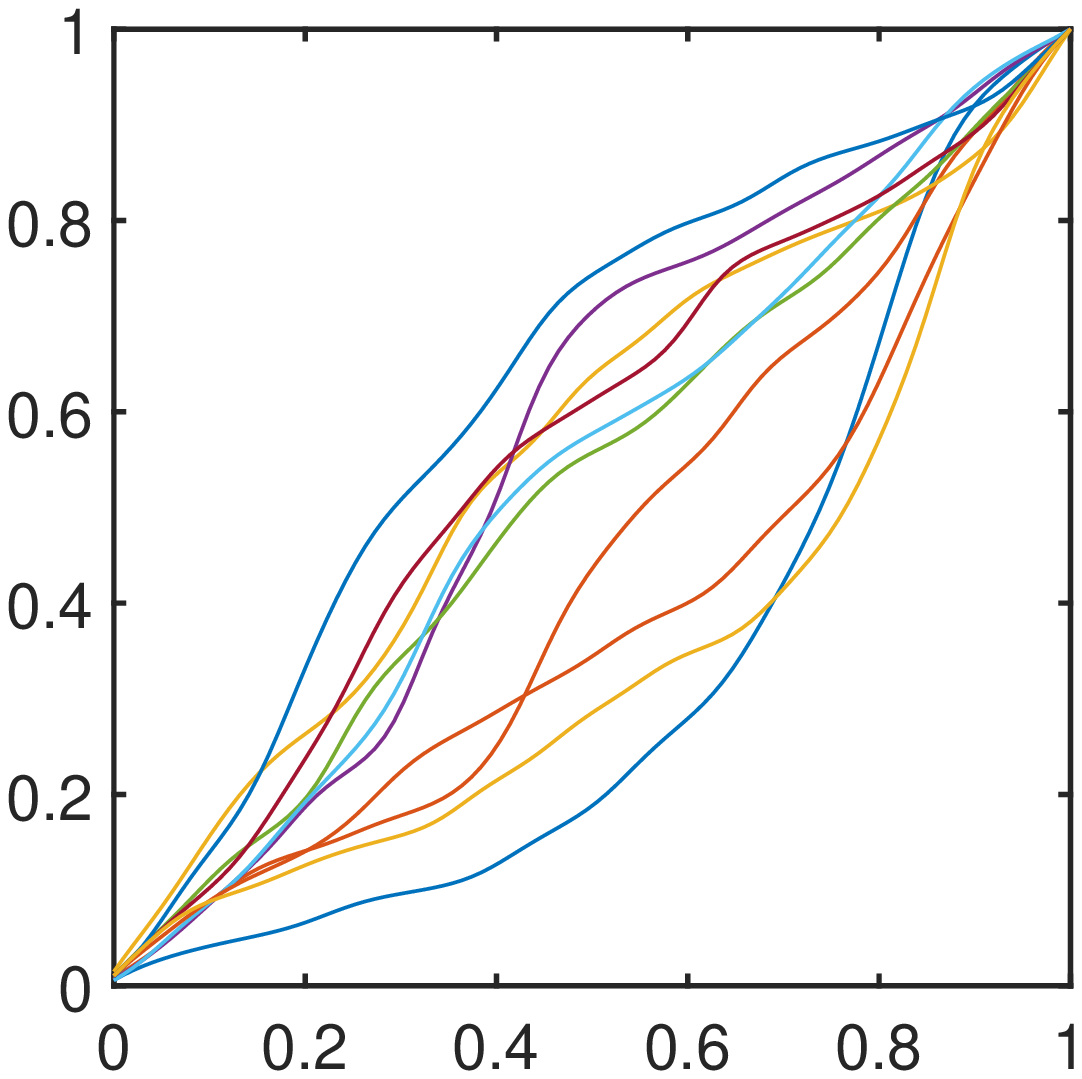}
		\caption{}
	\end{subfigure}
	\hspace{-0.7cm}
	\begin{subfigure}[h]{0.22\textwidth}
		\includegraphics[width=\textwidth]{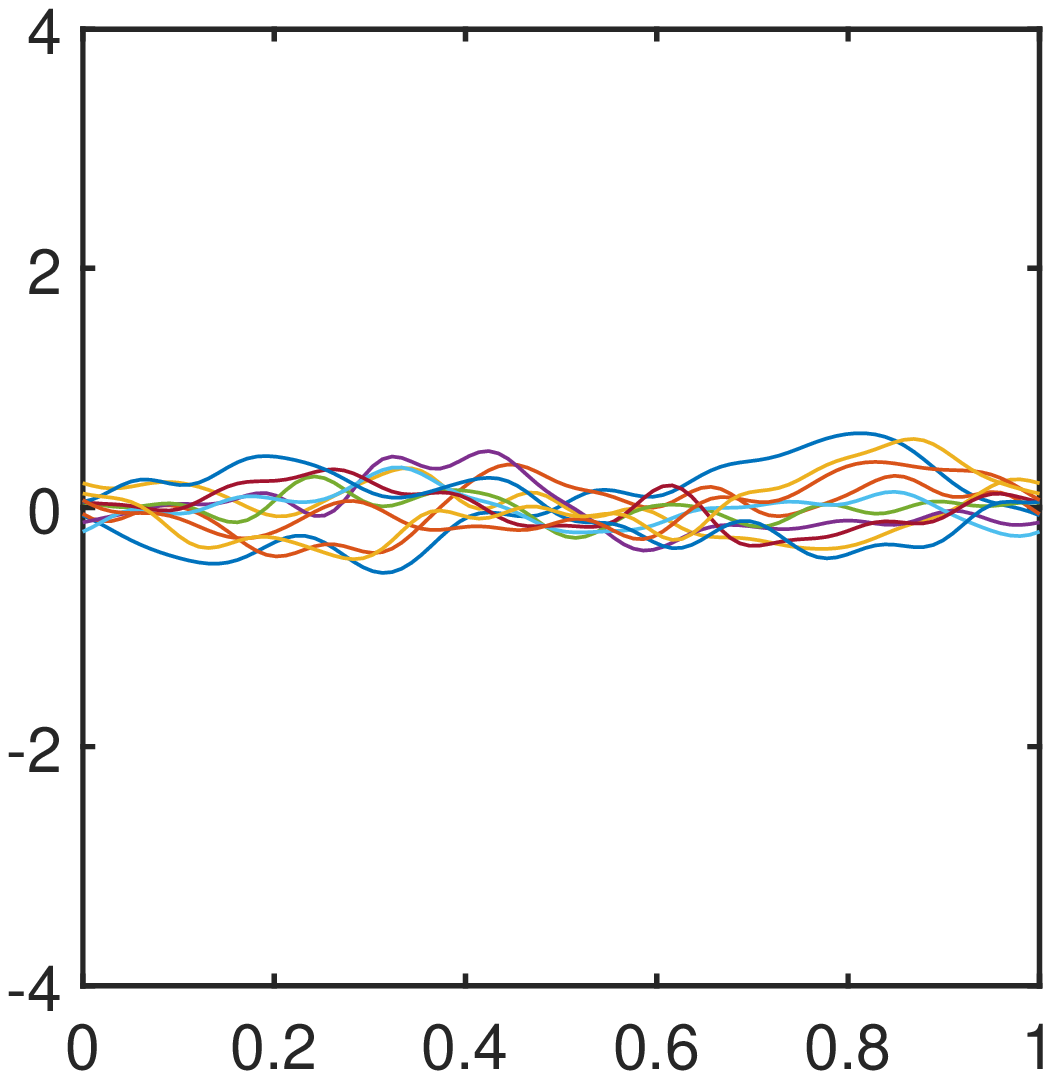}
		\vfill
		\includegraphics[width=\textwidth]{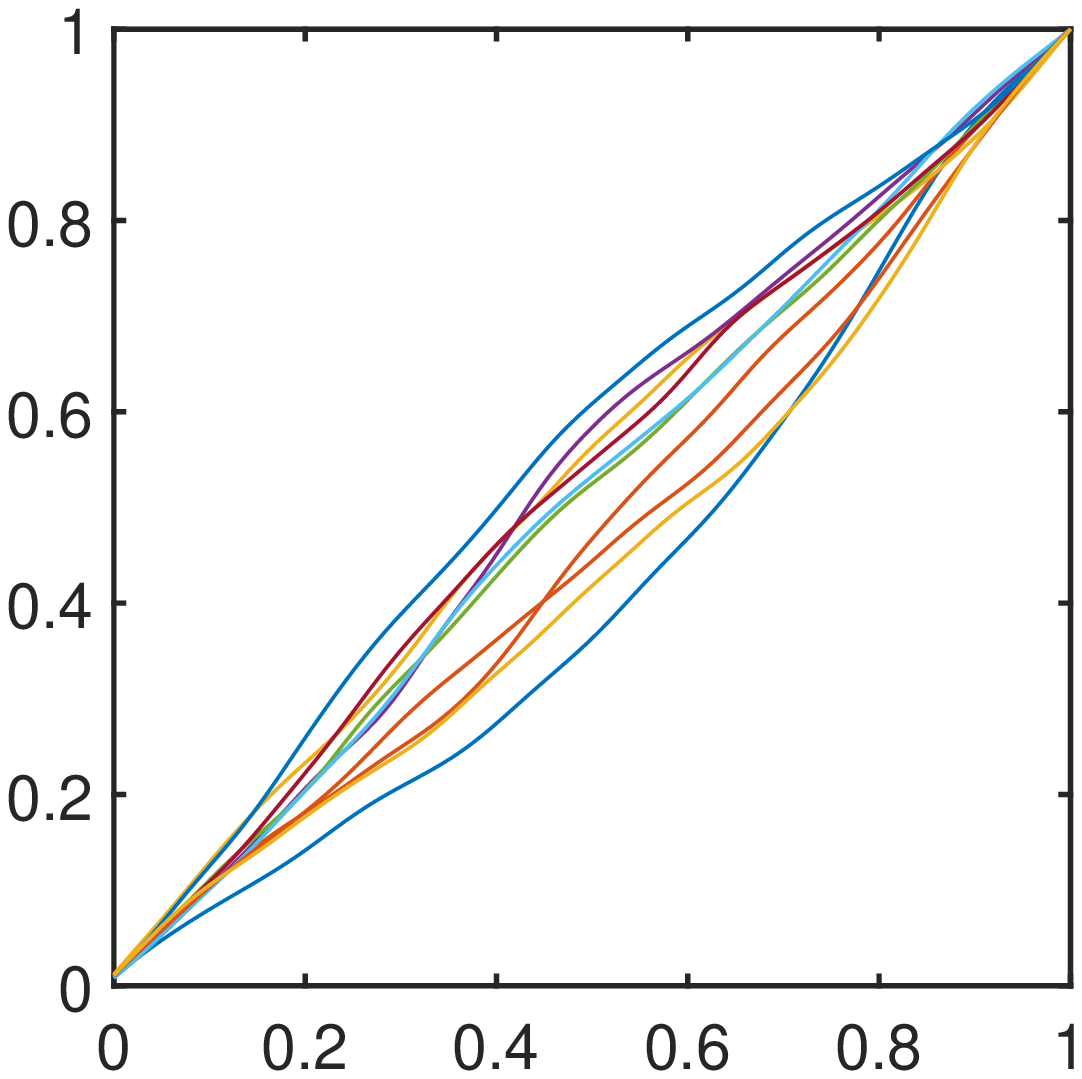}
		\caption{}
		\label{e}
	\end{subfigure}	
	\caption{Simulation examples using Algorithm \ref{alg:MBF}, with 10 stochastic processes in $H(0,1)$ in the top row and the corresponding 10 time warping functions in $\Gamma_1$ in the bottom row.  Column (a): $G_i\sim N(0, (1/i)^2)$, Column (b) $G_i\sim La(0, 1/(\sqrt{2}i))$, Column (c) $G_i\sim U(-\sqrt{3}/i, \sqrt{3}/i)$, Column (d) $G_i\sim N(0,(1/(2i))^2)$, Column (e) $G_i\sim N(0,(1/(5i))^2)$. }
	\label{fig: generative model}
\end{figure}

We now illustrate Algorithm \ref{alg:MBF} with $m = 20$, where the coefficients $\{G_i\}_{i=1}^{20}$ are from each of the following 5 different distributions:
\begin{enumerate}[(a)]
	\item $G_i\sim N(\mu_i,\sigma_i^2)$, where $\mu_i=0$, and $\sigma_i=\frac{1}{i}$, i.e., normal distribution with mean 0 and variance $\frac{1}{i^2}$ 
	
	\item $G_i\sim La(\mu_i,b_i)$, where $\mu_i=0$, $b_i=\frac{1}{\sqrt{2}i}$, i.e., Laplacian distribution with mean 0 and variance $\frac{1}{i^2}$ 
	
	\item $G_i\sim U(a_i,b_i)$, where $a_i=-\frac{\sqrt{3}}{i}$, $b_i=\frac{\sqrt{3}}{i}$, i.e., uniform distribution with mean 0 and variance $\frac{1}{i^2}$ 
	
	\item $G_i\sim N(\mu_i,\sigma_i^2)$, where $\mu_i=0$, and $\sigma_i=\frac{1}{2i}$, i.e., normal distribution with mean 0 and variance $\frac{1}{(2i)^2}$ 
	
	\item $G_i\sim N(\mu_i,\sigma_i^2)$, where $\mu_i=0$, and $\sigma_i=\frac{1}{5i}$, i.e., normal distribution with mean 0 and variance $\frac{1}{(5i)^2}$ 
\end{enumerate}
In each of these 5 cases, we generate 10 stochastic processes.  The results are shown in Figure \ref{fig: generative model}.  It is easy to see that the simulated warping functions have more variabilities than the previous example $\gamma(t) = \frac{e^{at}-1}{e^a-1}$ in Figure \ref{fig.simpleGamma}(b).
The first three columns show time warpings and their corresponding functions in $H(0,1)$ from one Gaussian process (Column (a)) and two non-Gaussian processes (Columns (b) and (c)).  The warping functions exhibit different types of variabilities, whereas the degrees of warping look similar as they share the same variances for the coefficients $G_i$. As a comparison, we also show two other Gaussian processes ((Columns (d) and (e)) with smaller variances.  It is obvious that when the variance gets smaller, the corresponding warping functions stay closer to the identity function $\gamma_{id}(t) = t$. 

\subsection{Estimation of basis functions on given observations}
\label{fPCA}

In Section \ref{sec: MBF}, we have developed a stochastic process framework to represent the time warping functions.  This representation is based on a complete orthonormal system in the $\mathbb L^2$ space.  In practice, we may look for an alternative basis with the given observations.  In this section, we will explore modeling with the functional principal component analysis (fPCA) method.

\subsubsection{Modeling and resampling via fPCA}
fPCA is a basis representation method in the Euclidean space.  When time warping observations in $\Gamma_1$ are given, we may transform them (stated in Theorem \ref{thm:iso1}) into the $H(0,1)$ space and then estimate orthonormal basis via the fPCA method.  This fPCA method has been exploited in \citep{happ2019general}, where the goal was dimension reduction on functional data.  In this paper, we will extend the idea to model and resample warping functions. 

Suppose we have $N$ independent and identically distributed time warping functions in $\Gamma_1$.  By applying the mapping defined in Theorem \ref{thm:iso1}, we can get $N$ continuous functions, denoted as $\{X_n\},\, n = 1,2,\cdots,N$, on $H(0,1)$. We can at first calculate the sample mean $m =\frac{1}{N}\sum_{k=1}^{N}X_n$ and the sample covariance $K(s,t)=\frac{1}{N-1}\sum_{n=1}^{N}(X_n(s)-m(s))(X_n(t)-m(t))$ of these $N$ functions. Then, we apply the eigen-decomposition to the covariance kernel $K$, and find the corresponding eigenpairs $(\lambda_i, e_i)_{i=1}^N$. According to Karhunen-Lo\`eve Expansion, $X_n$ can be represented in the following finite linear combination form: 
\[X_n=m +\sum_{i=1}^{N} Z_{ni} e_i\] 
where coefficients $Z_{ni}=\langle (X_n-m),e_i\rangle = \int_0^1(X_n(t)-m(t))e_i(t) dt$. 

Instead of reconstructing the time warping functions by transforming these reconstructed second-order processes back to $\Gamma_1$ as in \citep{happ2019general}, we will further analyze the coefficients $Z_{\cdot i}$ and estimate their distribution, denoted as $D_i$. Then, we can do resampling of functions in $H(0,1)$ as follows: 
\begin{equation}
	X_{new}= m + \sum_{i=1}^{N} G_i e_i.
	\label{equ: truKLpca}
\end{equation}
where $G_i$ are independent random samples from $D_i$.  Note that we can often take only the first few terms of the linear combinations, where the variance in $D_i$ are still significantly important (e.g. with a 95\% cutoff on cumulative variance in fPCA). 
Finally, by the isometric isomorphism between $H(0,1)$ and $\Gamma_1$, we can transform $X_{new}$ to get a warping function. 
In summary, the fPCA modeling and resampling of time warping function in $\Gamma_1$ is given in Algorithm \ref{alg:MPCA}.

\begin{algorithm}[h]
	\caption{Modeling and resampling with fPCA}
	\begin{algorithmic} 
		\Require $N$ observed warping functions $\gamma_n$ in $\Gamma_1$
		\State Transform the warping functions $\gamma_n$ in to $H(0,1)$: $X_n(t) =\log(\dot{\gamma_n}(t))-\int_{0}^{1} \log(\dot{\gamma_n}(s))\,ds $ 
		\State Calculate the mean $\hat{\mu}(t) = \frac{1}{N} \sum_{n=1}^{N}X_n(t)$ and the covariance $\hat{K}(s,t) = \frac{1}{N-1}\sum_{n=1}^{N}(X_n(s)-\hat{\mu}(s))(X_n(t)-\hat{\mu}(t))$ 
		\State Apply spectral decomposition to $\hat{K}$ to get the eigen sequence $\{(\lambda_i, e_i)\}_{i=1}^N$.  
		\State Find cutoff threshold $\delta$, let $m = \max \{i | \lambda_i > \delta\}$.
		\For {$k = 1: m$}
		\State Calculate the coefficients $Z_{nk}= \int_0^1 (X_n(t)-\hat{\mu}(t)) e_k(t) dt, n=1,2,\cdots N.$
		\State Use the sample $\{Z_{nk}\}_{n=1}^N$ to estimate their distribution $D_k$.
		\EndFor		
		\For {$r = 1: R$} \ \ (resample warping functions)
		\State Simulate coefficient $G_k$ using the estimated distribution $D_k$. 
		\State $X_{r}(t) = \hat{\mu}(t) + \sum_{k=1}^{m} G_k e_k(t)$	
		\State $\gamma_{r}(t) =\frac{\int_{0}^{t}\exp(X_{r}(s))\,ds}{\int_{0}^{1}\exp(X_{r}(\tau))\,d\tau}.$ 
		\EndFor
		\State Output $\{\gamma_{r}(t)\}_{r=1}^R$  
	\end{algorithmic}
	\label{alg:MPCA}
\end{algorithm}

\noindent {\bf Remark:} We simplify the resampling process in Algorithm \ref{alg:MPCA} by assuming the coefficients $\{G_i\}_{i=1}^N$ are independent and then generate samples independently.  This is true if the process is a Gaussian process.  However, in the framework of  Karhunen-Lo\`eve expansion, they are, in general, only uncorrelated. In practical use, we may need to model the coefficients simultaneously for a more appropriate resampling.

\subsubsection{Resampling examples}
We will use two examples to illustrate Algorithm \ref{alg:MPCA}.  In the first example, only one component is significant.  In the second one, there are multiple significant components. 

{\bf Simulation 1:} In this example, 500 warping functions are generated as: $\gamma_i(t)=\frac{e^{a_it}-1}{e^{a_i}-1},\,i=1,2,\cdots 500$, where $a_i$ is a random variable following each of the 3 different distributions (i.e., exponential, Laplacian, and piecewise uniform) given below:
\[
\begin{array}{llllll}
	(a) \ Exp(2)  & (b) \ La(2, 2\sqrt{2}) & (c) \ U((-2,-1)\cup(1,2)). 
\end{array}
\]

\begin{figure}[h]
	\centering
	\begin{subfigure}[h]{1\textwidth}
		\centering
		\begin{subfigure}[h]{0.24\textwidth}
			\raisebox{-\height}{\includegraphics[width=\textwidth]{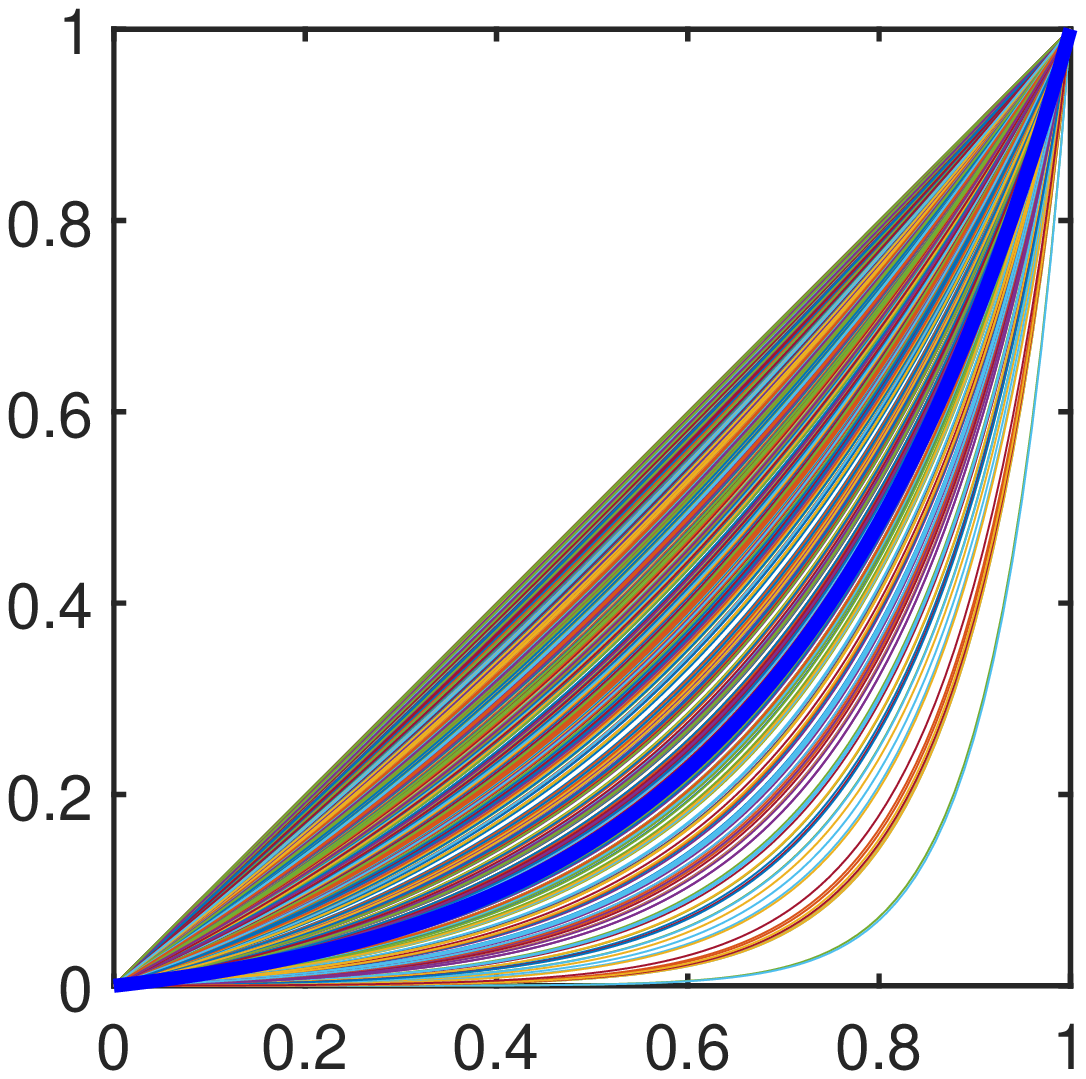}}
		\end{subfigure}%
		\begin{subfigure}[h]{0.24\textwidth}
			\raisebox{-\height}{\includegraphics[width=\textwidth]{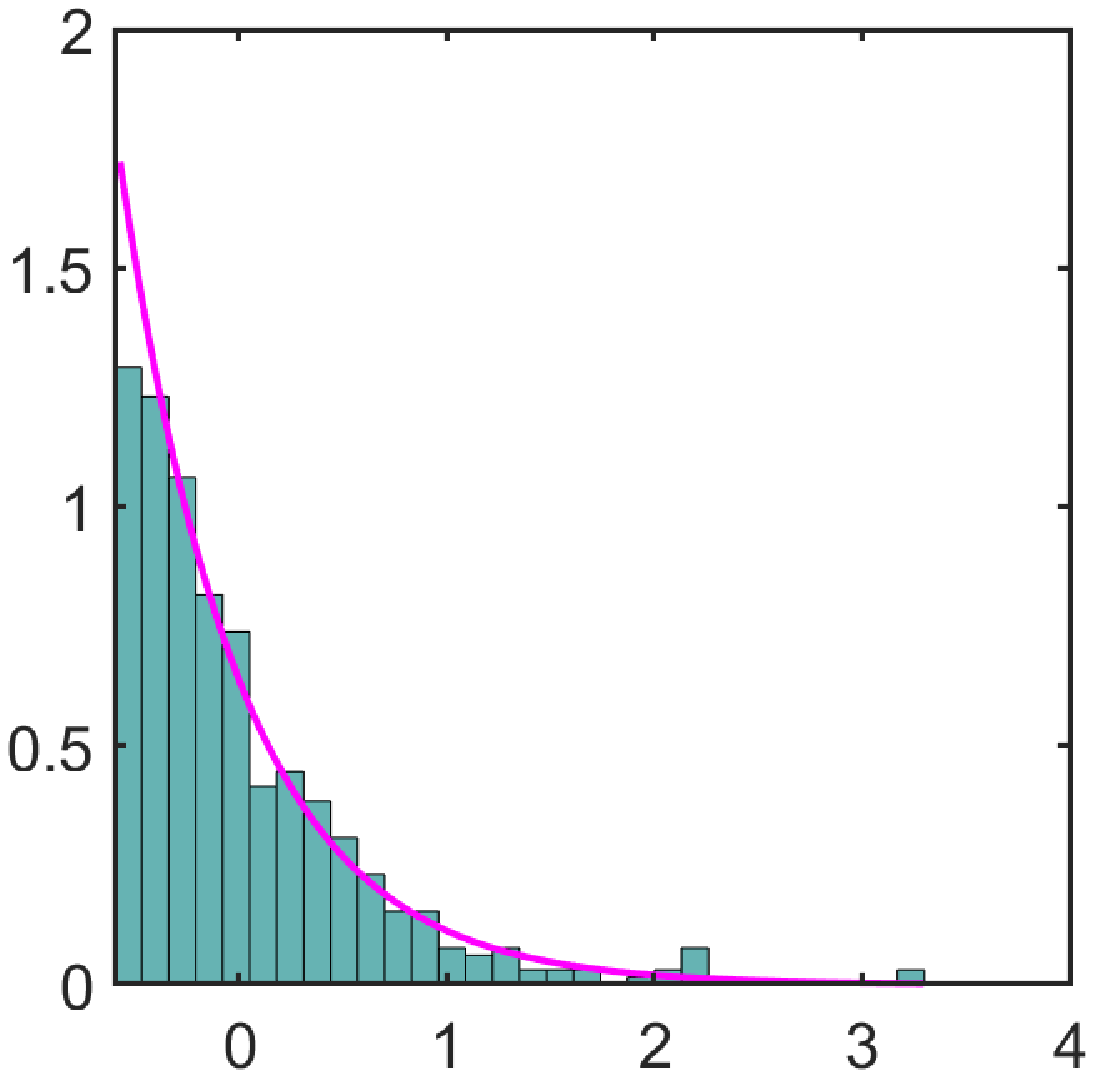}}
		\end{subfigure}%
		\begin{subfigure}[h]{0.24\textwidth}
			\raisebox{-\height}{\includegraphics[width=\textwidth]{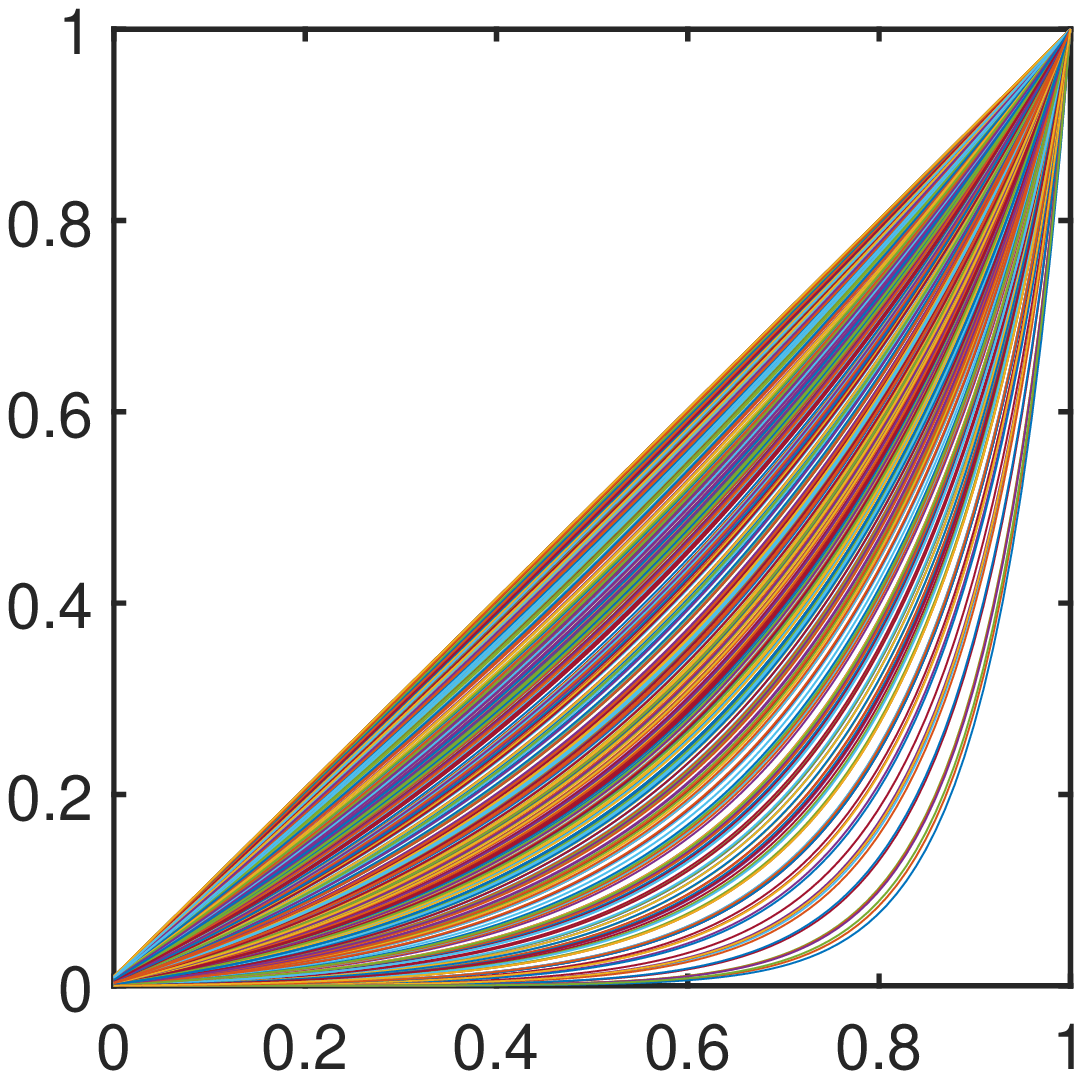}}
		\end{subfigure}%
		\caption{$a_i\sim Exp(2)$}
		\label{fig:s1exp}
	\end{subfigure}%
	\qquad
	\begin{subfigure}[h]{1\textwidth}
		\centering
		\begin{subfigure}[h]{0.24\textwidth}
			\raisebox{-\height}{\includegraphics[width=\textwidth]{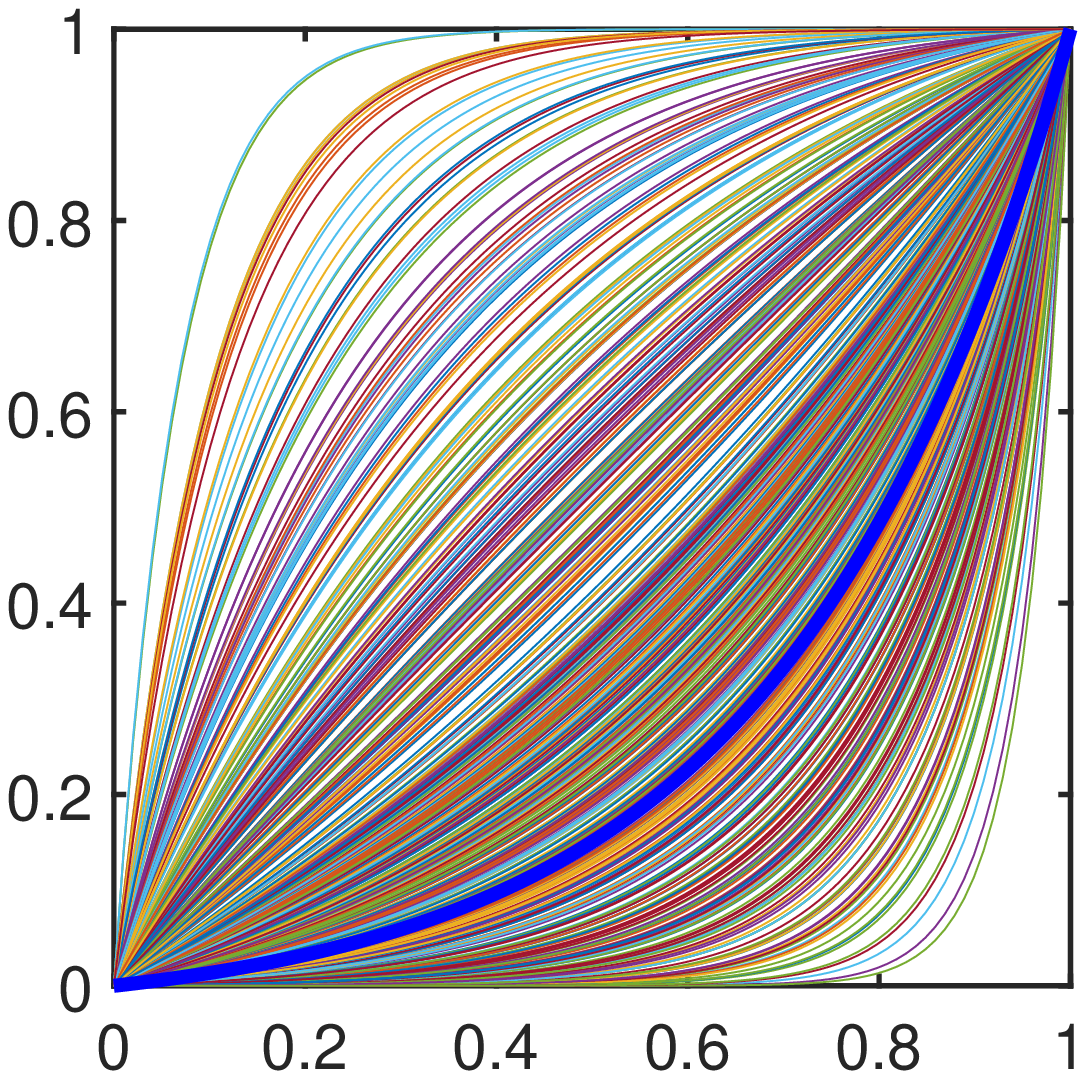}}
		\end{subfigure}%
		\begin{subfigure}[h]{0.24\textwidth}
			\raisebox{-\height}{\includegraphics[width=\textwidth]{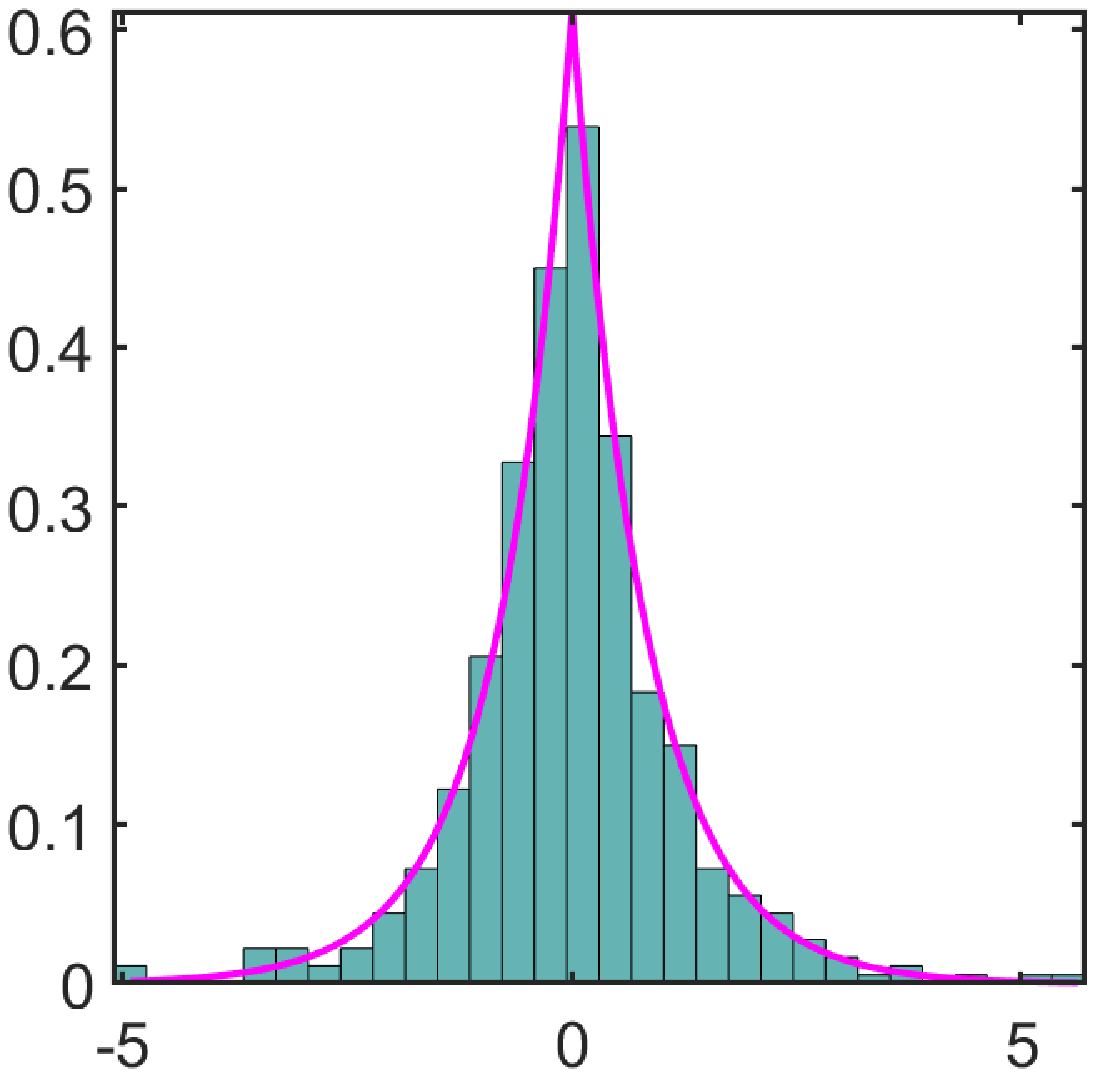}}
		\end{subfigure}%
		\begin{subfigure}[h]{0.24\textwidth}
			\raisebox{-\height}{\includegraphics[width=\textwidth]{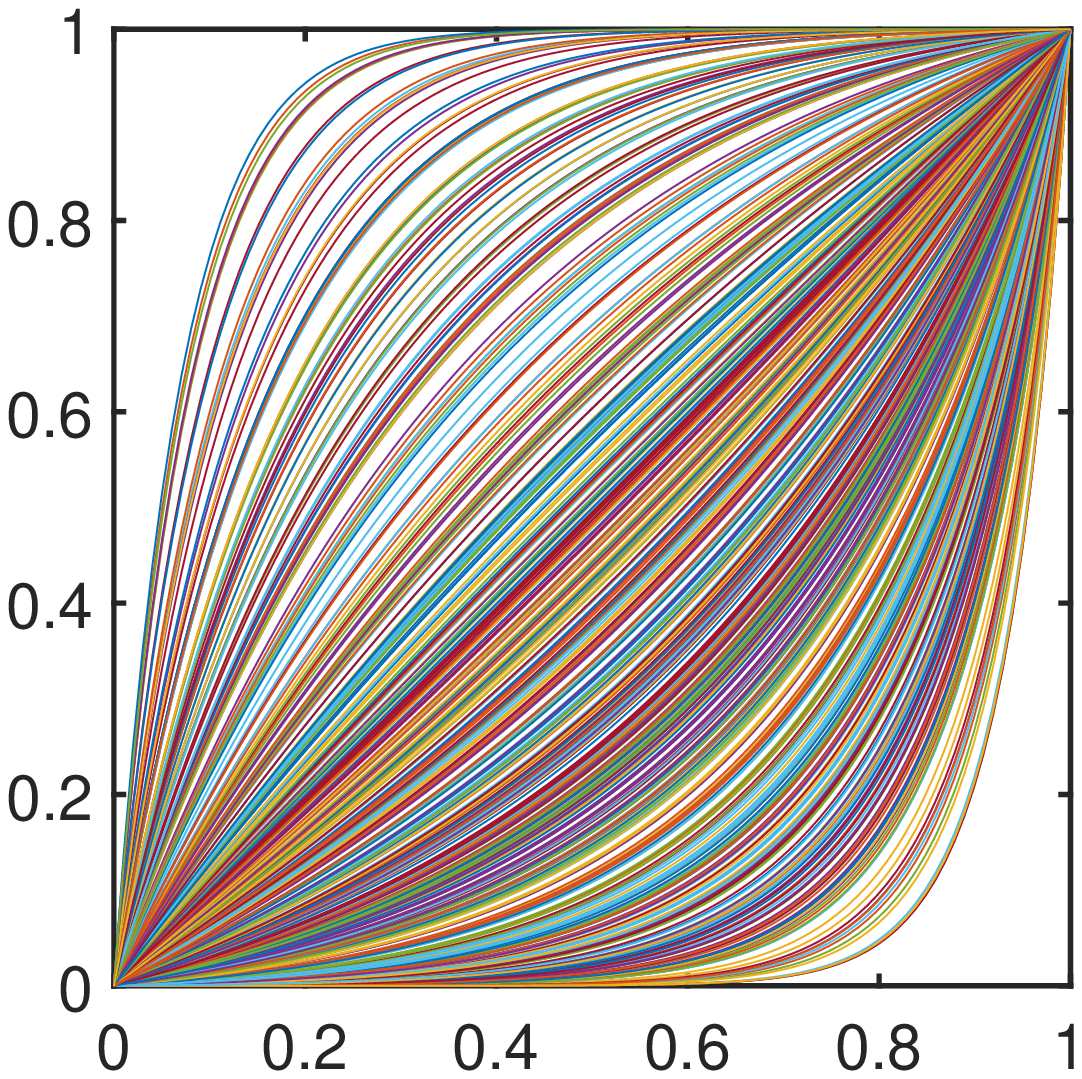}}
		\end{subfigure}%
		\caption{$a_i\sim La(2,2\sqrt{2})$}
	\end{subfigure}%
	\qquad
	\begin{subfigure}[h]{1\textwidth}
		\centering
		\begin{subfigure}[h]{0.24\textwidth}
			\raisebox{-\height}{\includegraphics[width=\textwidth]{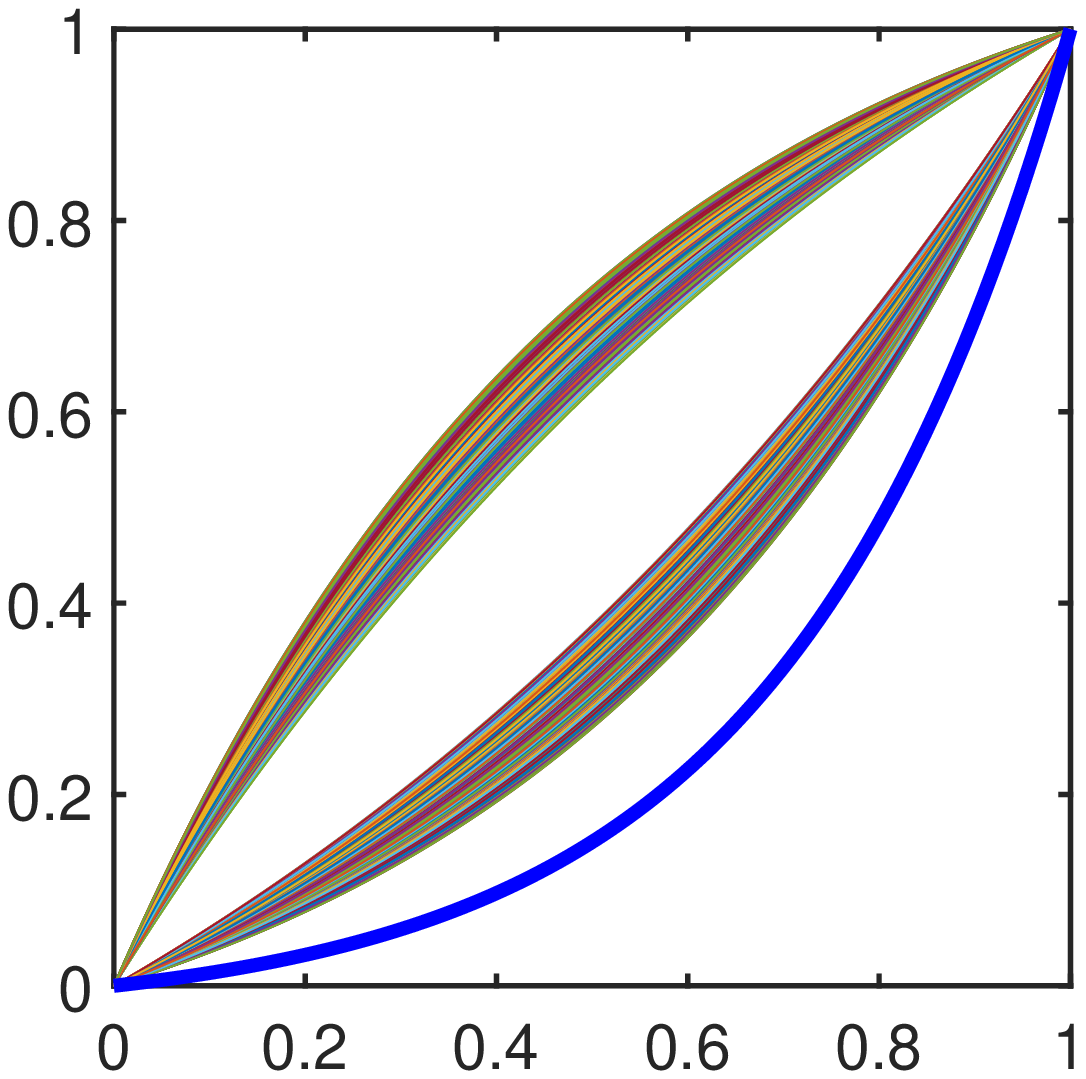}}
		\end{subfigure}%
		\begin{subfigure}[h]{0.24\textwidth}
			\raisebox{-\height}{\includegraphics[width=\textwidth]{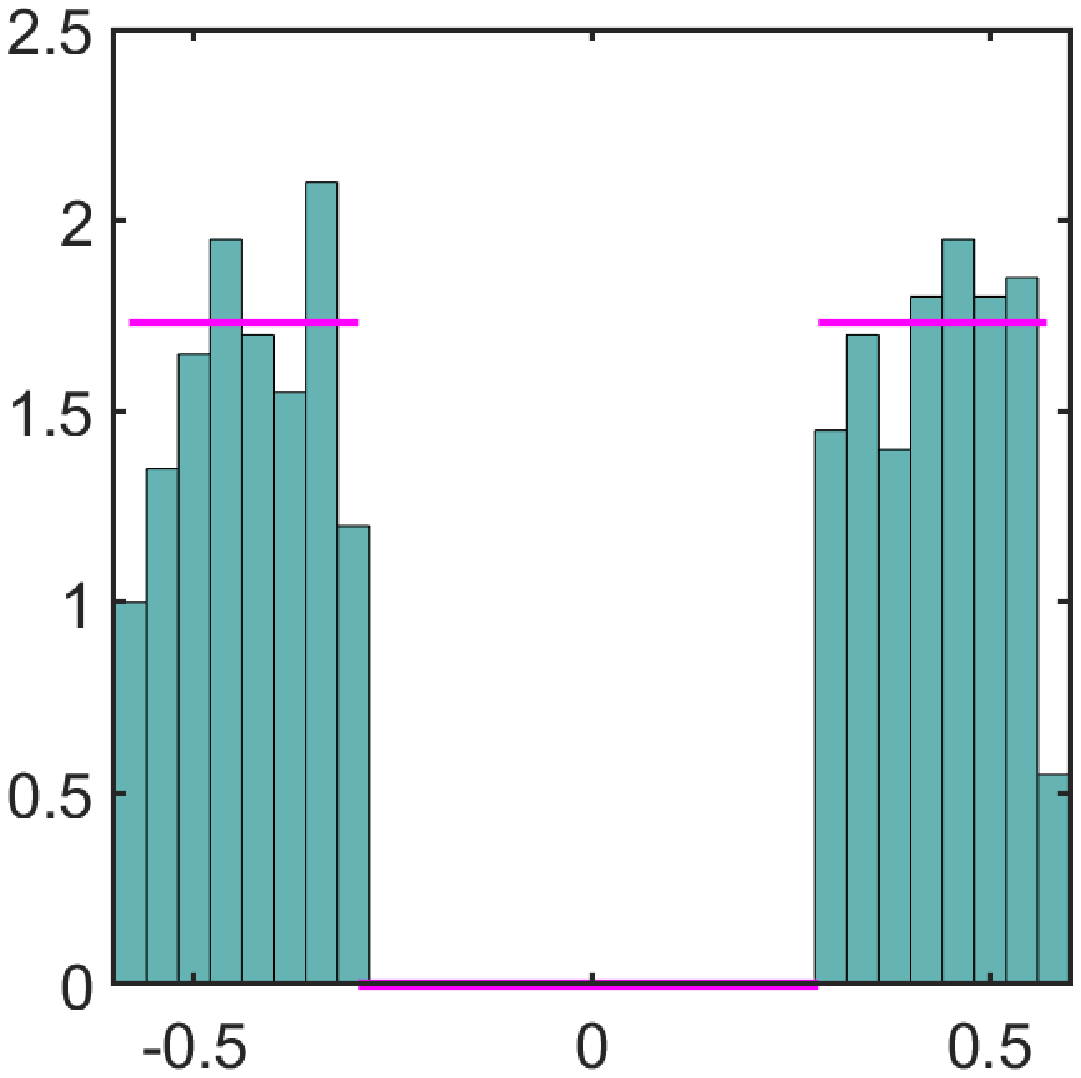}}
		\end{subfigure}%
		\begin{subfigure}[h]{0.24\textwidth}
			\raisebox{-\height}{\includegraphics[width=\textwidth]{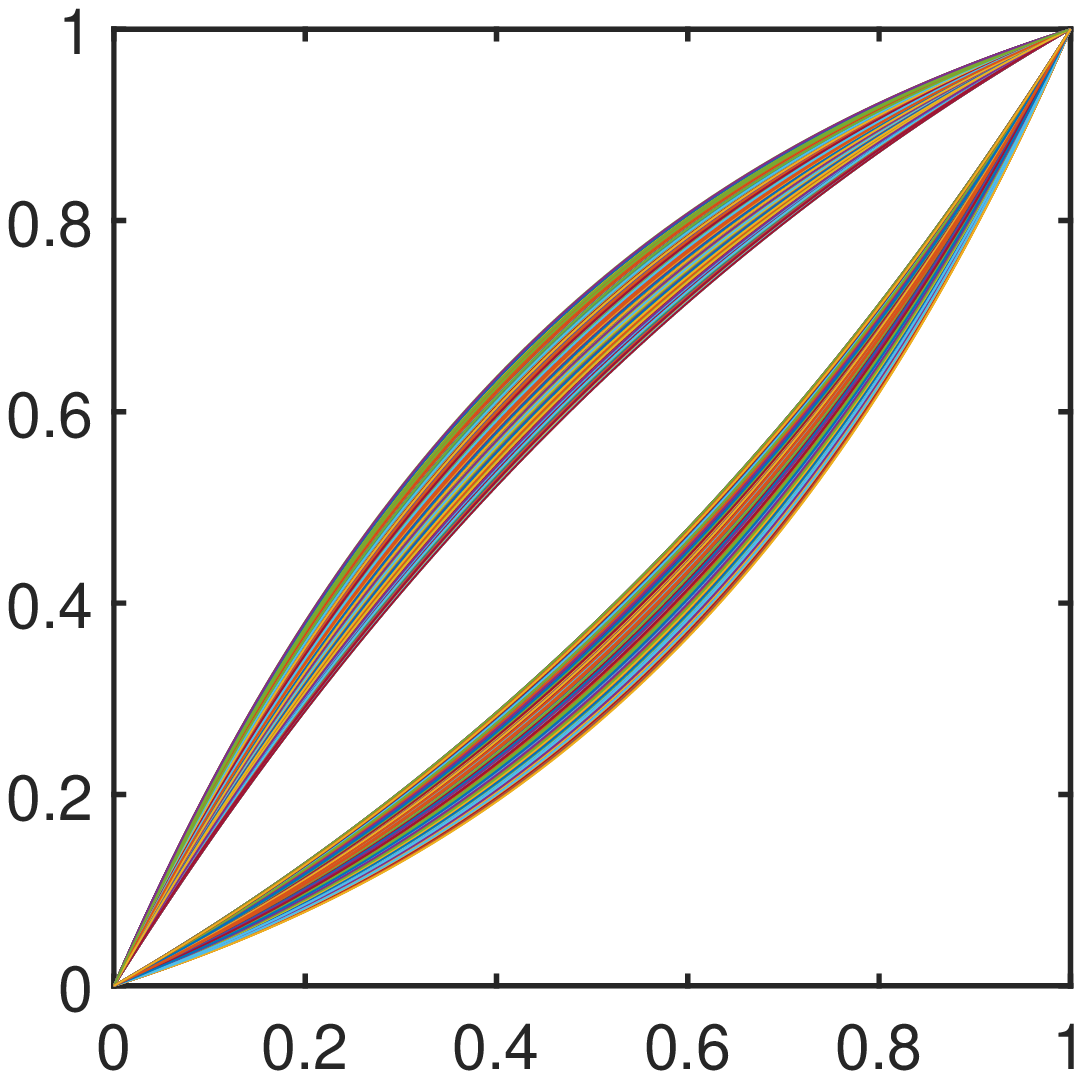}}
		\end{subfigure}%
		\caption{$a_i\sim U((-2,-1)\cup(1,2))$}
	\end{subfigure}%
	\caption{Result on Simulation 1.  The parameter $a_i$ follows each of three different distributions in three rows, respectively. Within each row, the left plot shows simulated 500 time warping functions and the first eigenfunction (in bold blue), the middle one shows the histogram of the first principal component and its true distribution curve (in magenta), and the right one shows the 500 resampled warping functions with the estimated model.}
	\label{fig: pca model 1}
\end{figure}

By Equation \eqref{eq: twclt}, each warping function $\gamma_i(t)=\frac{e^{a_it}-1}{e^{a_i}-1}$ can be transformed into: $h_i(t)=\log(\frac{a_i}{e^{a_i}-1})+a_it-\int_{0}^{1}\log(\frac{a_i}{e^{a_i}-1})+a_it\,dt = a_i(t-\frac{1}{2})$. The mean function of $h_i(t)$ is $m(t) = E(a_i)(t-\frac{1}{2})$, and the covariance function of $h_i(t)$ is: $Cov(s,t) =E(h_i(t)-m(t))(h_i(s)-m(s))= E(a_i-E(a_i))^2(t-\frac{1}{2})(s-\frac{1}{2})=Var(a_i)(t-\frac{1}{2})(s-\frac{1}{2})$.  By Mercer's theorem, it can be easily seen that there is only one nonzero eigenvalue, which is equal to $\frac{Var(a_i)}{12}$. The corresponding eigenfunction can be written as $f_1(t) = 2\sqrt{3}(t-\frac{1}{2})$, and the corresponding coefficient $Z_{i1}$ in Algorithm \ref{alg:MPCA} is equal to  $\langle(a_i-E(a_i))(t-1),2\sqrt{3}(t-\frac{1}{2})\rangle=\int_{0}^{1}(a_i-E(a_i))2\sqrt{3}(t-\frac{1}{2})^2dt=\frac{(a_i-E(a_i))}{2\sqrt{3}}$. Thus, for each of these 3 cases, there is only one significant component, and it follows the same type distribution as $a_i$, i.e. $2\sqrt{3}Z_{i1}+E(a_i)$ follows same distribution as $a_i$. 

Figure \ref{fig: pca model 1} illustrates 500 warping functions, together with the histogram of the first principal component $\{Z_{i1}\}_{i=1}^{500}$ and the resampling result for each of these 3 cases, respectively. It can be easily seen that the coefficient follows the true distribution of the first component so that we fit a parametric distribution for the estimated principal component for each case. For case (a), coefficient follows centralized exponential distribution with parameter $\beta=0.57$; In case(b), coefficients follow $La(0, 0.82)$, and in case (c), coefficients follow a piecewise uniform distribution $U(-0.58, -0.29)\cup (0.29, 0.58)$. In each case, we resampled 500 functions respectively. All these resampled functions (shown in the right column) look very similar to the original warping functions (shown in the left column).

{\bf Simulation 2:} In this example, 500 warping functions are given by: $\gamma_i(t)=\alpha_{1i}\frac{e^{a_it}-1}{e^{a_i}-1}+\alpha_{2i}\frac{e^{b_it}-1}{e^{b_i}-1}+(1-\alpha_{1i}-\alpha_{2i})\frac{e^{d_i\Big(\frac{e^{-c_it}-1}{e^{-c_i}-1}\Big)}-1}{e^{d_i}-1}$, where $a_i,\,d_i\sim  Exp(\frac{1}{3})$ (i.e., exponential distribution with mean $\frac{1}{3}$), $b_i\sim \chi^2(3)$ (i.e., Chi-square distribution with mean 3), and $c_i\sim \Gamma(0.5,2)$ (i.e., gamma distribution with mean 1). In addition, we have $x_i, y_i \sim U(0,1)$, and set $\alpha_{1i}=x_i, \alpha_{2i} = max(y_i-x_i,0)$. \\

\begin{figure}[h]
	\centering
	\begin{subfigure}[h]{0.24\textwidth}
		\includegraphics[width=\textwidth]{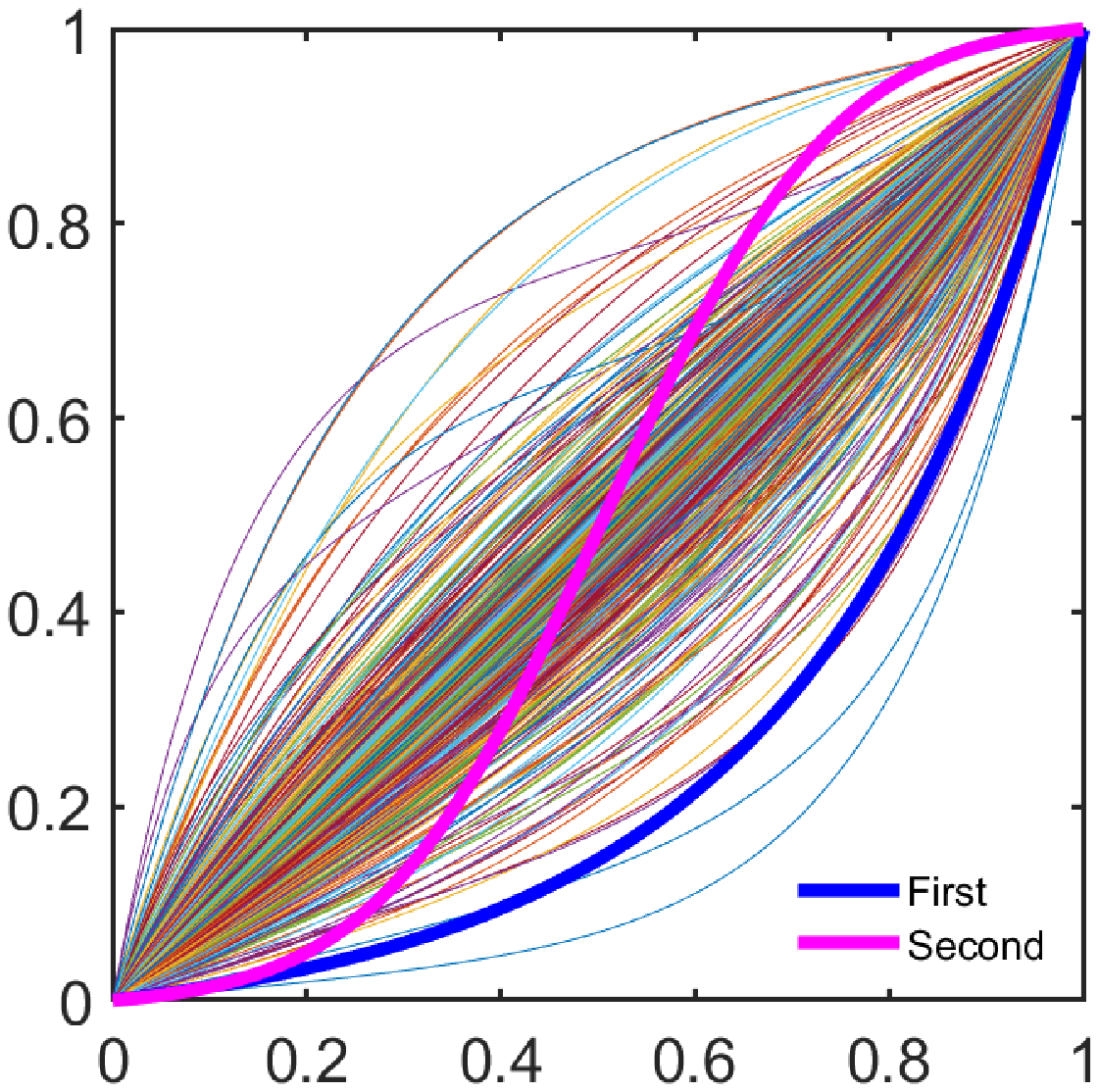}
		\caption{Observations}
	\end{subfigure}
	\begin{subfigure}[h]{0.24\textwidth}
		\includegraphics[width=\textwidth]{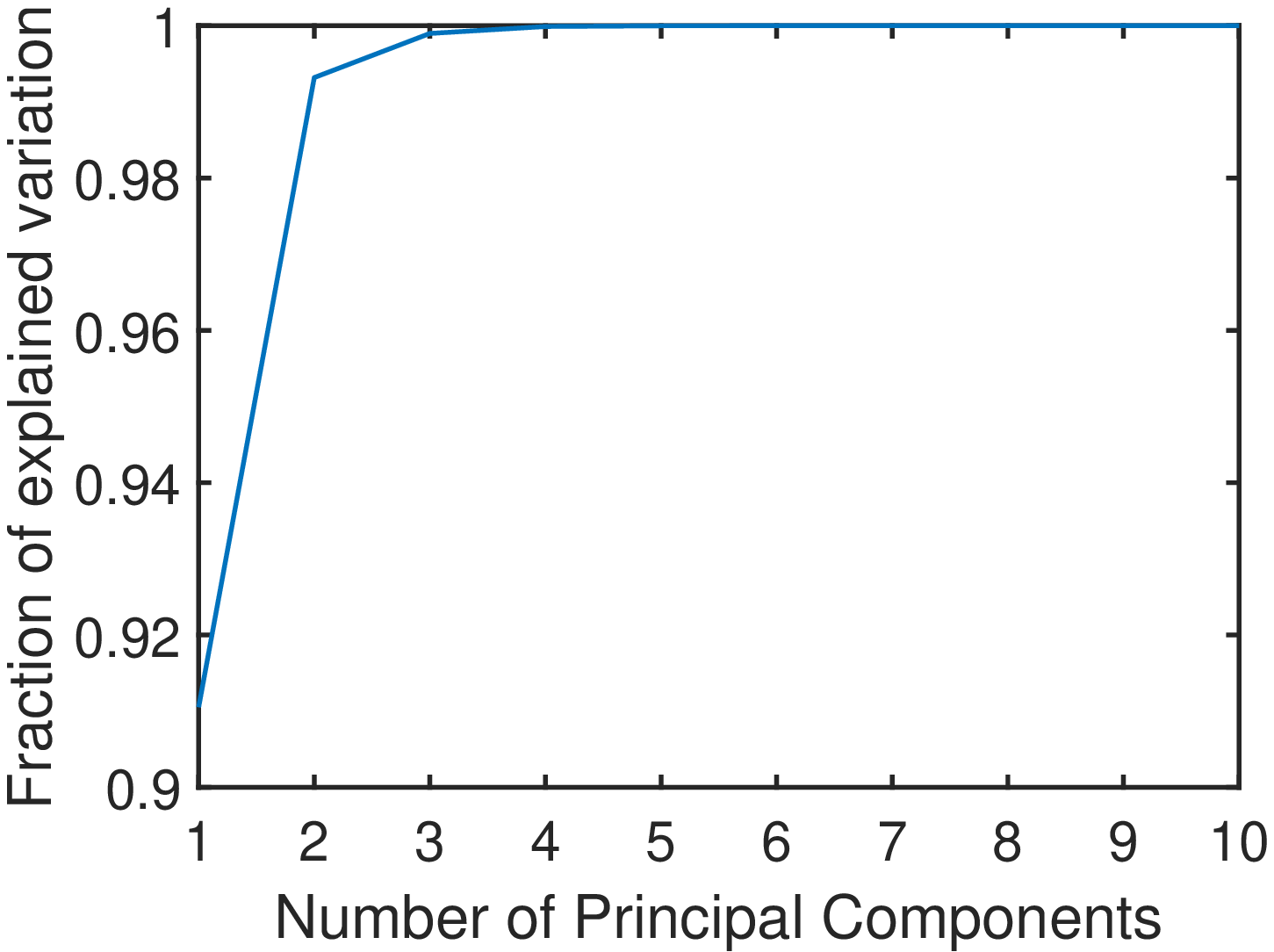}
		\caption{Eigenvalues}
	\end{subfigure}
	\vfill
	\begin{subfigure}[h]{0.24\textwidth}
		\includegraphics[width=\textwidth]{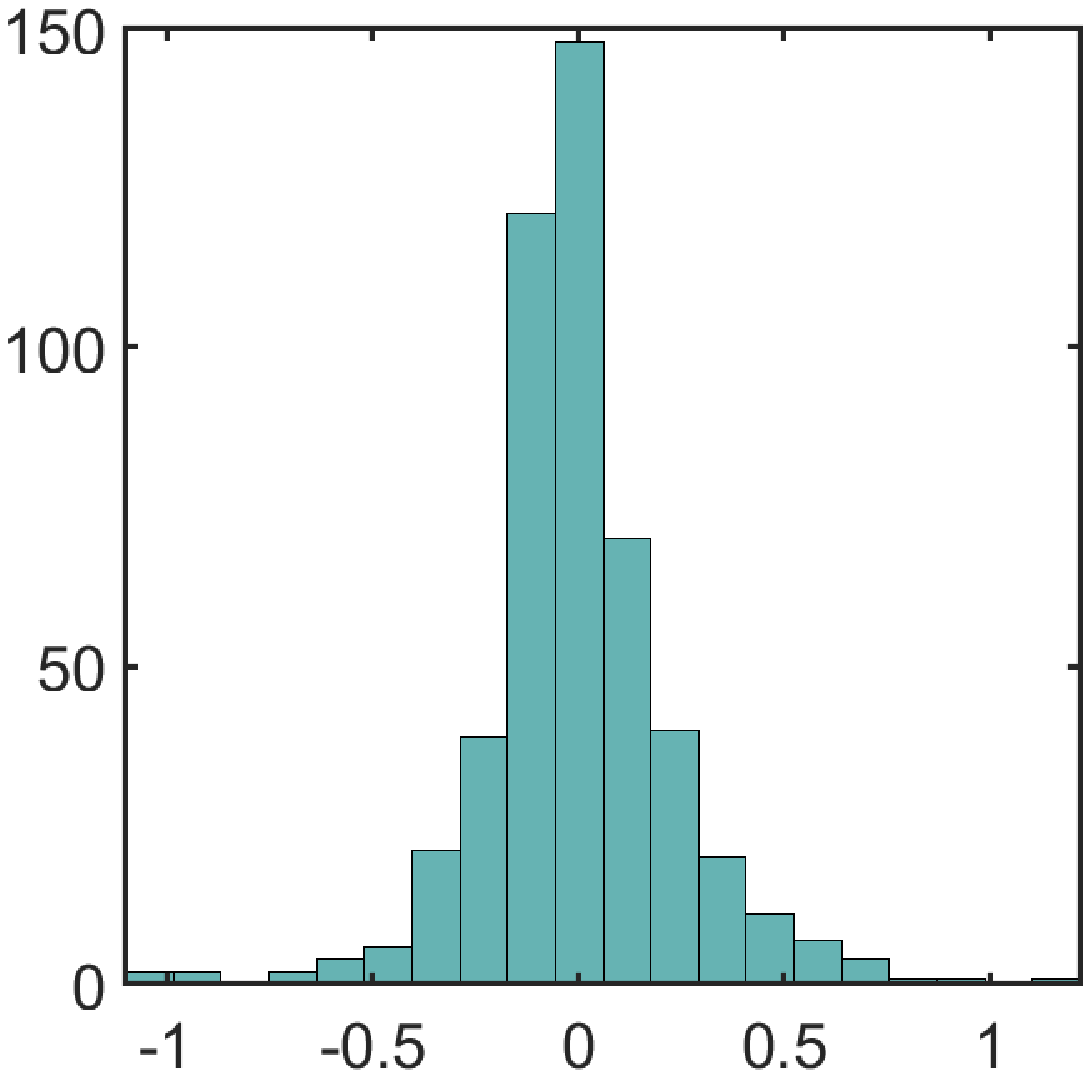}
		\caption{Coefficient 1}
	\end{subfigure}
	\begin{subfigure}[h]{0.24\textwidth}
		\includegraphics[width=\textwidth]{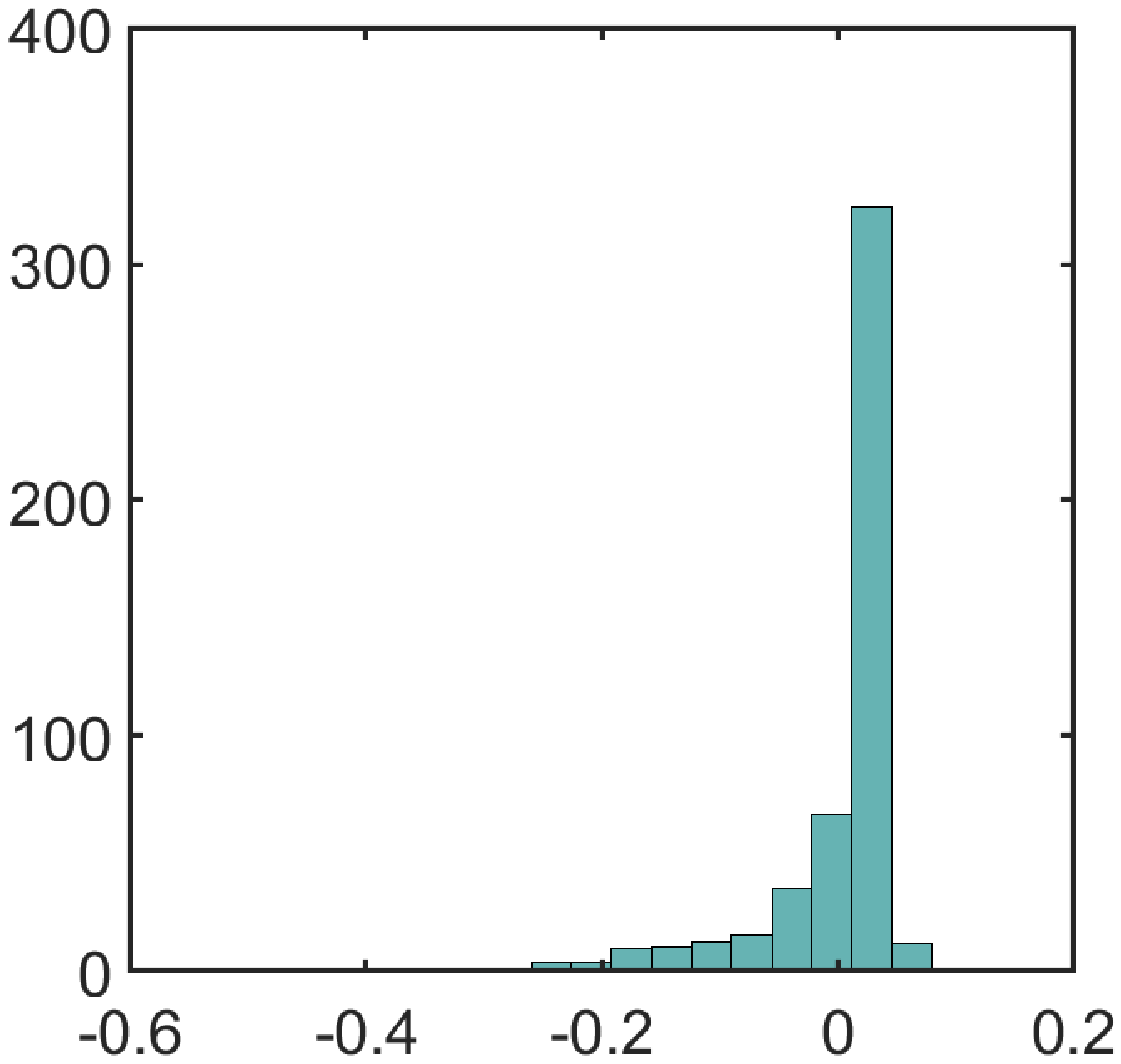}
		\caption{Coefficient 2}
	\end{subfigure}
	\begin{subfigure}[h]{0.24\textwidth}
		\includegraphics[width=\textwidth]{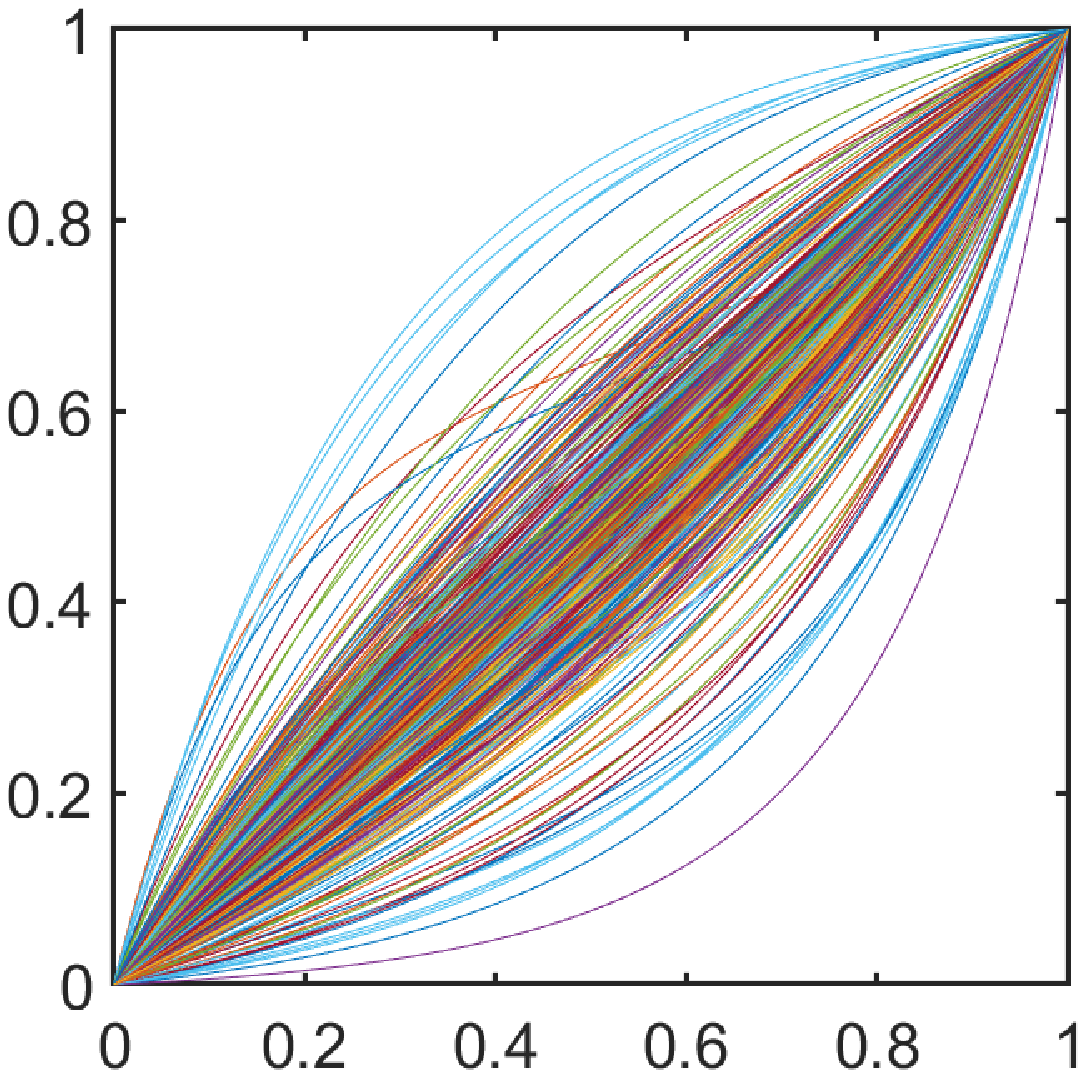}
		\caption{Resampling}
	\end{subfigure}
	
	\caption{Result on Simulation 2. (a) The curves represent 500 simulated time warping functions, and the bold blue and magenta curves represent the first and second eigenfunctions, respectively. (b) Fraction of variance explained by the first $n$ principal components.  (c) Histogram of the first principal component. (d) Histogram of the second principal component. (e) 500 resampled functions with the estimated model.}
	\label{fig:pca model 2}
\end{figure}

The principal component analysis and resampling result are shown in Figure \ref{fig:pca model 2}.  At first, the 500 warping functions are shown in Panel (a). The top 10 eigenvalues are shown in Panel (b).  We can see that the first two principal components explain over 99\% of the total variance, and thus the analysis is conducted only on the these two components.  To visualize the variability, we superimpose the first two eigenfunctions in Panel (a).  The distributions of the first two principal components are shown in Panels (c) and (d), respectively. There is no simple parametric form to describe the distributions, and we choose to adopt the conventional Gaussian kernel method to estimate distribution functions. Based on the estimated two distributions, we can use Algorithm \ref{alg:MPCA} to resample 500 warping functions, and the result is shown in Panel (e). It can be easily seen that the resampled curves also look very similar to the original time warping functions in Panel (a), which indicates the effectiveness of the fPCA modeling procedure. 

\subsection{Statistical inferences under the new framework}
\label{sec: Infces}
Using the CLR transformation, we can transform the warping space $\Gamma_1$ into a subspace of $\mathbb L^2$. In this way, we can perform statistical inferences on time warping functions by using conventional methods in the Euclidean space.  The procedures are straightforward and we briefly describe each method below.   
\begin{enumerate}
\item
{\bf Summary statistics:} Based on the bijective mapping $\psi_B$, we can easily define summary statistics in the time warping space $\Gamma_1$.  For example, the  {\em sample mean} of the time warping functions can be defined as follows: 
\begin{definition}
	Given $n$ time warping function in $\Gamma_1$, denoted as $\gamma_i(t),\, i=1,2,...,n$, their sample mean is defined as 
	\[\mu(t)=\psi_B^{-1}\Big(\frac{1}{n} \sum_{i=1}^n \psi_B(\gamma_i(t))\Big).\]
	\label{def:mean}
	\end{definition}
	The population mean can also be easily defined via the transformation.  Moreover, the high-order statistics such as covariance can be defined  via tensor operations in the  transformed space. In addition, we can define the median of warping functions by the notion of functional depth \citep{liu2017generalized, Qi2021ejs, zhou2022statistical}.  We note that the depth functions are not unique, which may result in different median warping functions. 
	
\item
{\bf fANOVA for time warping functions:} 
When a sample of warping functions is composed of two or more groups, interest may lie in finding out whether there is a real difference among these groups. We can directly apply the fANOVA to the CLR-transformed warping functions.
Suppose we have $k$ independent samples: $\gamma_{i1}(t), \gamma_{i2}(t),\cdots,\gamma_{in_i}(t),\,i=1,\cdots,k$. These $k$ samples satisfy $\psi_B(\gamma_{ij})(t)=\psi_B(\mu_i(t))+v_{ij}(t), v_{ij}(t)\sim GP, j=1,2,\cdots,n_i, i=1,2,\cdots,k,$
where $\mu_i(t)$ are the unknown group mean functions of the $k$ samples, and $v_{ij}(t)$ are the subject-effect functions, $j=1,2,\cdots,n_i, i=1,2,\cdots,k$. We can then do the following one-way ANOVA testing problem using the conventional fANOVA method  \citep{zhang2013analysis}:
$$H_0: \mu_1(t)\equiv\mu_2(t)\equiv \cdots\equiv\mu_k(t)  \text{ vs. } H_A: \text{at least one equality does not hold.} $$

\item 
{\bf Regression methods on time warping functions:} We can conduct two types of regression: 
\begin{itemize}
\item
\textbf{Linear Regression:} The model with time warping functions as predictor appears when the goal is to predict one external variable as a function of warpings. Assume a set of time warping functions $\gamma_1, \gamma_2, \cdots,\gamma_n$ is available, and that $i$-th function $\gamma_i$ is associated with an observation $y_i$ of an external response variable. By applying the mapping defined in Theorem \ref{thm:iso1} on the $n$ time warping functions, we can get $n$ corresponding functions $f_i=\psi_B(\gamma_i)$ on $H(0,1) \subset  \mathbb L^2$. The regression model is given as 
\begin{equation}
	y_i = \alpha + \langle f_i, \beta \rangle, \ \ \ i = 1, \cdots, n. 
\end{equation}
where $\langle f_i, \beta \rangle =\int_{0}^{1} f_i(t)\beta(t)\,dt$ is the $\mathbb{L}^2$ inner product, $\beta \in H(0,1)$ is the regression coefficient function, and $\alpha$ is the bias. 
We can then use the ordinary least square method to estimate the parameters.

\item
\textbf{Logistic Regression:} We can also use time warping function as predictor to predict an external binary variable $z_i, i = 1, \cdots, n$. This is a natural extension of the above linear regression model with the logistic link function. The logistic regression model is defined as 
\[z_i=h\Big(\alpha + \langle f_i, \beta \rangle\Big),  \ \ \ i = 1, \cdots, n.  \]
where $h(t)=\frac{1}{1+\exp (-t)}$ is the logistic link function. 
The model parameters $\alpha$ and $\beta$ can be estimated by maximizing the log-likelihood.
\end{itemize}
\end{enumerate}

\section{Bayesian Registration }
\label{Sec: BayesReg}

In this section, we will utilize the proposed framework on time warping to provide a new approach for Bayesian registration. Bayesian registration is a relatively new paradigm that incorporates the prior information of warping function to conduct function registration \citep{cheng2014analysis, cheng2016bayesian, lu2017bayesian, kurtek2017geometric,tucker2021multimodal,matuk2021bayesian}. Majority of these approaches are based on the SRVF (Square Root Velocity Function) transformation and explore appropriate representation of the warping functions on the corresponding tangent space, where a Gaussian process was used to model the inverse exponential transformed warping function. However, it was pointed out that this model is restricted to a bounded region of the positive orthant of the tangent space and linear operations on this region may get out of it and result in undesirable nonincreasing warping functions \citep{happ2019general}.   

Our registration is still based on the Bayesian framework in \citep{cheng2016bayesian}, whereas we propose a new prior on the CLR transformed warping space as a penalty term to control the degree of phase variation. The optimal warping is estimated by the maximum a posteriori (MAP) with a gradient method instead of MCMC simulation of the posterior distribution \citep{cheng2016bayesian,kurtek2017geometric}.  Unlike previous isotropic covariance representation\citep{cheng2016bayesian}, our full covariance can characterize nonuniform temporal variance as well as correlated relationship in the time domain.  We emphasize that our covariance is well defined in the CLR transformed Euclidean space so that we can use 2nd order stochastic process (e.g. a Gaussian process) as the prior term.

\subsection{New prior on time warping}
  
Let $f$ be an absolutely continuous function on the interval $[0,1]$. Its SRVF is defined as $q:[0,1]\rightarrow \mathbb{R}$, $q(t)=\dot{f}(t)/\sqrt{|\dot{f}(t)|}$  \citep{srivastava2011registration}. For $\gamma \in \Gamma_1$, the SRVF of $f \circ \gamma$ is given by: $(q,\gamma)=\sqrt{\dot{\gamma}(t)}q(\gamma(t))$. For two function $f_1,\, f_2$, we assume a zero mean Gaussian process for the difference of the corresponding SRVF functions $q_1,\,q_2$ , i.e., ${q_1-(q_2,\gamma)|\gamma}\sim GP$. 
If we use $q_1([t])$ and $(q_2,\gamma)([t])$ to denote vectors evaluated at the same finite points on the domain of $q_1(t)$ and $(q_2,\gamma)(t)$, respectively, then the joint distribution of these finite differences $q_1([t])-(q_2,\gamma)([t])|\gamma$ is a multivariate normal distribution based on the Gaussian process assumption, i.e., $\Big\{q_1([t])-(q_2,\gamma)([t])|\gamma\Big\}\sim N_k(0_k,\Sigma_{k\times k})$, where $k$ is the number of points.  Assuming $\Sigma_{k\times k}=\frac{1}{2\kappa} I_{k\times k}$ \citep{cheng2016bayesian}, the likelihood is given as: 
\begin{eqnarray}
	\pi(q_1,q_2|\gamma) &\propto& \exp\Big(-\kappa \|q_1-(q_2,\gamma)\|^2\Big). \nonumber 
\end{eqnarray}

In \cite{cheng2016bayesian}, a Dirichlet prior is assigned to model the warping $\gamma$. Here, we propose to use a Gaussian process prior to model the transformed warping functions in the $\mathbb L^2$ space, i.e.,$\log(\dot{\gamma}(t))-\int_{0}^{1}\log(\dot{\gamma}(s))ds\sim GP$. If we also discretize it into $k$ time points, we will have $\log(\dot{\gamma}([t]))-\int_{0}^{1}\log(\dot{\gamma}(s))ds\sim N_k(0_k,\Sigma^{\gamma}_{k\times k})$, where $(\Sigma^{\gamma}_{k\times k})^{-1}=h([s],[t])$ and $h(s,t): [0,1]\times [0,1] \rightarrow \mathbb R$ is a symmetric positive definite kernel.
Using Bayes theory, the posterior distribution for $\gamma([t])$ given $\Big(q_1([t]),q_2([t])\Big)$ is approximately 
\begin{eqnarray}
	\begin{split}
		&\pi(\gamma|q_1,q_2)\\
		&\propto \exp(-\kappa \parallel q_1-(q_2,\gamma)\parallel^2)\exp\bigg(-\int_{0}^{1}\int_{0}^{1} \Big(\log (\dot{\gamma}(s))-\int_{0}^{1}\log (\dot{\gamma}(u))\,du \Big)h(s,t)\Big(\log (\dot{\gamma}(t))-\int_{0}^{1}\log (\dot{\gamma}(u))\,du\Big) \,ds\,dt\bigg) \nonumber 
	\end{split}		
\end{eqnarray}
The optimal time warping is obtained by maximizing $\pi(\gamma|q_1,q_2)$ given above, which is equivalent to minimizing the following penalized form:
\[\parallel q_1-(q_2,\gamma)\parallel^2+\lambda\int_{0}^{1}\int_{0}^{1} \Big(\log (\dot{\gamma}(s))-\int_{0}^{1}\log (\dot{\gamma}(u))\,du \Big)h(s,t)\Big(\log (\dot{\gamma}(t))-\int_{0}^{1}\log (\dot{\gamma}(u))\,du\Big) \,ds\,dt,\]
where $\lambda =-\frac{1}{\kappa}$.  As $\parallel q_1-(q_2,\gamma)\parallel^2 = \parallel q_1\parallel^2+\parallel q_2\parallel^2 - \int_0^1 2 q_1(t) (q_2,\gamma)(t) dt$, we can get the loss function with respect to $\gamma$ as follows:
\begin{equation}
	\begin{split}
		J(\gamma) &= \int_{0}^{1}-2q_1(t)q_2(\gamma(t))\sqrt{\dot{\gamma}(t)}\,dt\\
		&+\lambda\int_{0}^{1}\int_{0}^{1} \Big(\log (\dot{\gamma}(s))-\int_{0}^{1}\log (\dot{\gamma}(u))\,du \Big)h(s,t)\Big(\log (\dot{\gamma}(t))-\int_{0}^{1}\log (\dot{\gamma}(u))\,du\Big) \,ds\,dt. 
	\end{split}
	\label{eq:loss}
\end{equation}

{\bf Remark:} The above penalized form is for any symmetric positive definite kernel $h(s,t)$.  There are two important special cases: isotropic and diagonal.
\begin{enumerate}
\item   
If the covariance is isotropic, i.e., $\Sigma^{\gamma}_{k\times k}=\frac{1}{2a}I_{k\times k}$, we just have to set $h(s,t)=a\delta (s-t)$ in Equation \eqref{eq:loss} to get the corresponding loss function.  
\item If the covariance is diagonal, i.e., $(\Sigma^{\gamma}_{k\times k})^{-1}=diag\{d_1, \cdots, d_k\}$, we can set $h(s,t)=r(t)\delta (s-t)$ in Equation \eqref{eq:loss}, where $r(t)$ is a function with $r([t]) = (d_1, \cdots, d_k)$, to get the corresponding loss function.
\end{enumerate}
Note that we have only shown one covariance-based process for the time warping model here. In general, we can assign other stochastic process priors under our framework, and they do not have to be a Gaussian process.

\subsection{Optimization and the alignment algorithm}
When there is no prior term, the loss function is an integration with respect to the warping function and a dynamic programming procedure can be applied to get the optimal warping function, albeit on a discrete grid \citep{srivastava2011registration}. However, with the prior term, the dynamic programming cannot be used because the loss function in Equation \eqref{eq:loss} can no longer be written under one integration. To deal with this problem, we propose to conduct the optimization via a gradient-based method.  Note that the time warping function is in a non-vector space with conventional $\mathbb L^2$ metric, and the gradient on warping cannot be used for optimization.  Analogous to the CLR transformation, we let $\phi(t) = \log (\dot{\gamma}(t))\in \mathbb L^2([0,1])$, and then we can get the new loss function of $\phi$ in the following form:
\begin{equation}
	\begin{split}
		J(\phi) &= \int_{0}^{1}-2q_1(t)q_2\Big(\int_{0}^{t}\exp (\phi(s))\,ds\Big)\sqrt{\exp(\phi(t))}\,dt\\
		&+\lambda \int_{0}^{1}\int_{0}^{1} \Big(\phi(s)-\int_{0}^{1}\phi(u)\,du \Big)h(s,t)\Big(\phi(t)-\int_{0}^{1}\phi(u)\,du\Big) \,ds\,dt
	\end{split}
	\label{eq: Losspsi}
\end{equation}
Note that we still have one constraint on $\phi(t)$, i.e., $\int_{0}^{1}\exp (\phi(t))\,dt = 1$. So when we apply the gradient descend, we will need to conduct this normalization to update time warping function in each iteration. Using the variational method, we can calculate the gradient of the loss function as follows (see details in Appendix B): 
\begin{equation}
	\begin{aligned}
		\frac{\partial J}{\partial \phi}(t)
		&=-2\exp(\phi(t))\int_{t}^{1}q_1(\mu)\dot{q}_2\Big(\int_{0}^{\mu}\exp(\phi(s))ds\Big)\sqrt{\exp(\phi(\mu))}\,d\mu-q_1(t)q_2\Big(\int_{0}^{t}\exp(\phi(s))\,ds\Big)\sqrt{\exp(\phi(t))}\\ 
		&+\lambda\bigg(\int_{0}^{1} h(t,s)\phi(s)\,ds + \int_{0}^{1} \phi(s)h(s,t)\,ds - \int_{0}^{1}\int_{0}^{1} \phi(s)h(s,u)\,ds\,du -\int_{0}^{1}\phi(u)\,du\int_{0}^{1}h(t,s)\,ds \\ 
		&-\int_{0}^{1}\int_{0}^{1} h(s,u)\phi(u)\,ds\,du-\int_{0}^{1} h(s,t)\,ds\int_{0}^{1}\phi(u)\,du
		+2\int_{0}^{1}\phi(u)\,du\int_{0}^{1}\int_{0}^{1}h(s,v)\,ds\,dv
		\bigg) 
		\label{eq: GD}
	\end{aligned}
\end{equation}
The gradients on the two special cases (isotropic covariance and diagonal covariance) are also given in Appendix B, where the calculations are more efficient because of the simplified structures on the covariance. 

Based on the gradient function in Equation \eqref{eq: GD}, we can apply the gradient descent method.  We emphasize that this method has linear computational order w.r.t. the number fo discrete points, and is highly efficient in practical calculation.  In contrast, the dynamic programming is in the quadratic order and can be very time-consuming when the number of discrete points is large.  In summary, the overall alignment process is given in the following algorithm:

\begin{algorithm}[h]
	\caption{Alignment with Bayesian Registration}
	\begin{algorithmic} 
		\Require Two real valued functions $f_1,\, f_2$ on interval $[0,1]$, initial warping $\gamma_0$, learning rate $\epsilon$, threshold $\delta$, tuning parameter $\lambda$.
		\State Calculate the SRVF functions $q_1(t), q_2(t)$ of ${f}_1(t), {f}_2(t)$, respectively.  
		\State Let $\phi(t) = \log(\dot{\gamma_0}(t))$, and estimate the loss function $J(\phi)$ using Equation \eqref{eq: Losspsi}. 
		\State Calculate derivative $\frac{\partial J}{\partial \phi}(t)$ of the loss function using Equation \eqref{eq: GD}.
		\While {$\|\frac{\partial J}{\partial \phi}\| > \delta$}
		\State $\phi(t) \leftarrow \phi(t) -\epsilon\frac{\partial J}{\partial \phi} $. 
		\State $\phi(t) \leftarrow \phi(t) - \log \big(\int_{0}^{1}\exp(\phi(s))\,ds\big)$.
		\State Recalculate the loss function with the new $\phi(t)$.
		\EndWhile
		\State Let $\phi_{new}$ be the last $\phi$ in the while loop.  Then the optimal warping is: $\gamma_{new}(t) =\int_{0}^{t}\exp(\phi_{new}(s))ds$. 
		\State Output $\gamma_{new}(t)$.
	\end{algorithmic}
	\label{alg:AligBayes}
\end{algorithm}

{\bf Remark:}  In addition to normalizing $\phi$ at each iteration, we can also use the Lagrange multiplier technique to solve the optimization with constraint $\int_{0}^{1}\exp (\phi(t))\,dt = 1$.  It is found that this method provides similar optimization performance as that in Algorithm \ref{alg:AligBayes} and is therefore omitted in this paper.

\subsection{Alignment illustrations}

We will now illustrate the new Bayesian registration with three examples.  The first one is based on isotropic covariance and has been studied in previous methods.  The other two focus on diagonal and full covariances, which describe nonuniform and correlated constraints in the time domain.  To the best of our knowledge, such studies have not been well explored in Bayesian registration. 

\subsubsection{Isotropic covariance}
\begin{figure}[h]
	\centering
	\begin{subfigure}[h]{0.5\textwidth}
		\includegraphics[width=\textwidth]{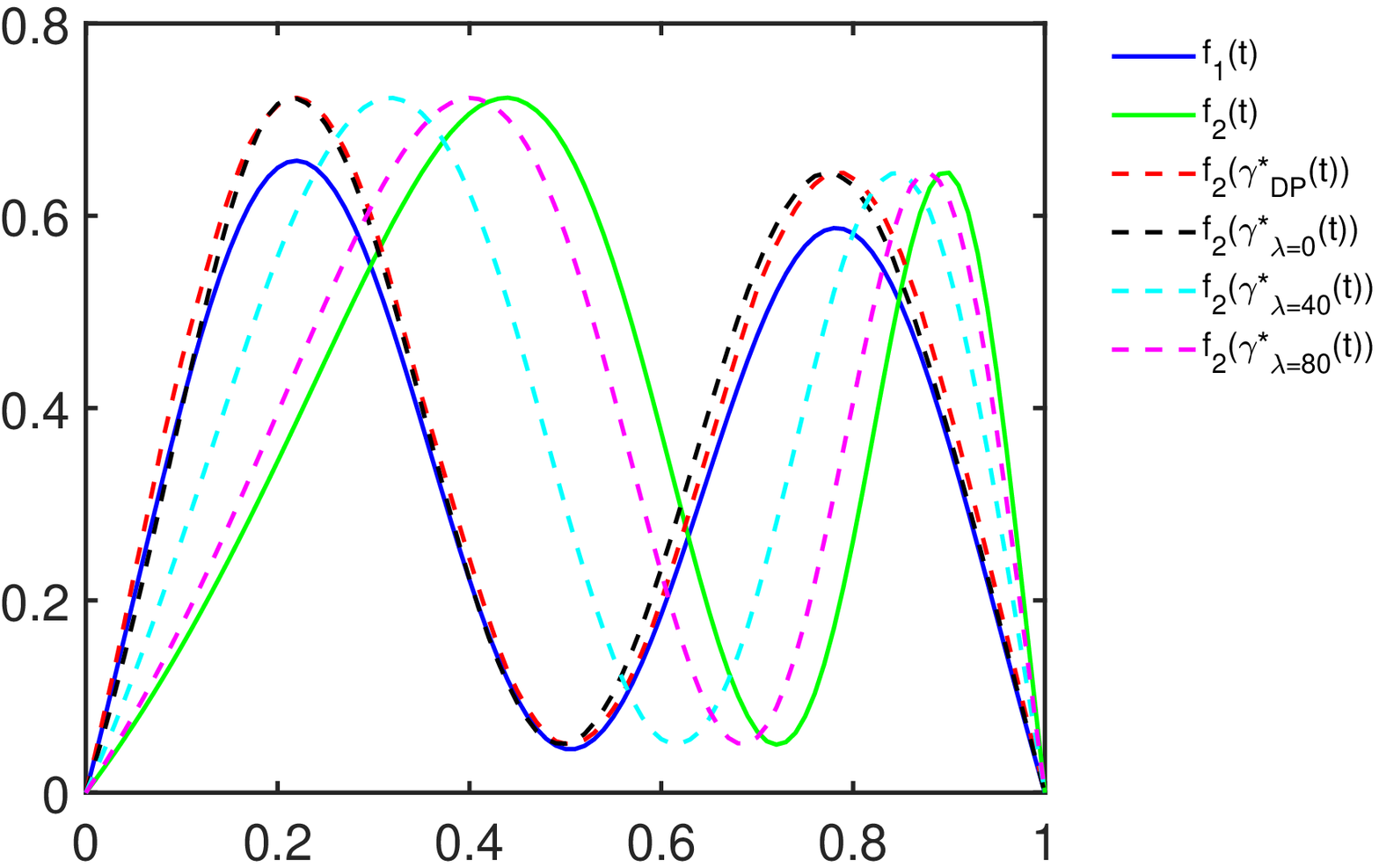}
		\caption{Alignment result}
	\end{subfigure}\hspace{-0.2cm}
	\begin{subfigure}[h]{0.4\textwidth}
		\includegraphics[width=\textwidth]{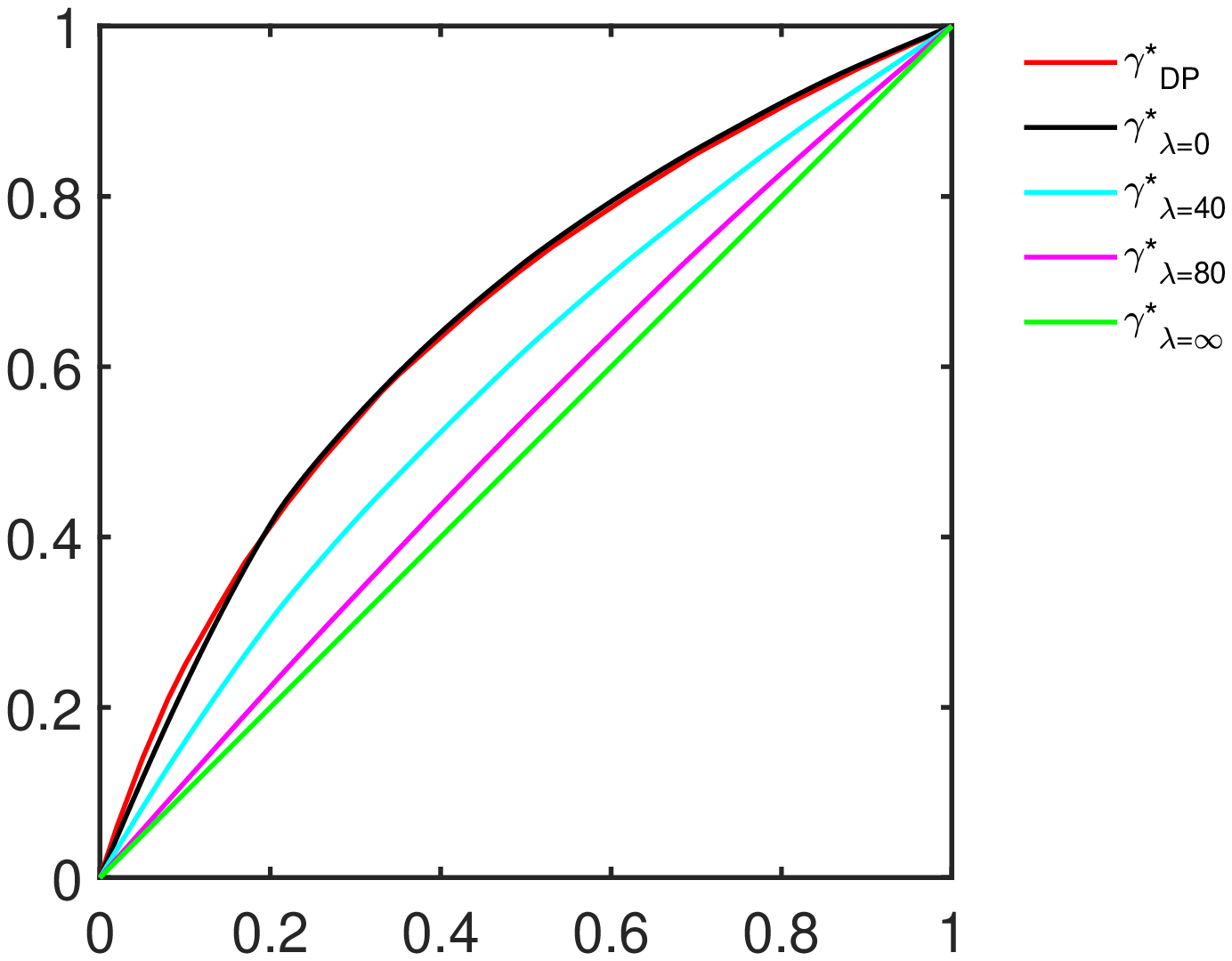}
		\caption{Optimal warpings}
	\end{subfigure}
	\caption{Bayesian registration illustration with isotropic covariance kernel. (a) Original functions and alignment functions. The blue and green solid curves are the two given functions $f_1$ and $f_2$, respectively, the red dotted curve is the aligned $f_2$ using dynamic programming, and the black, cyan, and magenta dotted curves are the aligned $f_2$ using Algorithm \ref{alg:AligBayes} with $\lambda$ equal to $0,\, 40,\, 80,$ respectively. (b) Optimal warping functions in the alignment. The red curve is the optimal warping function from dynamic programming. The black, cyan, magenta, and green curves are the optimal warping functions from Algorithm \ref{alg:AligBayes} with $\lambda$ equal to $0,\, 40,\, 80$, and $\infty$, respectively.}
	\label{fig.reg}
\end{figure}

We here use one example to illustrate the Bayesian alignment with isotropic covariance kernel. We at first simulate one bimodal function, i.e., $f(t)=z_1e^{-(t-0.22)^2/2}+z_2e^{-(t-0.78)^2/2}$, where $z_1,\, z_2 \sim U(0.75, 1.25), t\in [0,1]$. Then we obtain two functions $f_i(t) = f(\gamma_i(t))$ with warping functions $\gamma_i(t)=\frac{e^{a_it}-1}{e^{a_i}-1},\, i=1,\, 2$, where $a_1=-0.5$ and $a_2 = 2$. In addition, we scale $f_2(t)$ up by 1.1 for better visualization. The functions $f_1(t)$ and $f_2(t)$ are shown as blue and green solid curves, respectively, in Figure \ref{fig.reg}(a).  Our goal is find optimal warping function $\gamma^*$ to minimize the loss function in Equation \eqref{eq:loss}.  When $\lambda = 0, 40,$ and $80$, the optimal warping functions are calculated using Algorithm \ref{alg:AligBayes} and shown in Figure \ref{fig.reg}(b).  We can see that when $\lambda = 0$, the optimal warping is very close to the one estimated using dynamic programming; the difference is only about numerical errors.  When $\lambda$ gets larger, the optimal warping is closer to the identity warping $\gamma_{id}(t) = t$ (optimal warping when $\lambda = \infty$).  The aligned functions $f_2(\gamma^*(t))$ are also shown in  Figure \ref{fig.reg}(a).  We can see $f_2(\gamma^*(t))$ is right on the top of $f_1(t)$ when $\lambda = 0$, and only slightly shift from $f_2(t)$ when $\lambda = 80$.  These results clearly demonstrate the effectiveness of the prior model in the Bayesian alignment process.  

\subsubsection{Diagonal covariance}
\begin{figure}[h]
	\centering
	\begin{subfigure}[h]{0.3\textwidth}
		\includegraphics[width=\textwidth]{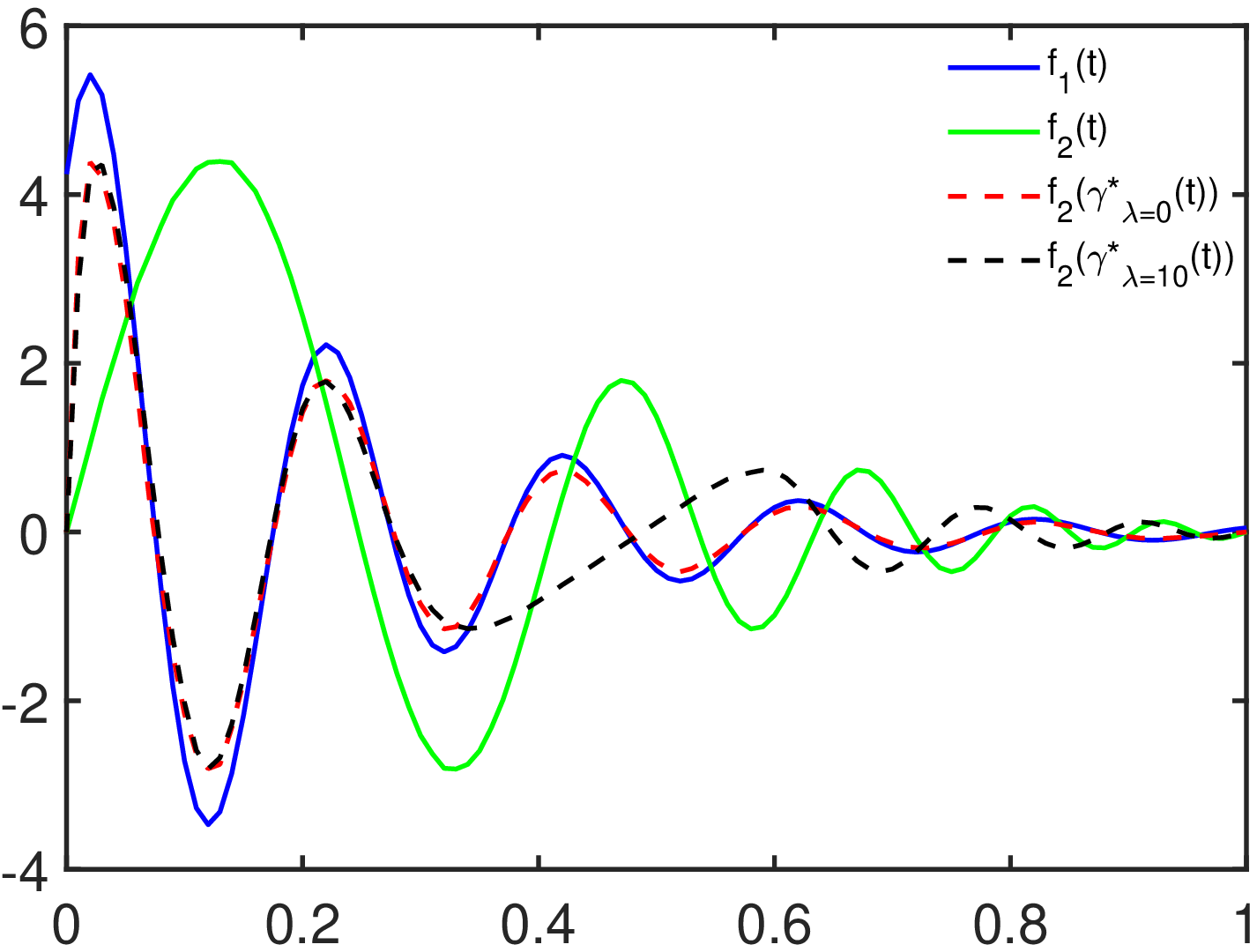}
		\caption{Alignment result}
	\end{subfigure}
    \quad
	\begin{subfigure}[h]{0.3\textwidth}
		\includegraphics[width=\textwidth]{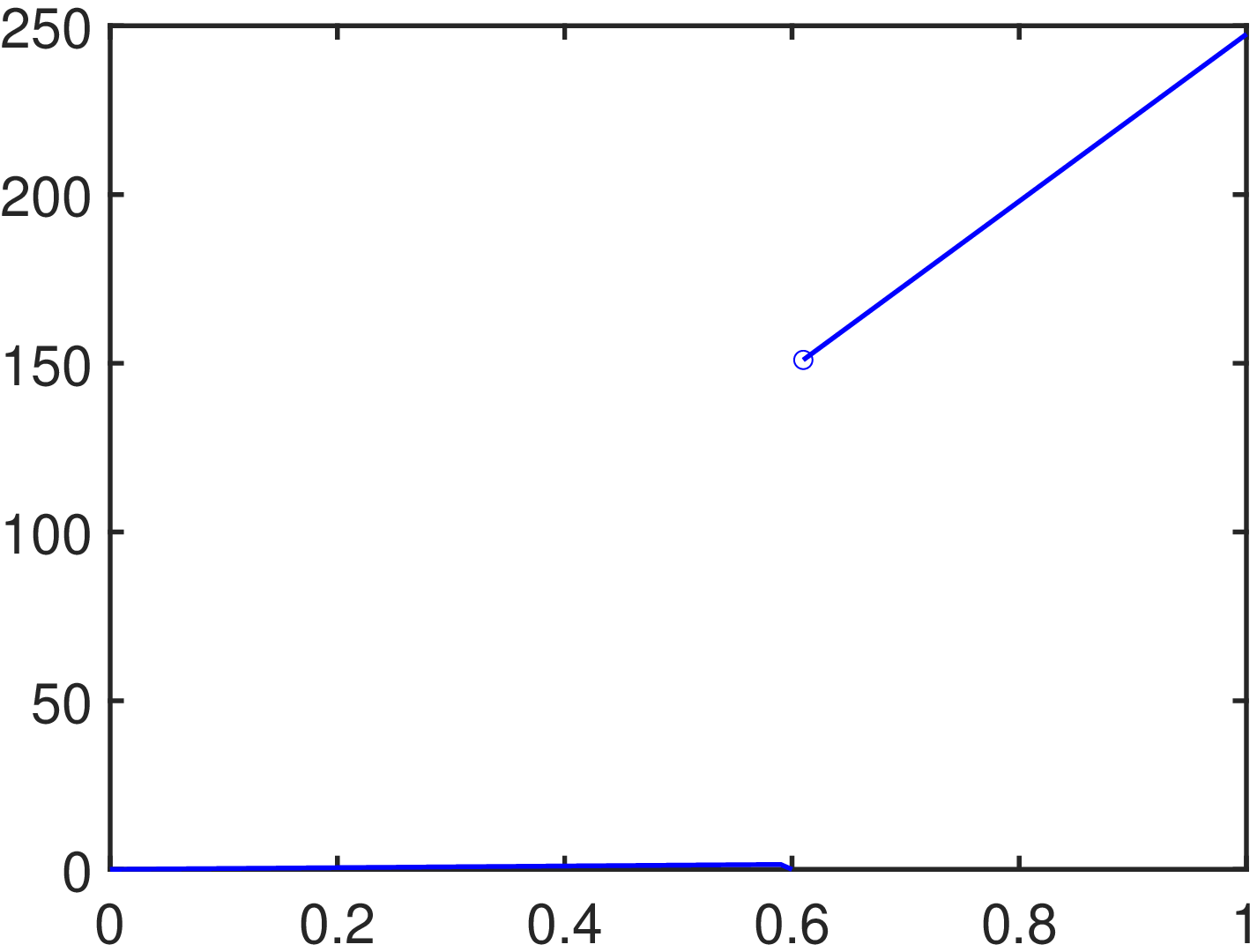}
		\caption{$r(t)$}
	\end{subfigure}
    \quad
	\begin{subfigure}[h]{0.3\textwidth}
		\includegraphics[width=\textwidth]{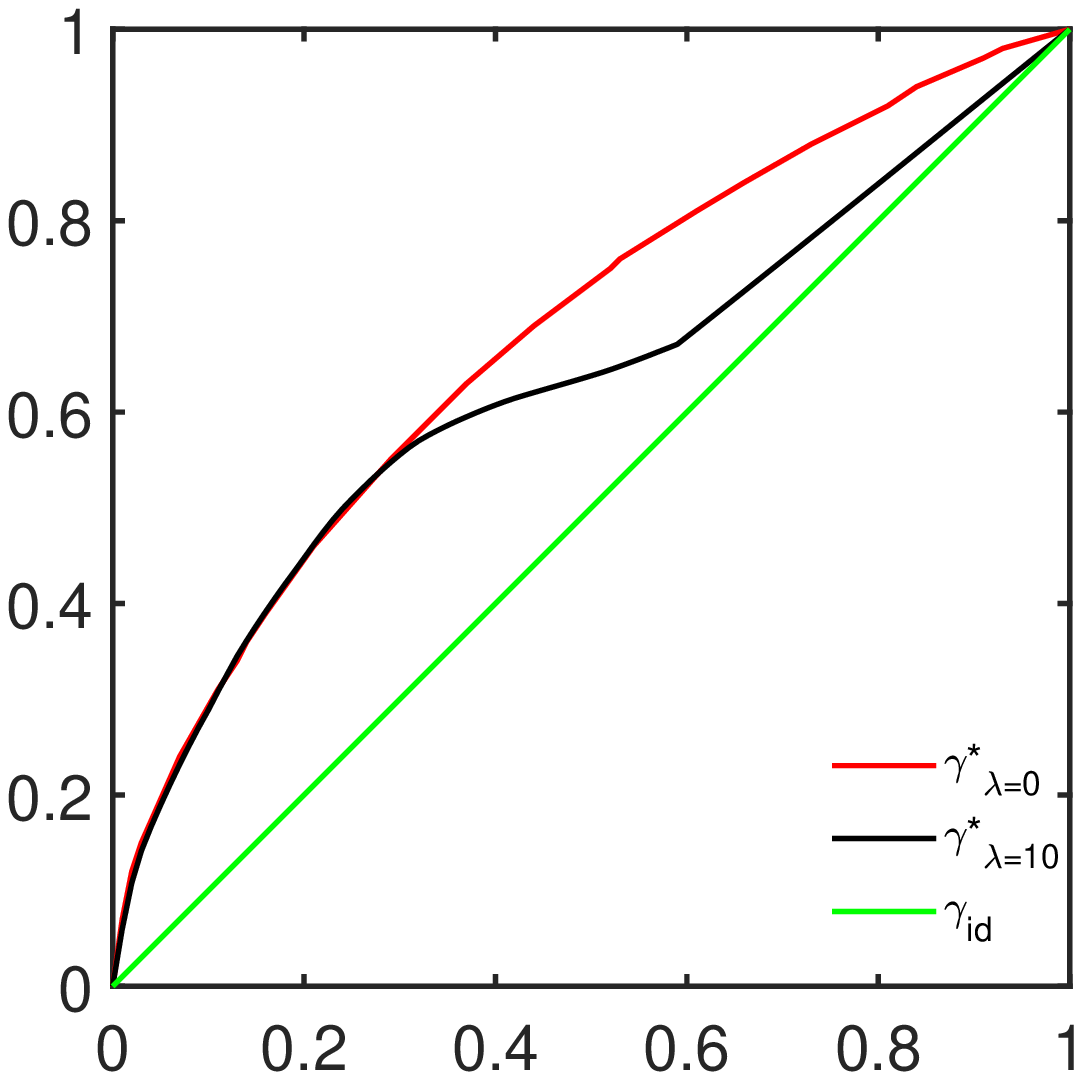}
		\caption{Optimal warpings}
	\end{subfigure}
	\caption{Bayesian registration illustration with diagonal covariance kernel. (a) Original functions and alignment functions. The blue and green solid curves are the two given functions $f_1$ and $f_2$, the red and  black dotted curves are the aligned $f_2$ using Algorithm \ref{alg:AligBayes} with $\lambda$ equal to $0$ and $10$, respectively. (b) The $r(t)$ function which indicates nonuniform penalty in the time domain. (c) Optimal warping functions in the alignment. The red and black curve are the optimal warping functions from Algorithm \ref{alg:AligBayes} with $\lambda$ equal to $0$ and $10$, respectively. The green curve is the identity warping function.}
	\label{fig.reg1}
\end{figure}

We now use one example to illustrate the Bayesian alignment with diagonal covariance kernel. We at first simulate two multimodal functions, i.e., $f_1(t)=6 \cdot 0.8^{20t} \cdot \cos (10\pi t-\frac{\pi}{4})$ and $g(t)=5 \cdot 0.8^{20t} \cdot \sin (10\pi t), t\in [0, 1]$. Then we generate a warped version of $g(t)$ by defining  $f_2(t) = g(\gamma(t))$ with warping functions $\gamma(t)=\frac{e^{2t}-1}{e^{2}-1}$. The functions $f_1(t)$ and $f_2(t)$ are shown as blue and green solid curves in Figure \ref{fig.reg1}(a), respectively. Moreover, we set $h(s,t)=r(t)\delta(s-t)$, where $r(t)=\begin{cases}
	0.025(t+0.1) & \text {if $0 \le t \le 0.6$} \\
	250t & \text {if $0.6 < t \le 1$} 
\end{cases}$.  
$r(t)$ is positive and piecewise linear on $[0, 1]$ (see its graph in Figure\ref{fig.reg1}(b)). Its function value is close to 0 on $[0, 0.6]$, and much larger in magnitude on $(0.6, 1]$, which indicates nonuniform penalty in the time domain.

Similar to the isotropic covariance case, our goal is find optimal warping function $\gamma^*$ to minimize the loss function in Equation \eqref{eq:loss}. The registration results are shown in Figure \ref{fig.reg1}.  When $\lambda = 0, $ and $10$, the optimal warping functions are calculated using Algorithm \ref{alg:AligBayes} and shown in Figure \ref{fig.reg1}(b).  We can see that when there is no penalty (i.e., $\lambda = 0$), the optimal warping can align $f_2$ to $f_1$ very well. The aligned functions $f_2(\gamma^*(t))$ are also shown in  Figure \ref{fig.reg1}(a).  We can see $f_2(\gamma^*(t))$ is right on the top of $f_1(t)$ when $\lambda = 0$. When there is a penalty, the optimal warping at the first part in the domain overlap the optimal warping when $\lambda = 0$, but the latter part gets closer to the identity warping $\gamma_{id}(t) = t$ (optimal warping when $\lambda = \infty$). Indeed, $f_2(\gamma^*(t))$ is right on the top of $f_1(t)$ when $\lambda = 10$ for $t\in [0,0.3]$ and start to be lagged compare to the $f_1(t)$ from $t=0.3$ when $\lambda = 10$.  These results clearly demonstrate the effectiveness of the prior model for the nonuniform constraint in the time domain in the Bayesian alignment process.

\subsubsection{Full covariance}
We have shown two examples to illustrate penalty on time warping using the diagonal terms on the covariance kernel, which essentially describes the variability at each time point.  Now we use another example to illustrate the Bayesian alignment with general non-diagonal covariance which takes into account co-variability between two different time points. In general, $h(s,t)$ in Equation \eqref{eq: Losspsi} can be any symmetric, positive definite kernel. To simplify the illustration, we here assume $h(s,t)$ has the following block form: 
\begin{equation}
h(s,t)=\begin{cases}
	a\,\delta(s-t), & \text {if $s,t\in [0,\frac{1}{2})\times[0,\frac{1}{2})$} \\
	b\,\delta(s-t), & \text {if $s,t\in [\frac{1}{2},1]\times[\frac{1}{2},1]$} \\
	c\,\delta(s-t-\frac{1}{2}), & \text {if $s,t\in [\frac{1}{2},1]\times[0,\frac{1}{2})$} \\
	c\,\delta(s-t+\frac{1}{2}), & \text {if $s,t\in [0,\frac{1}{2})\times[\frac{1}{2},1]$} 
\end{cases}, 
\label{eq:h}
\end{equation}

where $ a > 0, b > 0, \text{ and } ab > c^2$.  It is easy to verify that $h(s,t)$ is symmetric and positive definite. 

Let $\zeta(s)=\phi(s)-\int_{0}^{1}\phi(u)\,du$, the penalty term in the loss function in Equation \eqref{eq: Losspsi} with the given $h(s,t)$ can be rewritten as: 
\begin{equation}
	\begin{split}
                &  \int_{0}^{1}\int_{0}^{1} \Big(\phi(s)-\int_{0}^{1}\phi(u)\,du \Big)h(s,t)\Big(\phi(t)-\int_{0}^{1}\phi(u)\,du\Big) \,ds\,dt\\
		 = & \int_{0}^{\frac{1}{2}} \Big(\sqrt{a}\zeta(s)\,ds+\sqrt{b}\zeta(s+\frac{1}{2})\Big)^2\,ds+2(c-\sqrt{ab})\int_{0}^{\frac{1}{2}} \zeta(s)\zeta(s+\frac{1}{2})\,ds
		\nonumber
	\end{split}
\end{equation}

By fixing the diagonal coefficients $a$ and $b$, we focus on the penalty with respect to the off-diagonal coefficient $c$.  It is easy to see that 1)
if $\int_{0}^{\frac{1}{2}} \zeta(s)\zeta(s+\frac{1}{2})\,ds<0$, then the penalty is a decreasing function of $c$, and 2) if $\int_{0}^{\frac{1}{2}} \zeta(s)\zeta(s+\frac{1}{2})\,ds>0$, then the penalty is an increasing function of $c$,
We now use two simulations to illustrate these two cases, respectively. In each case, we let $a=b=1$ and set $c=0.9,0.2,-0.2,-0.9$ to see how the co-variates will affect the function alignment. 
\begin{enumerate}
	\item[Case 1.]  [$\int_{0}^{\frac{1}{2}} \zeta(s)\zeta(s+\frac{1}{2})\,ds<0$]: We first simulate one bimodal function $f_1(t)=2\sin (4\pi t), t\in [0,1]$, and then warp $f_1$  to get $f_2$ as follows: $$f_2(t)=\begin{cases}
		f_1(0.5\gamma_1(2t)), & \text {if $0 \le t \le 0.5$} \\
		f_1(0.5\gamma_2(2t-1)+0.5), & \text {if $0.5 < t \le 1$}
	\end{cases}, $$
where $\gamma_i(t)=\frac{e^{a_it}-1}{e^{a_i}-1},\, i=1,\, 2$, with $a_1 = -5 , a_2 = 5$. The functions $f_1(t)$ and $f_2(t)$ are shown as yellow and green solid curves, respectively, in Figure \ref{fig.reg2}(a). The optimal warping functions $\gamma^*(t)$ are shown in  Figure \ref{fig.reg2}(b).  We can see that when there is no penalty, the optimal warping can align $f_2$ to $f_1$ very well.  The corresponding $f_2(\gamma^*(t))$ in Figure\ref{fig.reg2}(a) stays right on the top of $f_1(t)$ when $\lambda = 0$ . When $\lambda > 0$, same as the previous two simulation examples, the optimal warping also gets closer to the identity warping $\gamma_{id}(t) = t$. In addition, it can be seen that as $c$ becomes smaller, the optimal warping gets closer to the identity warping, and the corresponding $f_2(\gamma^*(t))$ moves further away from $f_1$.  We point out that because $2(c-\sqrt{ab})\int_{0}^{\frac{1}{2}} \zeta(s)\zeta(s+\frac{1}{2})\,ds$ is always non-negative, the off-diagonal terms add more penalty to the warping than the diagonal terms only, given in the integration $\int_{0}^{\frac{1}{2}} \Big(\sqrt{a}\zeta(s)\,ds+\sqrt{b}\zeta(s+\frac{1}{2})\Big)^2\,ds$.

\begin{figure}[h]
	\centering
	\begin{subfigure}[h]{0.5\textwidth}
		\includegraphics[width=\textwidth]{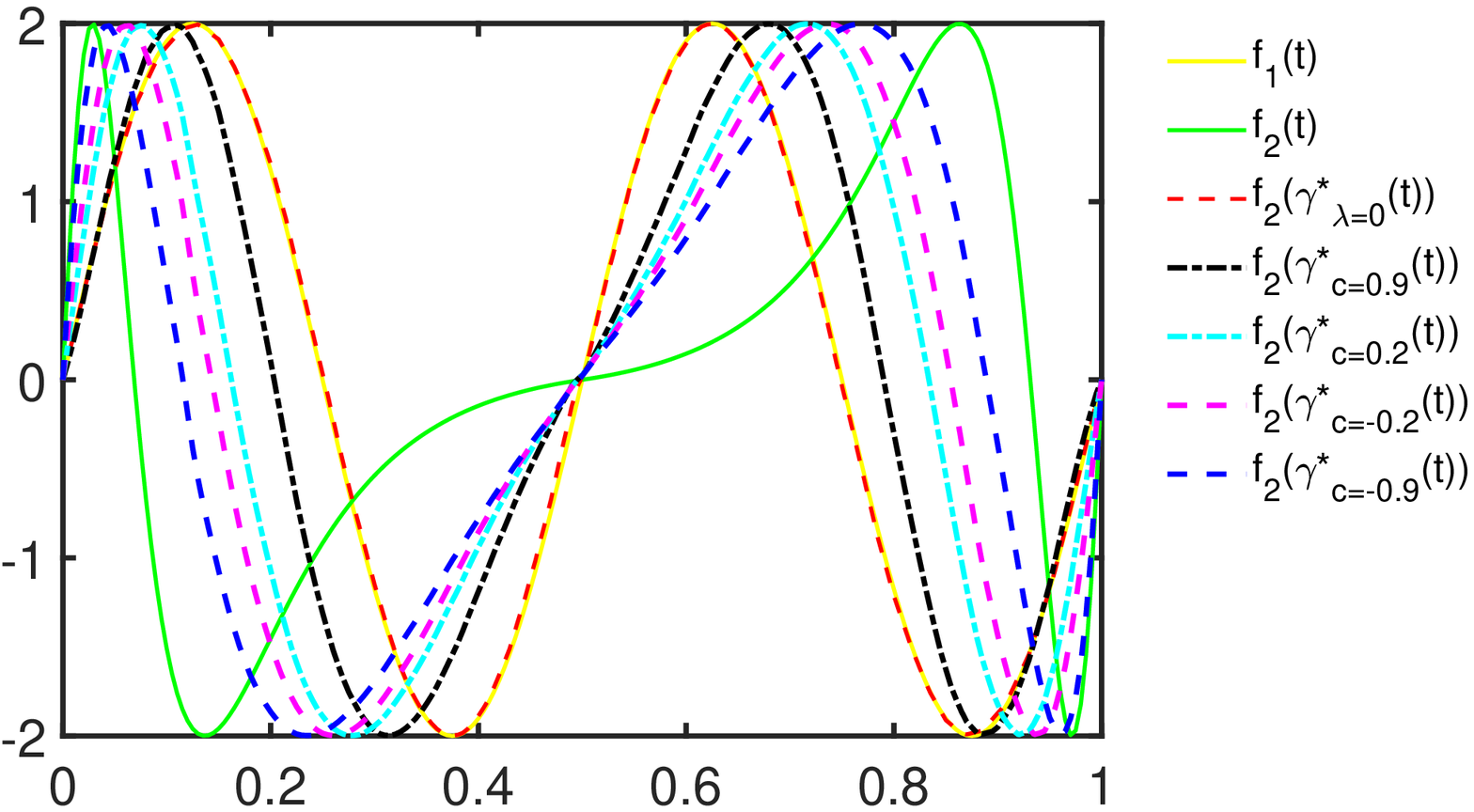}
		\caption{Alignment result}
	\end{subfigure}\hspace{-0.2cm}
	\begin{subfigure}[h]{0.35\textwidth}
		\includegraphics[width=\textwidth]{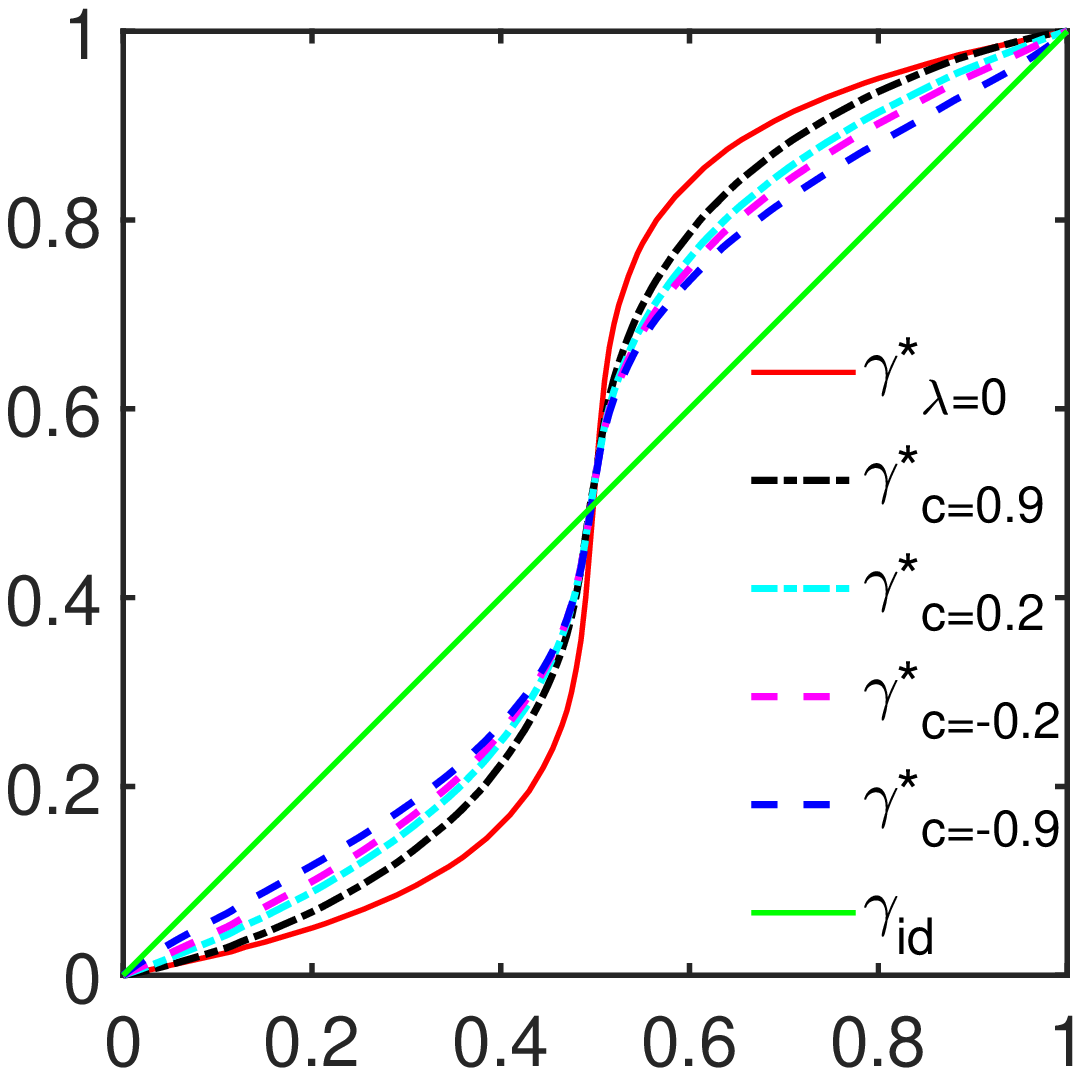}
		\caption{Optimal warpings}
	\end{subfigure}

	\caption{Bayesian registration illustration with full covariance for Case 1. (a) Original functions and alignment functions. The yellow and green solid curves are the two given functions $f_1$ and $f_2$, respectively, and the red dotted curve is the aligned $f_2$ without penalty.  The black, cyan, magenta, and blue dotted curve are the aligned $f_2$ with penalty term with $c$ in Equation \eqref{eq:h} equal to $0.9$, $0.2$, $-0.2$, and $-0.9$, respectively.  (b) Optimal warping functions in the alignment. The red curve is the optimal warping functions with $\lambda$ equal to $0$. The black, cyan, magenta, and blue dotted curve are the optimal warping functions with penalty term with $c$ equal to $0.9$, $0.2$, $-0.2$, and $-0.9$, respectively.}
	\label{fig.reg2}
\end{figure}

	\item[Case 2.]  [$\int_{0}^{\frac{1}{2}} \zeta(s)\zeta(s+\frac{1}{2})\,ds \ge 0$]: We first simulate one multi-modal function $f_1(t)=2\sin (8\pi t), t\in [0,1]$, and then warp $f_1$ to get $f_2$ as follows: 
	$$f_2(t)=\begin{cases}
		f_1(0.25\gamma_1(4t)), & \text {if $t \in [0,0.25)$} \\
		f_1(0.25\gamma_2(4t-1)+0.25), & \text {if $t \in [025,0.5)$} \\
		f_1(0.25\gamma_1(4t-2)+0.5), & \text {if $t \in  [0.5, 0.75)$} \\
		f_1(0.25\gamma_2(4t-3)+0.75), & \text {if $t\in [0.75, 1]$} 
	\end{cases},  $$ 
 where $\gamma_i(t)=\frac{e^{a_it}-1}{e^{a_i}-1},\, i=1,\, 2$, with $a_1 = -5 , a_2 = 5$.  The functions $f_1(t)$ and $f_2(t)$ are shown as yellow and green solid curves, respectively, in Figure \ref{fig.reg3}(a).  The optimal warping functions $\gamma^*(t)$ are shown in  Figure \ref{fig.reg3}(b).   Same as in case 1, when there is no penalty, the optimal warping can align $f_2$ to $f_1$ very well.  When $\lambda > 0$, same as in the previous two simulation examples, the optimal warping also gets closer to the identity warping $\gamma_{id}(t) = t$. However, in contrast to the result in Case 1, as $c$ becomes larger, the optimal warping gets closer to the identity warping $\gamma_{id}(t) = t$, and the corresponding $f_2(\gamma^*(t))$ moves further away from $f_1$.  Because now $2(c-\sqrt{ab})\int_{0}^{\frac{1}{2}} \zeta(s)\zeta(s+\frac{1}{2})\,ds$ is always non-positive, the off-diagonal terms reduce penalty from the diagonal terms.  This further demonstrates of the effect of the off-diagonal terms in Bayesian registration. 
 
\end{enumerate}


\begin{figure}[h]
	\centering
	\begin{subfigure}[h]{0.5\textwidth}
		\includegraphics[width=\textwidth]{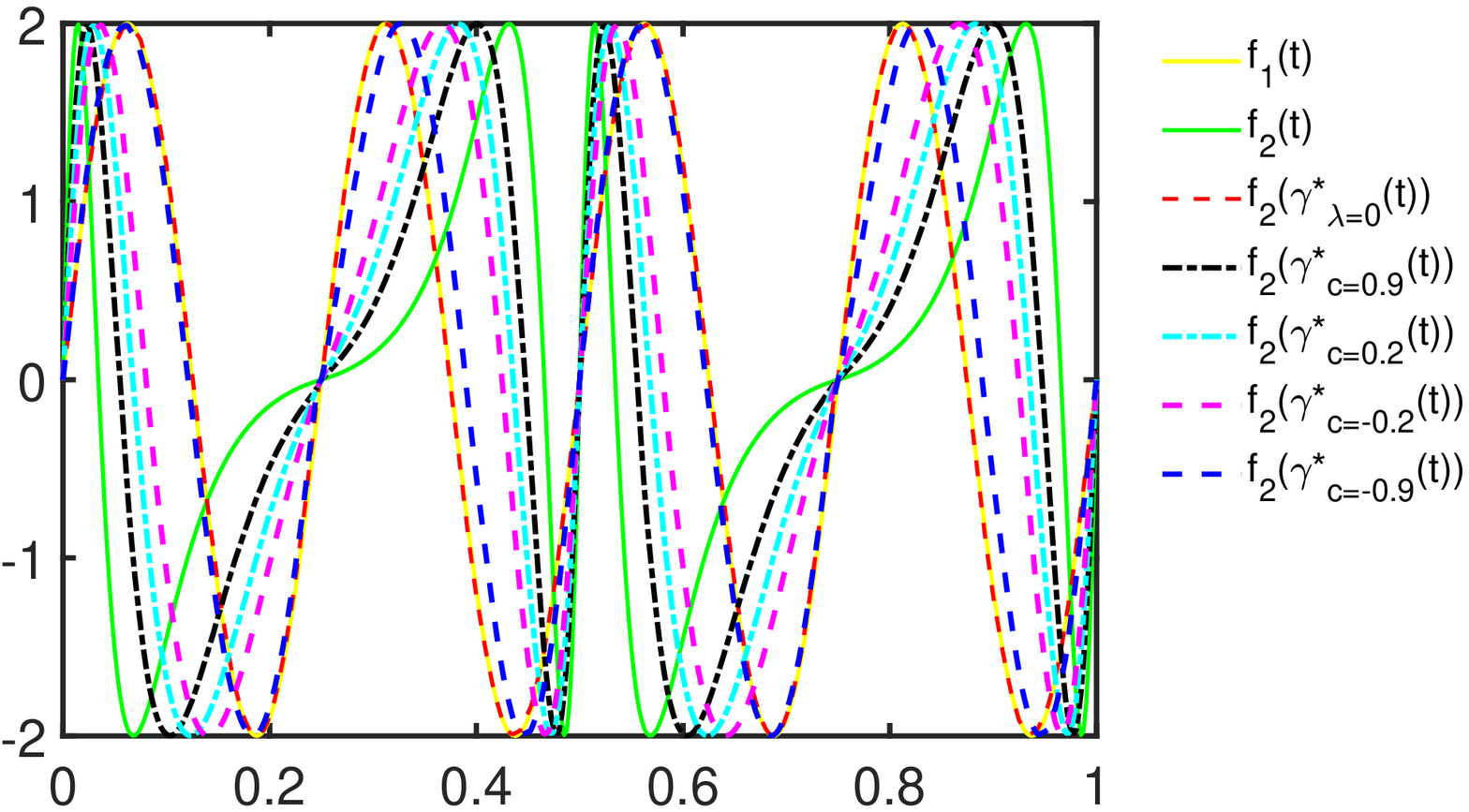}
		\caption{Alignment result}
	\end{subfigure}\hspace{-0.2cm}
	\begin{subfigure}[h]{0.35\textwidth}
		\includegraphics[width=\textwidth]{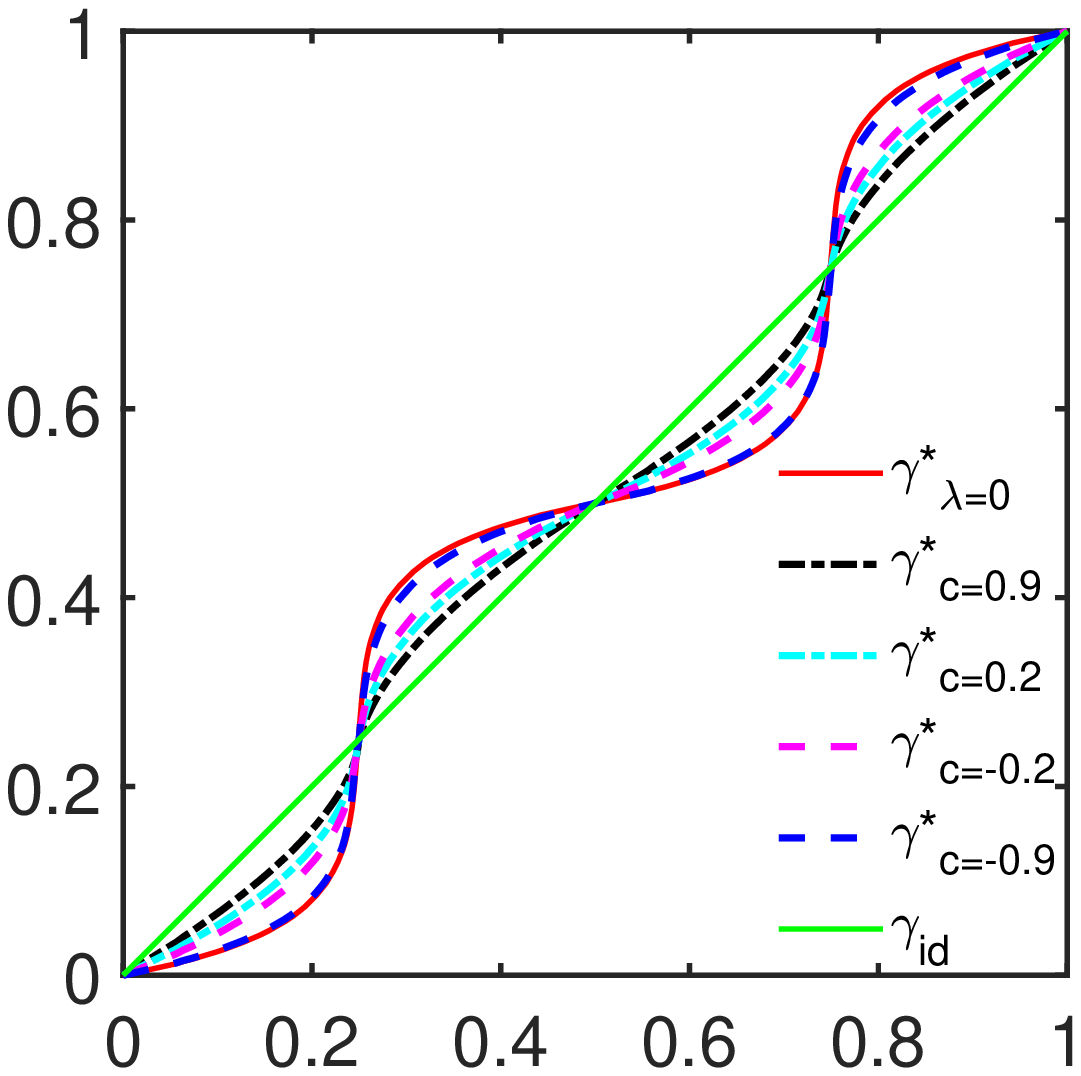}
		\caption{Optimal warpings}
	\end{subfigure}
	\caption{Same as Figure \ref{fig.reg2} except for illustration with full covariance for Case 2.} 
	\label{fig.reg3}
\end{figure}

\section{Real data application}
\label{sec: app}
In this section, we will apply our method to the well-known Berkeley Growth curve data (available at https://rdrr.io/cran/fda/man/growth.html), where the heights of 39 boys and 54 girls were recorded at thirty-one time points from age 1 to age 18 \citep{ramsay2006functional}.  As each growth curve is increasing in the age interval [1, 18], it is also a warping function, albeit with a different domain and range.  To examine the variability of time warping, we linearly transform the growth functions into standard warping functions from $[0, 1]$ onto $[0, 1]$.  The original recording time points are not evenly spaced, and we adopt a smoothing procedure using cubic splines. The smoothed curves in both male and female groups are shown in Column (a) of 
Figure \ref{fig.berkeley}.  We will use the proposed method in Section \ref{fPCA} to model and resample these observations. 
\subsection{Modeling and resampling}
\begin{figure}[h]
	\begin{adjustwidth}{-1cm}{-1cm}
		\centering
		\begin{subfigure}[t]{\dimexpr0.2\textwidth+20pt\relax}
			\makebox[20pt]{\raisebox{40pt}{\rotatebox[origin=c]{90}{Female}}}%
			\includegraphics[width=\dimexpr\linewidth-20pt\relax]
			{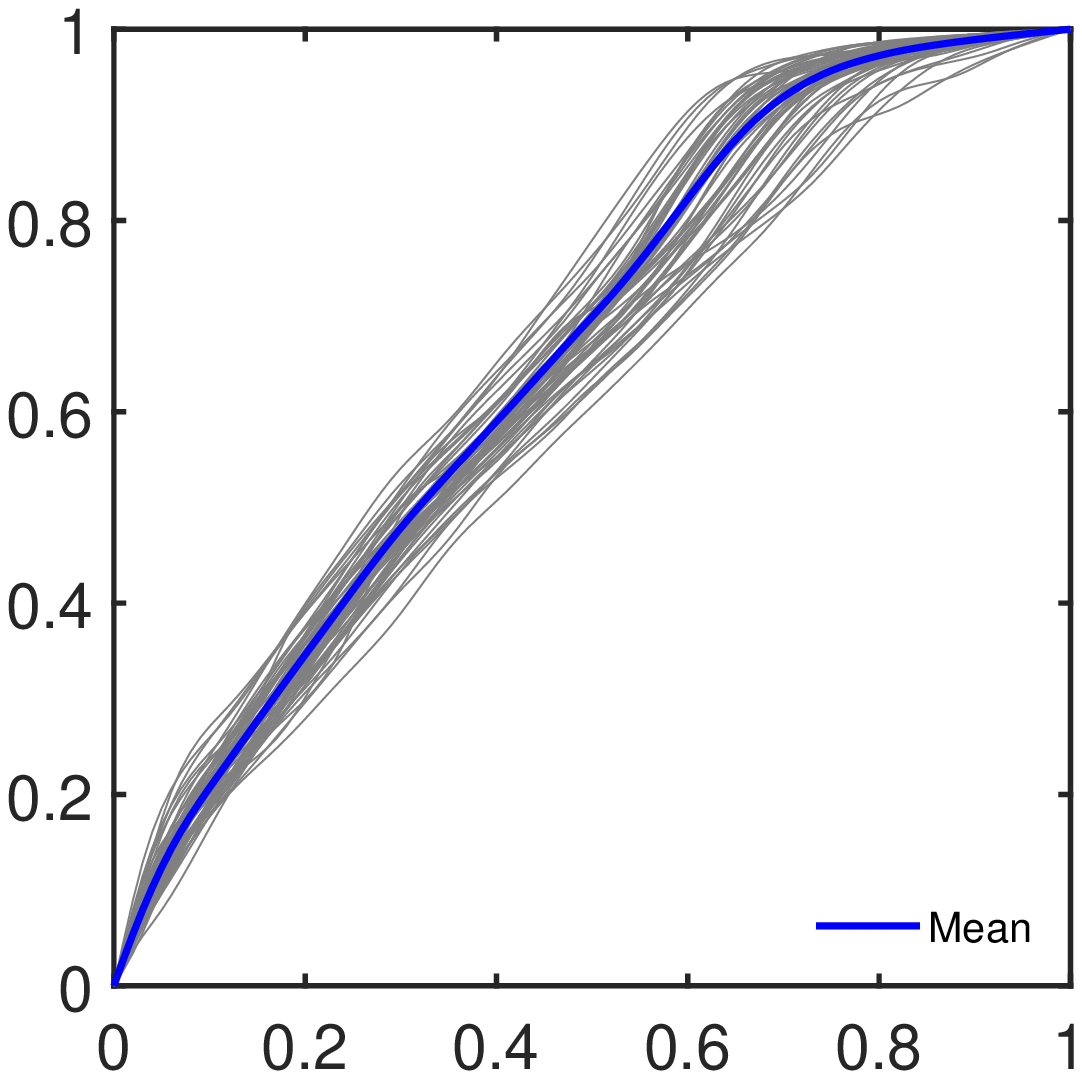}
			\makebox[20pt]{\raisebox{40pt}{\rotatebox[origin=c]{90}{Male}}}%
			\includegraphics[width=\dimexpr\linewidth-20pt\relax]
			{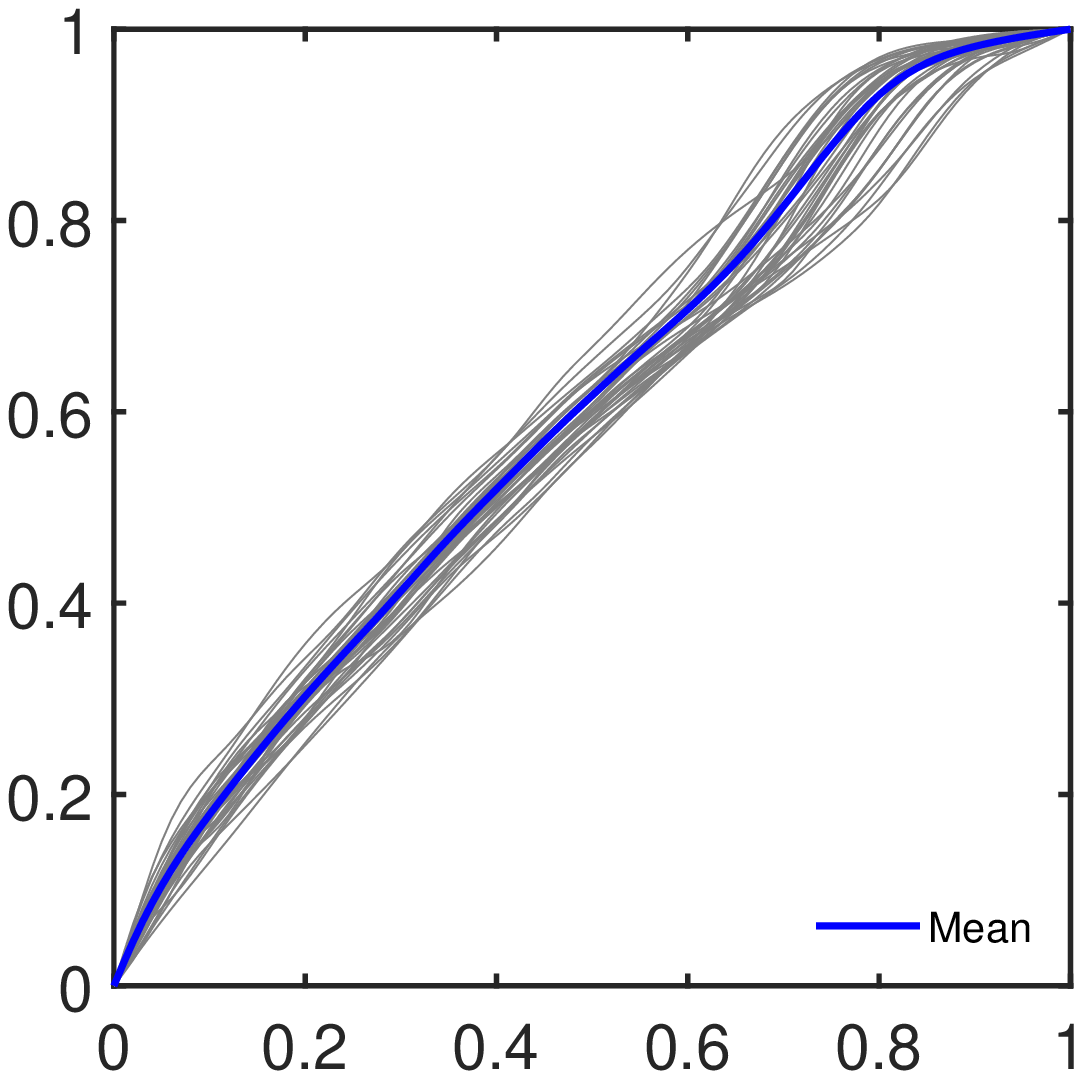}
			\caption{Original}
		\end{subfigure}	\hspace{-0.7cm}
		\begin{subfigure}[t]{0.2\textwidth}
			\includegraphics[width=\textwidth]  
			{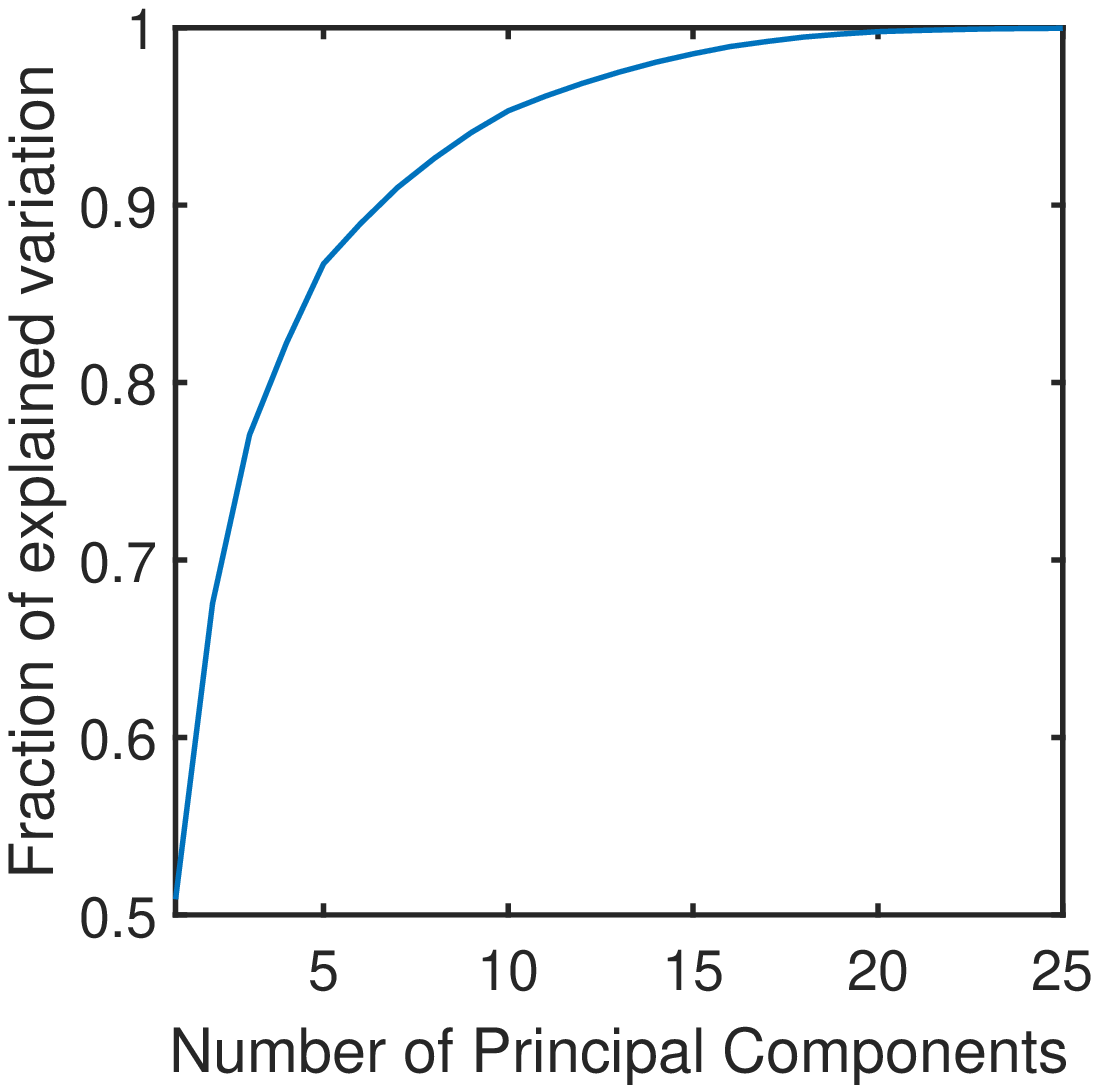}
			\includegraphics[width=\textwidth]
			{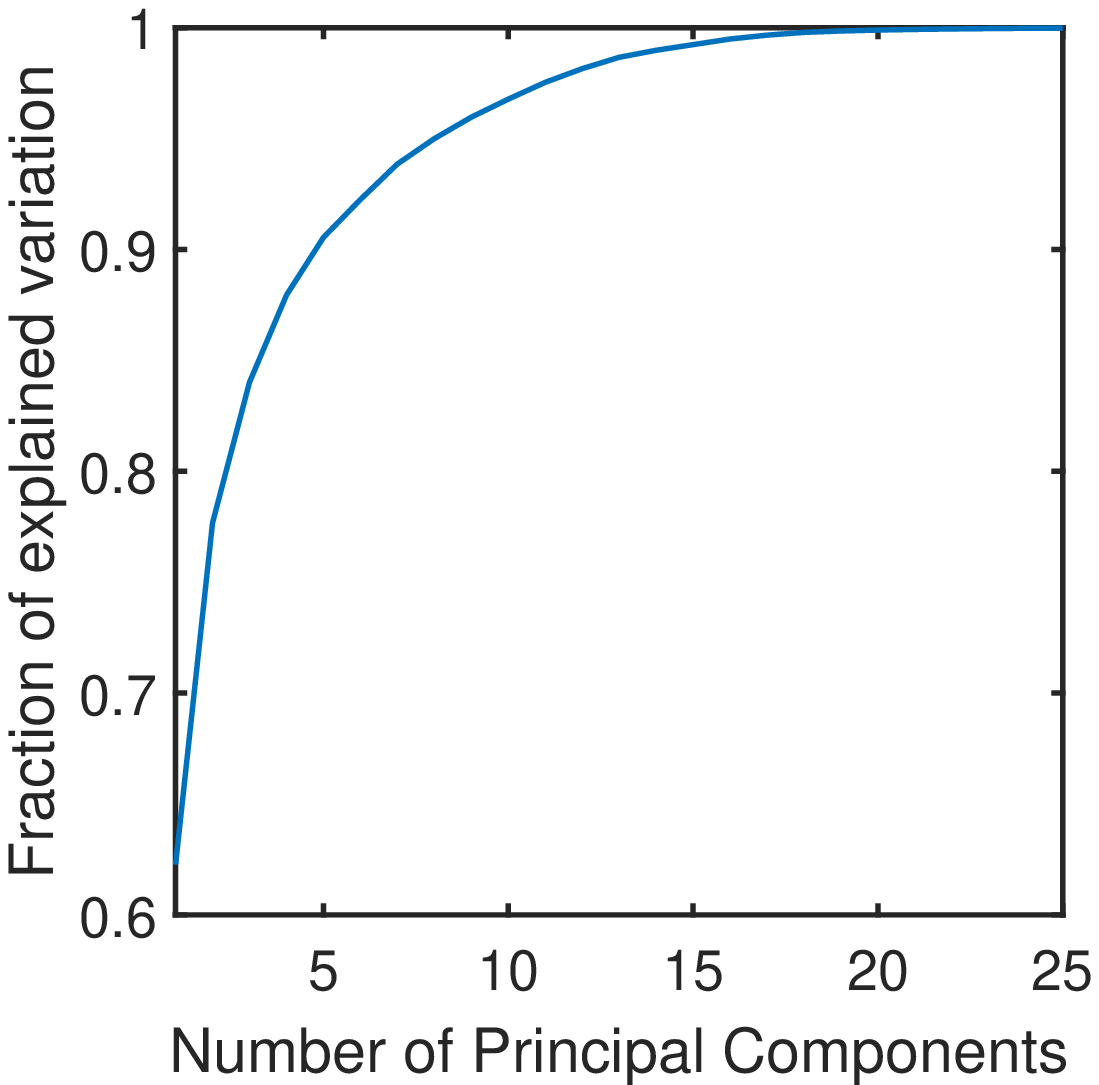}
			\caption{Eigenvalues}
		\end{subfigure}	\hspace{-0.7cm}
		\begin{subfigure}[t]{0.20\textwidth}
			\includegraphics[width=\textwidth]  
			{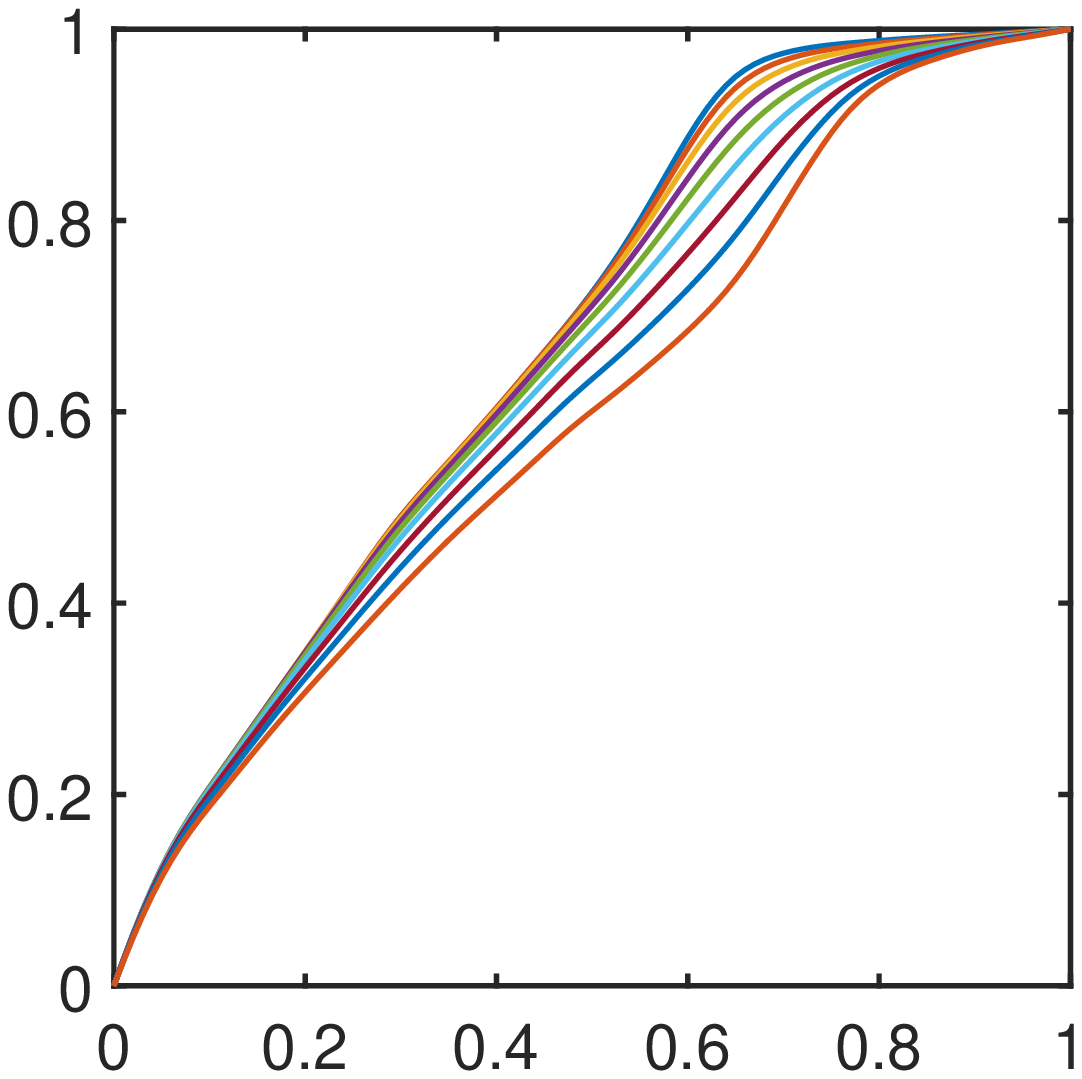}
			\includegraphics[width=\textwidth]
			{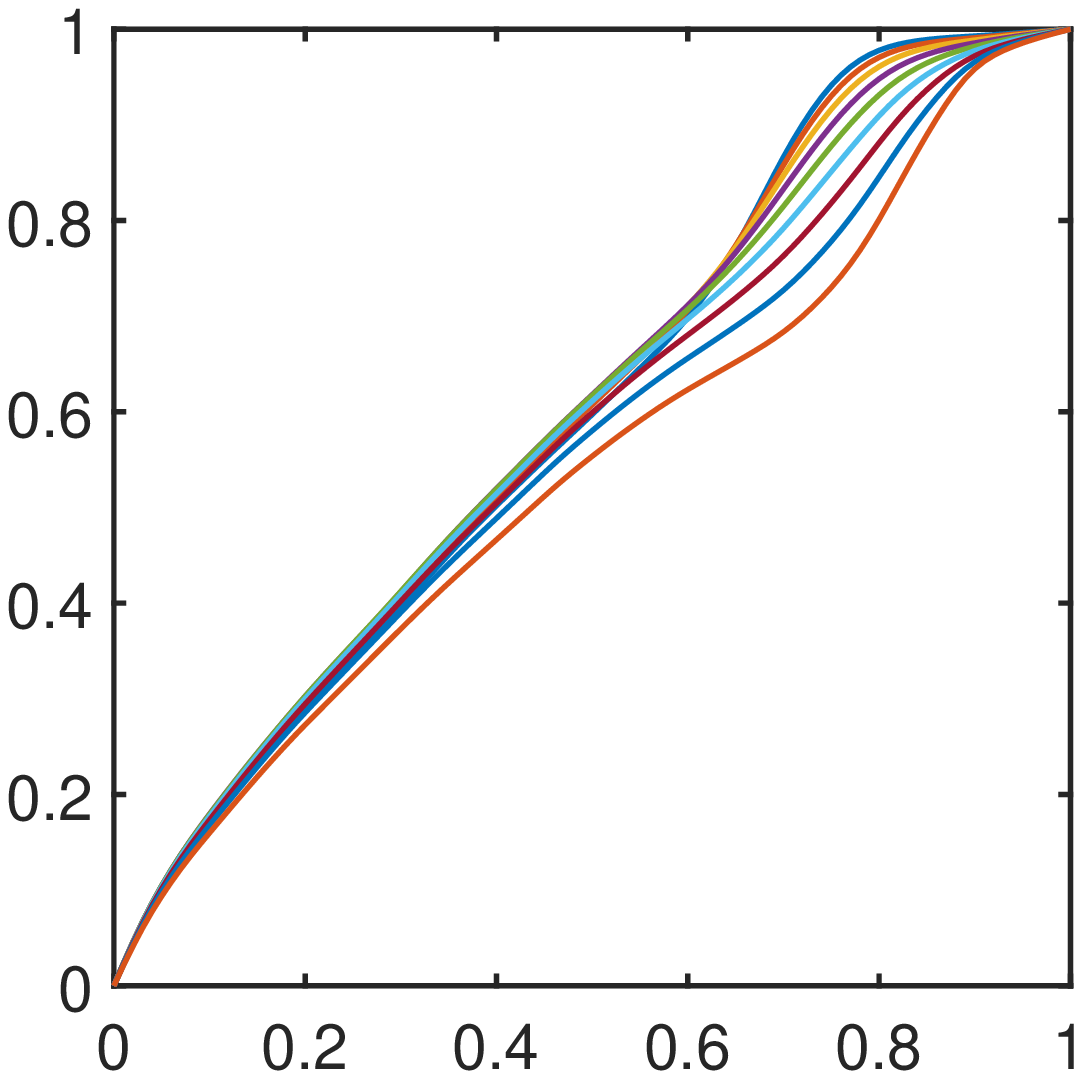}
			\caption{1st}
		\end{subfigure}	\hspace{-0.7cm}
		\begin{subfigure}[t]{0.20\textwidth}
			\includegraphics[width=\textwidth]  
			{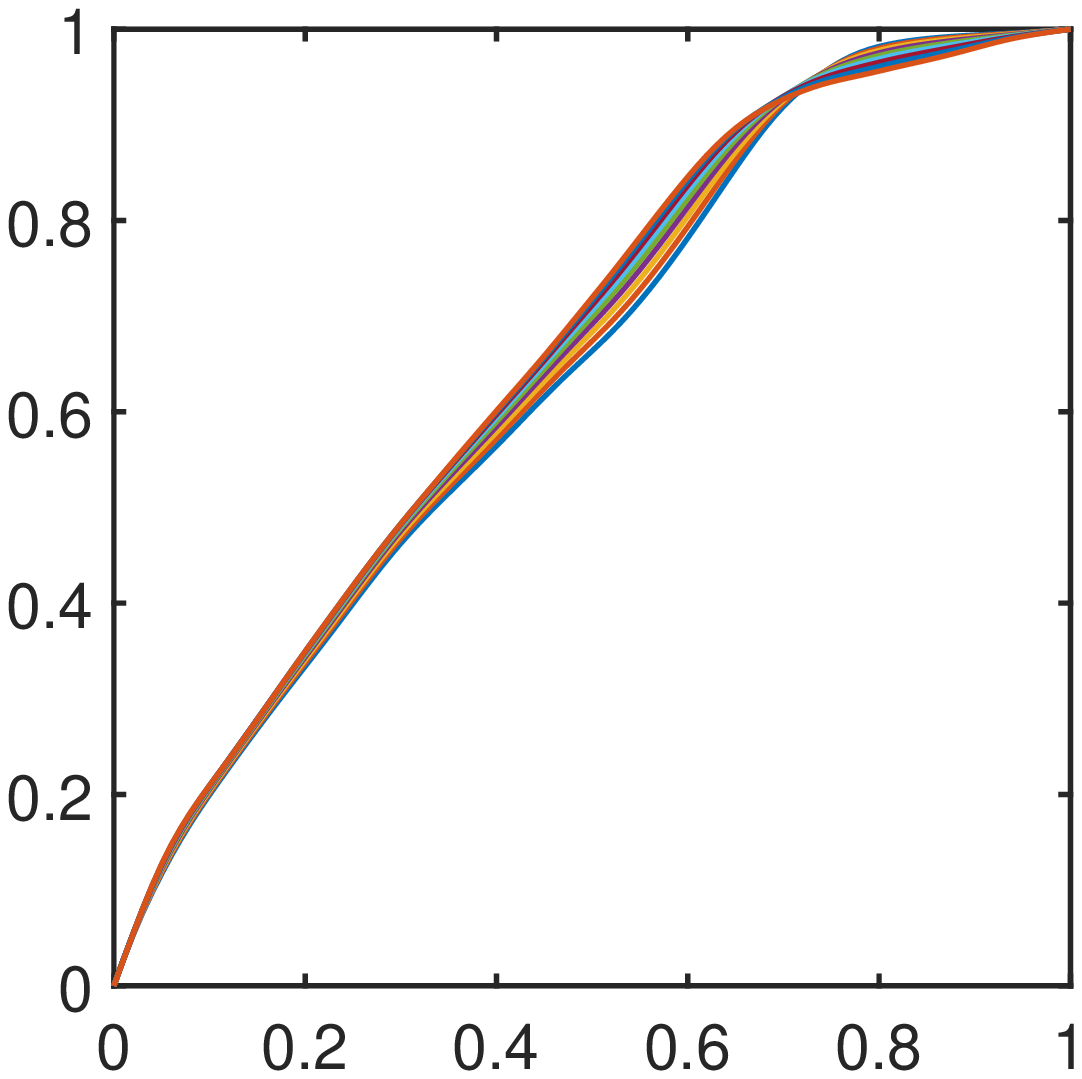}
			\includegraphics[width=\textwidth]
			{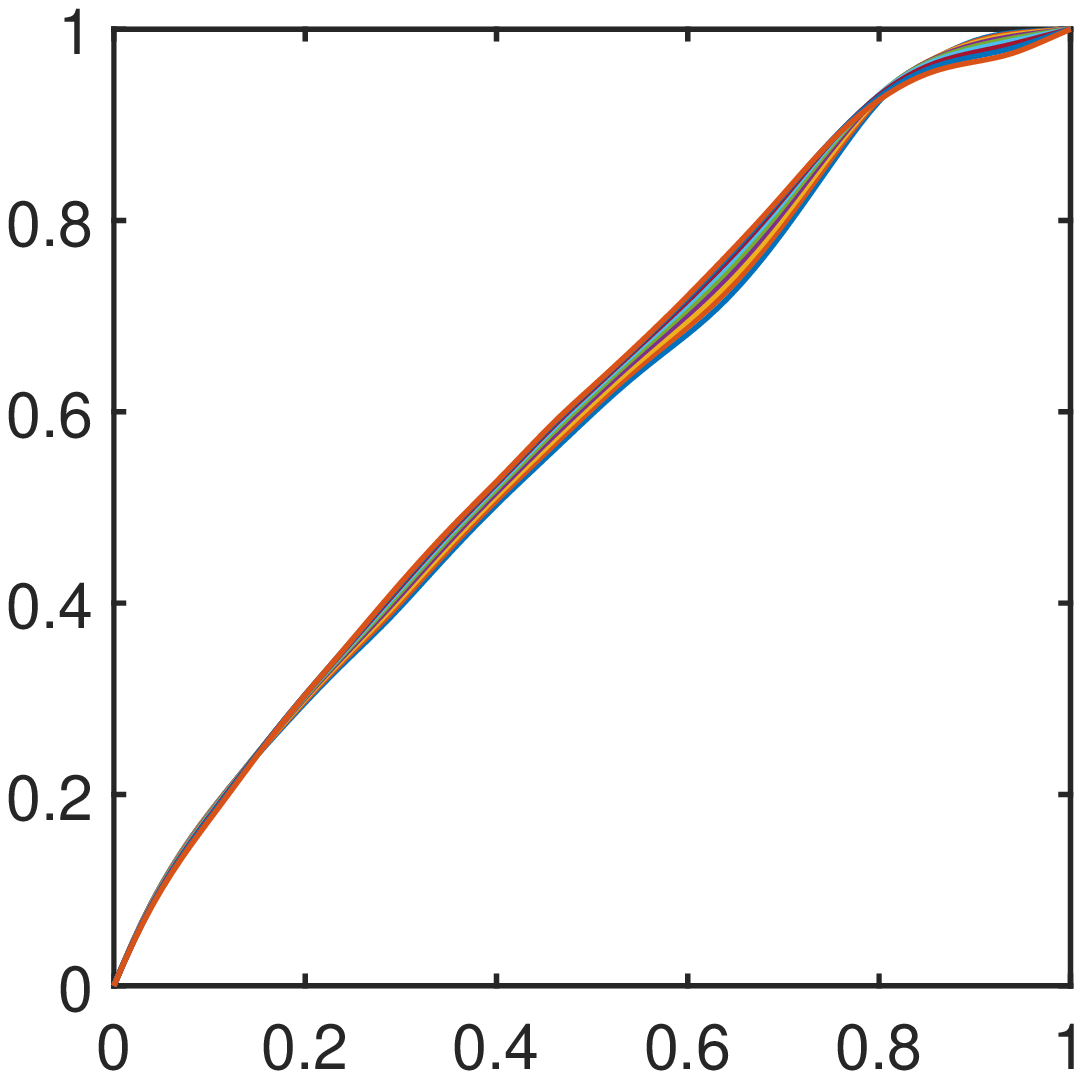}
			\caption{2nd}
		\end{subfigure}	\hspace{-0.7cm}
		\begin{subfigure}[t]{0.20\textwidth}
			\includegraphics[width=\textwidth]  
			{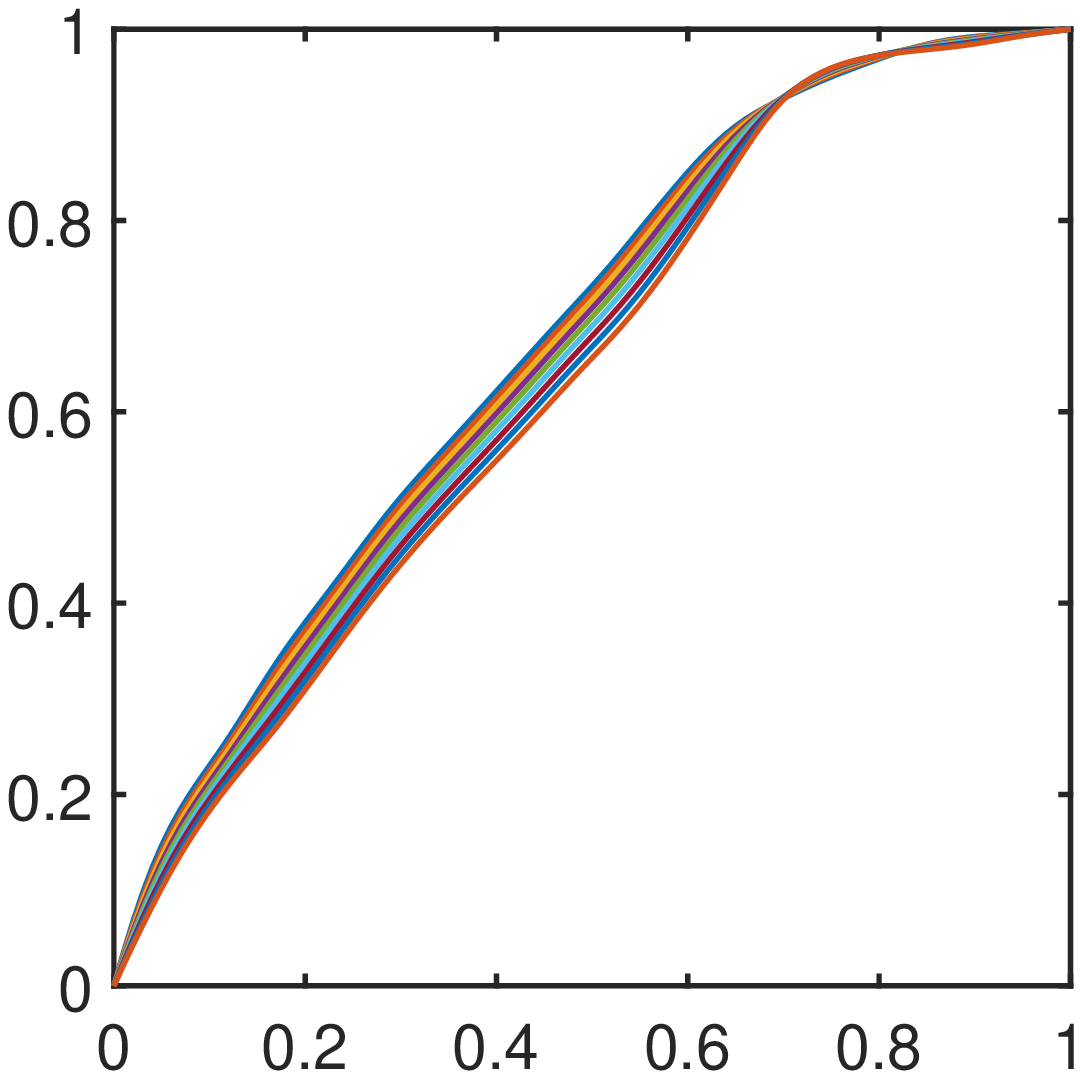}
			\includegraphics[width=\textwidth]
			{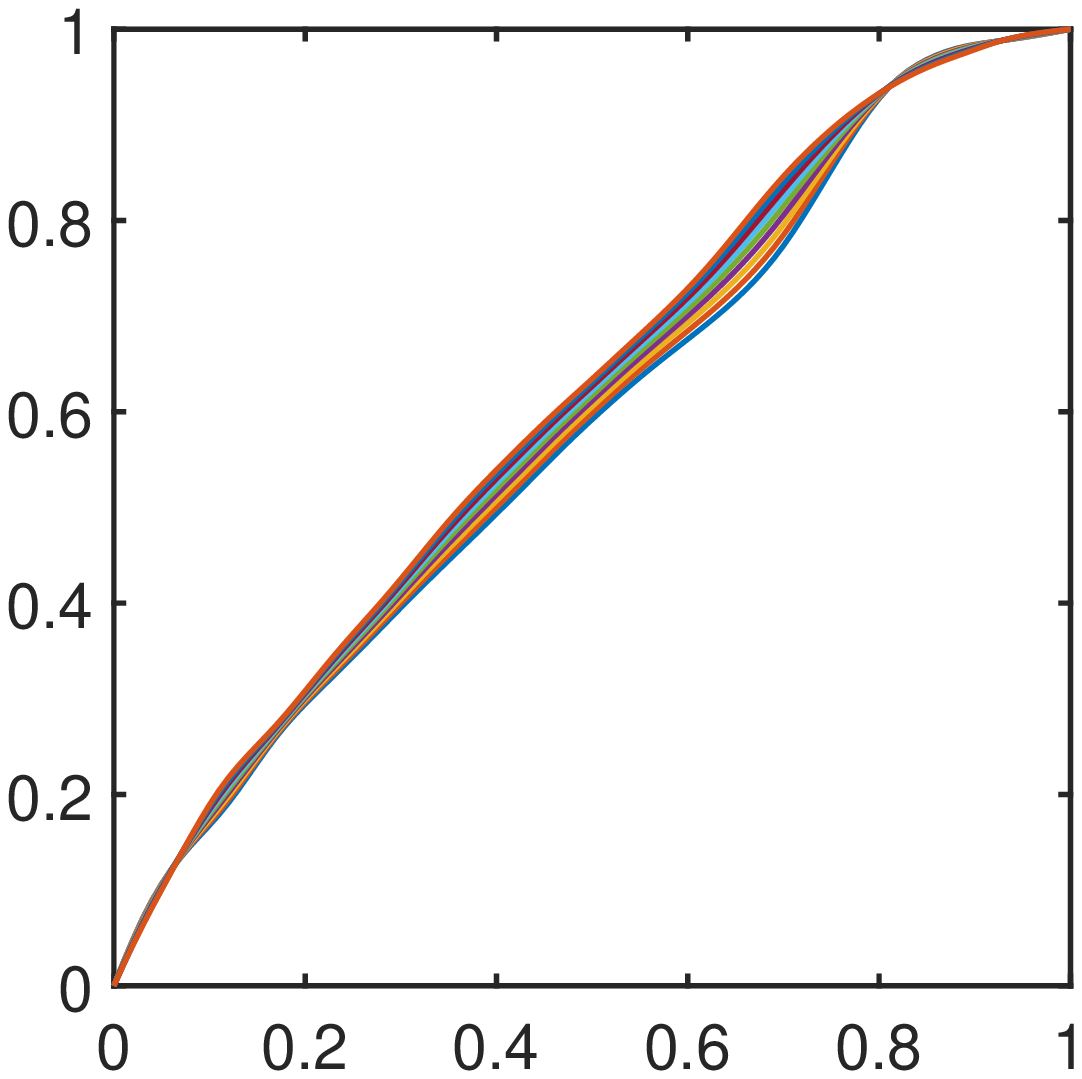}
			\caption{3rd}
		\end{subfigure}	\hspace{-0.7cm}
		\begin{subfigure}[t]{0.20\textwidth}
			\includegraphics[width=\textwidth]  
			{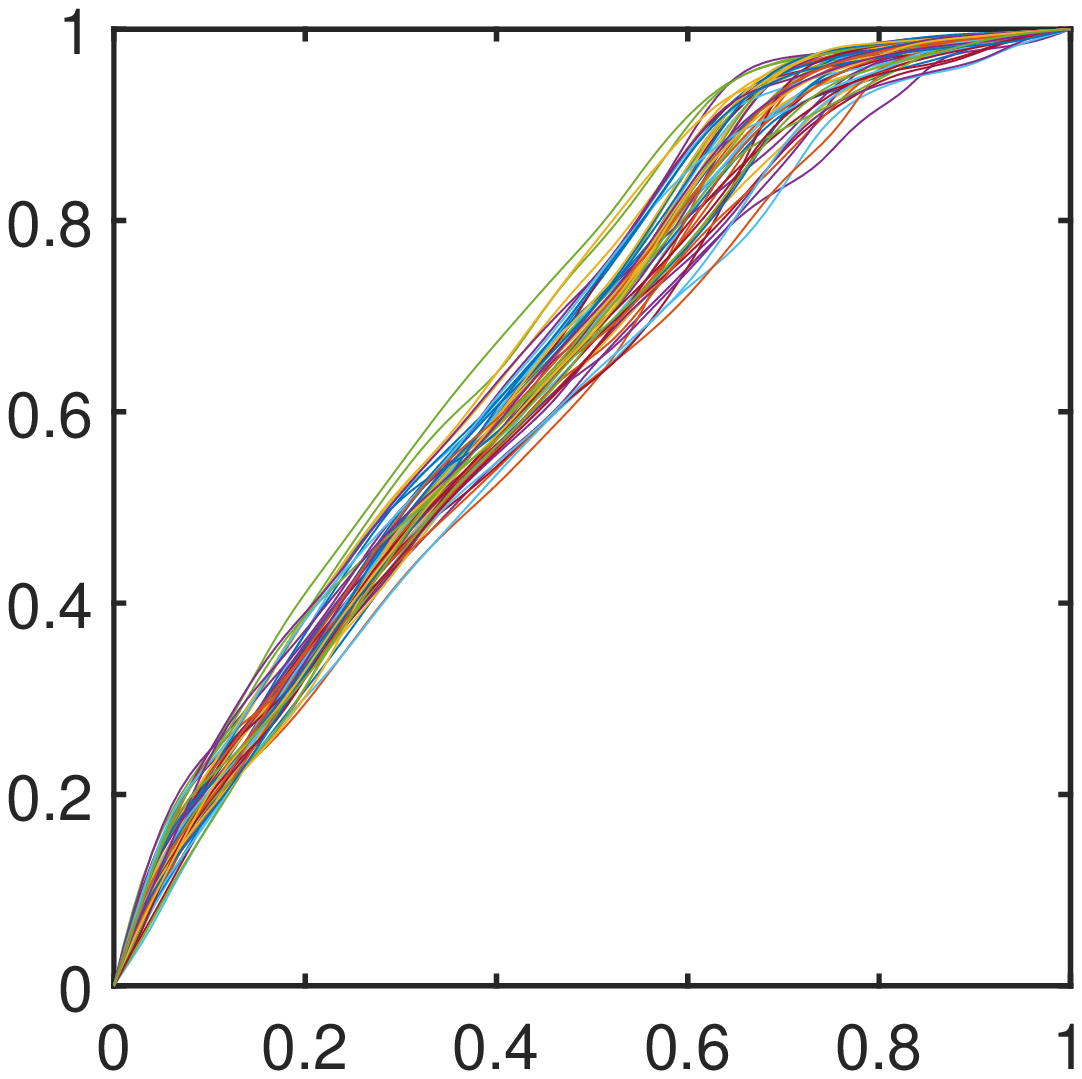}
			\includegraphics[width=\textwidth]
			{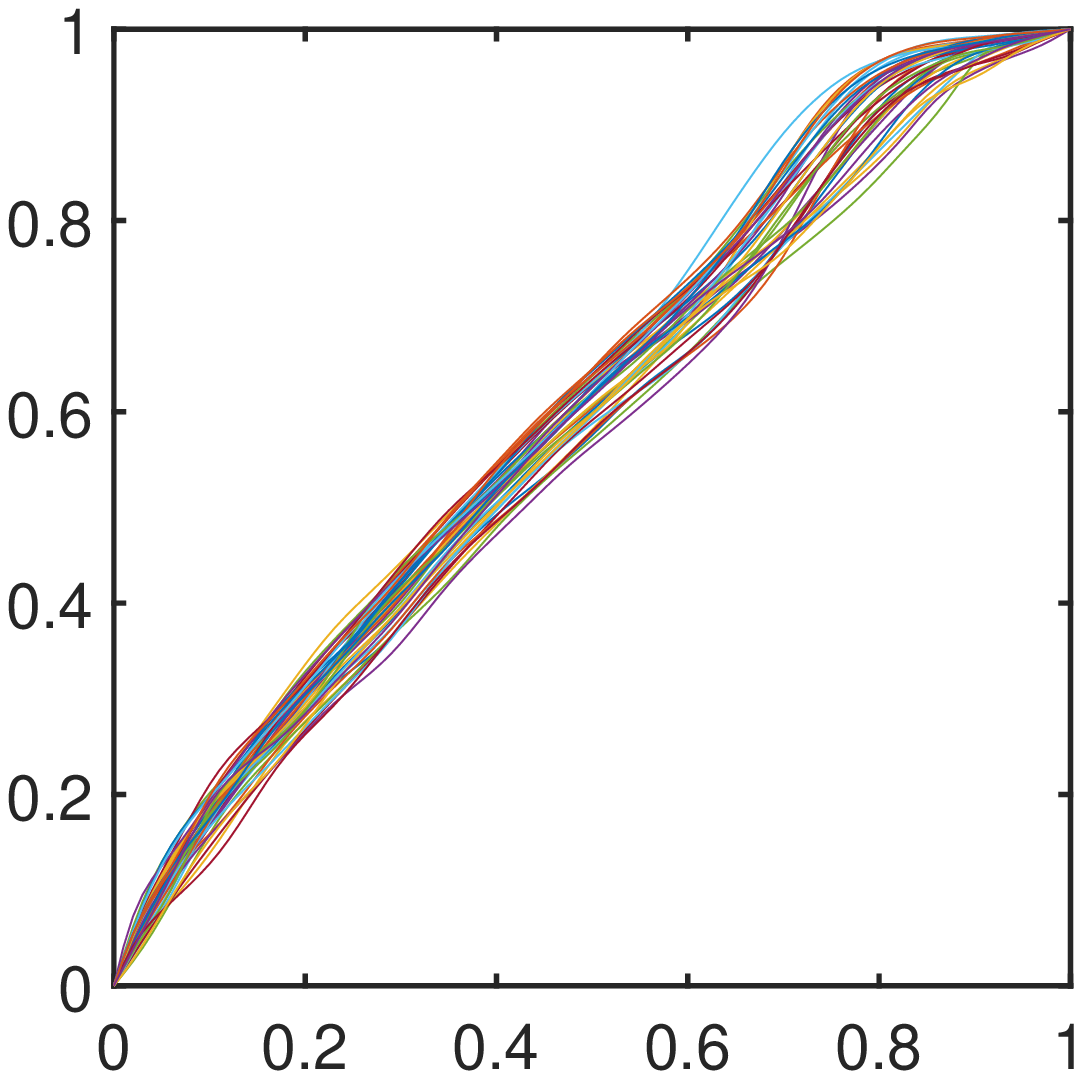}
			\caption{Resample}
		\end{subfigure}
	\end{adjustwidth}
	\caption{Results on Berkeley growth data for the female group (1st row) and male group (2nd row). (a) Grey curves represent original growth functions in the linearly-transformed space from [0, 1] to [0, 1]. The blue curve represents the mean function, given in Definition \ref{def:mean}. (b) Fraction of variance explained by the first $n$ principal components. (c) Function curves in the form of $\hat{\mu} + c\hat{\lambda}_1^{1/2}\hat{f}_1$ to visualize the effect of the first eigenfunction as perturbation from the mean, where $\hat \mu$ is the estimated mean function, $c$ ranges from -2 to 2 with a 0.5 step size, $\hat f_1$ is the estimated first eigenfunction, and $\hat \lambda_1$ is the estimated first eigenvalue. (d) and (e), same as (c) except for the second and third eigenfunctions and eigenvalues, respectively.  (f) resampled functions with the same sample size as the original ones.}
	\label{fig.berkeley}
\end{figure}

Figure \ref{fig.berkeley} also shows the principal component analysis and resampling result. The eigenvalues are shown in Column (b). We can see that the first twenty principal components explain almost 100\% of the total variance in both groups. To visualize the variability contributed by the first three eigenfunctions, we plot curves representing the effects of these three eigenfunctions as perturbation from the mean in Columns (c), (d) and (e), respectively. Moreover, we adopt the conventional Gaussian kernel method to estimate the first twenty coefficient distributions. Based on the estimated results, we use Algorithm \ref{alg:MPCA} to resample the same amount of warping functions as the original dataset, and the resampling results are shown in Column (e). It can be easily seen that the resampled curves also look very similar to the original curves in Column (a) in either group.  This indicates the effectiveness of the fPCA modeling procedure in practice. 

\subsection{fANOVA}
We then apply the proposed framework to test if there is any significant difference between the mean growth curves of male and female groups. We first use the CLR-transformation in Equation \eqref{eq: twclt} to convert growth warping curves to functions in a Euclidean space, this becomes a classical two-sample problem for functional data, and we adopt the functional ANOVA method for comparison \citep{zhang2013analysis}.
Without any Gaussian process assumption on the transformed data, we can use a bootstrap approach with 10000 replicates for comparison. The functional ANOVA has two types of bootstrap test statistics: $\mathbb L^2$-norm-based test statistic and the $F$-type test statistic. It is found that the corresponding test statistics are $572.9428$ and $49.7896$, respectively, and the associated p-values for both statistics are less than $10^{-4}$. This indicates a significant difference between the mean growth curves of females and males. 

\subsection{Classification with logistic regression}
We finally use the CLR-transformed warping functions as a predictor to classify whether the growth curve is male or female.
Table \ref{tab:cm} shows the classification confusion matrix calculated using the first two eigenfunctions. It turns out that 33 out of 39 male growth curves and 47 out of 54 female growth curves were correctly classified by the model. Table \ref{tab:classper} presents the corresponding classification performance with various criteria such as precision, sensitivity, specificity, accuracy, and F-measure. We can see that all these criteria have high values at around 0.85.  This desirable performance indicates that the logistic regression using transformed warping functions as an explanatory variable is an appropriate classification method.  
	
\begin {table}[H]
	\caption {Confusion Matrix} \label{tab:cm} 
	\begin{center}
		\begin{tabular}{l|l|c|c|c}
			\multicolumn{2}{c}{}&\multicolumn{2}{c}{True gender}&\\
			\cline{3-4}
			\multicolumn{2}{c|}{}&Male&Female&\multicolumn{1}{c}{Total}\\
			\cline{2-4}
			\multirow{2}{*}{Classification Result}& Male & $33$ & $7$ & $40$\\
			\cline{2-4}
			& Female & $6$ & $47$ & $53$\\
			\cline{2-4}
			\multicolumn{1}{c}{} & \multicolumn{1}{c}{Total} & \multicolumn{1}{c}{$39$} & \multicolumn{1}{c}{$54$} & \multicolumn{1}{c}{$93$}\\
		\end{tabular}
	\end{center}
\end {table}

\begin {table}[H]
	\caption {Classification Performance} \label{tab:classper} 
	\begin{center}
		\begin{tabular}{||c c c c c c c c c||} 
			\hline
			\rowcolor{Gray}
			TP & FP & FN & TN& precision & sensitivity & specificity& accuracy & F-Measure  \\ [0.5ex] 
			\hline\hline
			33 & 7 & 6 & 47 & 0.83 & 0.85 & 0.87 & 0.86 & 0.84 \\ [1ex] 
			\hline
			47 & 6 & 7 & 33 & 0.89 & 0.87 & 0.85 & 0.86 & 0.88 \\ [1ex] 
			\hline
		\end{tabular}
	\end{center}
\end {table}

\section{Summary}
\label{sec6}
In this paper, we have proposed a new framework to model time warping functions as a linear inner-product space, which is an apparent advantage over the previous nonlinear approximation methods.  The critical element of this process is a derivative operation of the warping function and then a centered logratio transformation to transform the warping functions into a Euclidean space. We have also defined two warping spaces to make the transformation mathematically precise. The first one, bounded warping space $\Gamma_1$, is isometrically isomorphic to the space of bounded, centered $\mathbb L^2$ functions. We extended this bounded $\mathbb L^2$ to a Hilbert space, mapping it to a more general warping space $\Gamma_2$. These two warping spaces provide sufficient representation for practical use. We then stated several statistical inferences under this new framework, including using fPCA to construct a model for functional warping observations, performing fANOVA for group comparisons of warpings, and performing regressions with the time warping functions as explanatory variables. We also applied our new framework in Bayesian registration to provide time-variant and temporally correlated constraints in function alignment. Finally, we illustrate the method in a real-world dataset and obtain reasonable result.  

We point out that $\Gamma_2$ is not a vector space, which limits its usefulness in the modeling process. We will aim to extend the warping space to a Hilbert space in the future. If this can be done, the warping will be fully described by a stochastic process in Euclidean space. In addition, we have used a Gaussian process prior for warping in the Bayesian registration. A more general non-Gaussian process will be explored to capture more complex variabilities in practical data.

\bibliographystyle{plainnat}  
\bibliography{references}  

\newpage

\section*{Appendices}
\small
{\bf Supporting Information: Additional information for this article is available} 

\subsection*{A. Proof of Proposition \ref{prop} }
\label{appa}

For any $n\in \mathbb{N}$,
\[\sum_{i=1}^{n}\bigg|\mu_i \phi_i(s) \phi_i(t)\bigg|\leq \sum_{i=1}^{n}\bigg|\mu_i \sqrt{2} \sqrt{2}\bigg|=2 \sum_{i=1}^{n} \mu_i \leq 2 \sum_{i=1}^{\infty} \mu_i<\infty\]
Then, 

\[\sum_{i=1}^{\infty}\bigg|\mu_i \phi_i(s) \phi_i(t)\bigg|= \lim_{n\to\infty} \sum_{i=1}^{n}\bigg| \mu_i \phi_i(s) \phi_i(t)\bigg|\leq \lim_{n\to\infty}2\sum_{i=1}^{n} \mu_i = 2 \sum_{i=1}^{\infty} \mu_i<\infty	, \quad \forall s,t\in [0,1]\]

\noindent Thus, $K(s,t)=\sum_{i=1}^{\infty}\mu_i \phi_i(s) \phi_i(t)$ converges absolutely.\\
For any $s,t\in [0,1]$, we have: 
\[\bigg|K(s,t)-\sum_{i=1}^{n}\mu_i \phi_i(s) \phi_i(t)\bigg|\leq \sum_{i=n+1}^{\infty}\bigg|\mu_i \phi_i(s) \phi_i(t)\bigg|\leq 2 \sum_{i=n+1}^{\infty} \mu_i\]
As $\sum_{i=1}^{\infty}\mu_i<\infty$, we can get: 
\[\lim_{n\rightarrow \infty}\bigg|K(s,t)-\sum_{i=1}^{n}\mu_i \phi_i(s) \phi_i(t)\bigg|=0. \]
Thus, $\sum_{i=1}^{\infty}\mu_i \phi_i(s) \phi_i(t)$ converges uniformly.	\\

We will then prove that $K$ is symmetric, non-negative definite, and continuous:
\begin{itemize}
	\item {\bf Symmetry:} It is easy to see that $K(s,t)=\sum_{i=1}^{\infty}\mu_i \phi_i(s) \phi_i(t)=\sum_{i=1}^{\infty}\mu_i \phi_i(t) \phi_i(s)=K(t,s), \quad \forall s,t\in [0,1]$.
	\item {\bf Non-negative definiteness:} $\forall f \in L^2([0,1])$, we have:
	\begin{eqnarray}
		\int_{0}^{1}\int_{0}^{1}f(s)K(s,t)f(t)dsdt
		&=& \int_{0}^{1}\int_{0}^{1}f(s)\sum_{i=1}^{\infty}\mu_i \phi_i(s) \phi_i(t) f(t)dsdt\nonumber \\
		&=& \sum_{i=1}^{\infty}\mu_i \int_{0}^{1}\int_{0}^{1} f(s) \phi_i(s) \phi_i(t) f(t)dsdt\nonumber \\
		&=&\sum_{i=1}^{\infty}\mu_i \bigg(\int_{0}^{1}f(s) \phi_i(s)ds\bigg)^2\geq 0 \nonumber
	\end{eqnarray}
	\item {\bf Continuity:} Define $K_n(s,t):=\sum_{i=1}^{n}\mu_i \phi_i(s) \phi_i(t)$.  As $K$ is uniformly convergent, for any $\epsilon>0$ there exists $n_\epsilon \in \mathbb{N}$ such that for any $s,t\in [0,1]$:
	\[\bigg|K(s,t)-K_{n_\epsilon}(s,t)\bigg|<\epsilon/3\]
	Because the basis function $\phi_i$ is uniformly continuous, there exists $\delta>0$ such that for any $s,s',t,t'\in [0,1]$:
	\[\bigg|K_{n_\epsilon}(s,t)-K_{n_\epsilon}(s',t')\bigg|=\bigg|\sum_{i=1}^{n_\epsilon}\mu_i \phi_i(s) \phi_i(t)-\sum_{i=1}^{n_\epsilon}\mu_i \phi_i(s') \phi_i(t')\bigg|<\epsilon/3,\]
	whenever $|s-s'|<\delta$ and  $|t-t'|<\delta$.
	Hence, 
	\[\bigg|K(s,t)-K(s',t')\bigg|\leq \bigg|K(s,t)-K_{n_\epsilon}(s,t)\bigg|+\bigg|K_{n_\epsilon}(s,t)-K_{n_\epsilon}(s',t')\bigg|+\bigg|K_{n_\epsilon}(s',t')-K(s',t')\bigg|<\epsilon.\]	  
\end{itemize}

\subsection*{B. Calculation of Gradient of $J(\phi)$}
To get $\nabla J(\phi)$, we define $\tilde{\phi} = \phi+\epsilon g$, where $\epsilon\in \mathbb{R}, \, g\in L^2(0,1)$, then,
\begin{eqnarray}
	\begin{split}
	J(\tilde{\phi}) &=\int_{0}^{1}-2q_1(t)q_2\bigg(\int_{0}^{t}\exp((\phi+\epsilon g)(s))\,ds\bigg)\sqrt{\exp((\phi+\epsilon g)(t))}\,dt \\
	 & + \lambda\int_{0}^{1}\int_{0}^{1} \Big(\phi(s)-\int_{0}^{1}\phi(u)\,du \Big)h(s,t)\Big(\phi(t)-\int_{0}^{1}\phi(u)\,du\Big) \,ds\,dt \nonumber \\
	&= J_1(\tilde{\phi})+\lambda J_2(\tilde{\phi}). \nonumber
	\end{split}
\end{eqnarray}

The directional derivative of $J$ in the direction $g$ is given by, $D_g J(\phi)= \langle\nabla J(\phi),g\rangle =\frac{d J(\tilde{\phi})}{d \epsilon}\bigg|_{\epsilon=0}$, we calculate it by two parts, the first part:
\begin{equation}
	\begin{split}
		\frac{d J_1(\tilde{\phi})}{d \epsilon}
		&=\int_{0}^{1}-2q_1(t)\dot{q}_2\bigg(\int_{0}^{t}\exp((\phi+\epsilon g)(s))\,ds\bigg)\int_{0}^{t}\exp((\phi+\epsilon g)(u)) g(u)\,du\sqrt{\exp((\phi+\epsilon g)(t))}\,dt\\
		\quad
		&-\int_{0}^{1}q_1(t)q_2\bigg(\int_{0}^{t}\exp((\phi+\epsilon g)(s))\,ds\bigg)\sqrt{\exp((\phi+\epsilon g)(t))}g(t)\,dt. \\
		\nonumber
	\end{split}
\end{equation}
The second part is:
\begin{equation}
	\begin{split}
		\frac{d J_2(\tilde{\phi})}{d \epsilon}
		& = \int_{0}^{1}\int_{0}^{1} (\phi+\epsilon g)(s)h(s,t)(\phi+\epsilon g)(t)\,ds\,dt-\int_{0}^{1}(\phi+\epsilon g)(u)\,du\int_{0}^{1}\int_{0}^{1}(\phi+\epsilon g)(s)h(s,t)\,ds\,dt\\
		& -\int_{0}^{1}(\phi+\epsilon g)(u)\,du\int_{0}^{1}\int_{0}^{1}h(s,t)(\phi+\epsilon g)(t)\,ds\,dt
		+\Big(\int_{0}^{1}(\phi+\epsilon g)(u)\,du\Big)^2\int_{0}^{1}\int_{0}^{1}h(s,t)\,ds\,dt.
		\nonumber
	\end{split}
\end{equation}

Let $\epsilon = 0$:
\begin{equation}
	\begin{split}
		D_g J_1(\phi)
		&=\int_{0}^{1}-2q_1(t)\dot{q}_2\bigg(\int_{0}^{t}\exp(\phi(s))\,ds\bigg)\int_{0}^{t}\exp(\phi (u)) g(u)\,du\sqrt{\exp(\phi(t))}\,dt- \int_{0}^{1}	q_1(t)q_2\bigg(\int_{0}^{t}\exp(\phi(s))\,ds\bigg)\sqrt{\exp(\phi(t))}g(t)\,dt\\
		&=-2\int_{0}^{1}\int_{0}^{t}q_1(t)\dot{q}_2\bigg(\int_{0}^{t}\exp(\phi(s))\,ds\bigg)\exp(\phi(u))g(u)\sqrt{\exp(\phi(t))}\,du \, dt- \int_{0}^{1}	q_1(t)q_2\bigg(\int_{0}^{t}\exp(\phi(s))\,ds\bigg)\sqrt{\exp(\phi(t))}g(t)\,dt\\
		&= -2\int_{0}^{1}\int_{u}^{1}q_1(t)\dot{q}_2\bigg(\int_{0}^{t}\exp(\phi(s))\,ds\bigg)\sqrt{\exp(\phi(t))}\,dt	\exp(\phi(u))g(u)\,du - \int_{0}^{1}	q_1(t)q_2\bigg(\int_{0}^{t}\exp(\phi(s))\,ds\bigg)\sqrt{\exp(\phi(t))}g(t)\,dt\\
		&= -2\int_{0}^{1}\int_{t}^{1}q_1(u)\dot{q}_2\bigg(\int_{0}^{u}\exp(\phi(s))\,ds\bigg)\sqrt{\exp(\phi(u))}\,du	\exp(\phi(t))g(t)\,dt - \int_{0}^{1}	q_1(t)q_2\bigg(\int_{0}^{t}\exp(\phi(s))\,ds\bigg)\sqrt{\exp(\phi(t))}g(t)\,dt\\
		&= \Bigg\langle g,\, -2\exp(\phi(t))\int_{t}^{1}q_1(u)\dot{q}_2\bigg(\int_{0}^{u}\exp(\phi(s))\,ds\bigg)\sqrt{\exp(\phi(u))}\,du	-q_1(t)q_2\bigg(\int_{0}^{t}\exp(\phi(s))\,ds\bigg)\sqrt{\exp(\phi(t))}\Bigg\rangle.
		\nonumber
	\end{split}
\end{equation}
And,

\begin{equation}
	\begin{split}
		D_g J_2(\phi)
		& = \int_{0}^{1}\int_{0}^{1} g(s)h(s,t)\phi(t)\,ds\,dt+\int_{0}^{1}\int_{0}^{1} \phi(s)h(s,t)g(t)\,ds\,dt -\int_{0}^{1}g(u)\,du\int_{0}^{1}\int_{0}^{1}\phi(s)h(s,t)\,ds\,dt\\
		&-\int_{0}^{1}\phi(u)\,du\int_{0}^{1}\int_{0}^{1}g(s)h(s,t)\,ds\,dt -\int_{0}^{1}g(u)\,du\int_{0}^{1}\int_{0}^{1}h(s,t)\phi(t)\,ds\,dt 
		-\int_{0}^{1}\phi(u)\,du\int_{0}^{1}\int_{0}^{1}h(s,t)g(t)\,ds\,dt\\
		& +2\int_{0}^{1}\phi(u)\,du\int_{0}^{1}g(v)\,dv\int_{0}^{1}\int_{0}^{1}h(s,t)\,ds\,dt\\
		& = \langle g, \int_{0}^{1} h(t,s)\phi(s)\,ds \rangle + \langle g, \int_{0}^{1} \phi(s)h(s,t)\,ds \rangle - \langle g, \int_{0}^{1}\int_{0}^{1} \phi(s)h(s,u)\,ds\,du \rangle- \langle g, \int_{0}^{1}\phi(u)\,du\int_{0}^{1}h(t,s)\,ds \rangle\\
		& - \langle g, \int_{0}^{1}\int_{0}^{1} h(s,u)\phi(u)\,ds\,du \rangle- \langle g, \int_{0}^{1} h(s,t)\,ds\int_{0}^{1}\phi(u)\,du \rangle +2\langle g,\int_{0}^{1}\phi(u)\,du\int_{0}^{1}\int_{0}^{1}h(s,v)\,ds\,dv\rangle.
		\nonumber
	\end{split}
\end{equation}

Thus, the gradient is given by 
\begin{equation}
	\begin{split}
		\nabla J(\phi)
		&=-2\exp(\phi(t))\int_{t}^{1}q_1(u)\dot{q}_2\bigg(\int_{0}^{u}\exp(\phi(s))\,ds\bigg)\sqrt{\exp(\phi(u))}\,du	-q_1(t)q_2\bigg(\int_{0}^{t}\exp(\phi(s))\,ds\bigg)\sqrt{\exp(\phi(t))}\\
		& +\lambda\bigg(\int_{0}^{1} h(t,s)\phi(s)\,ds + \int_{0}^{1} \phi(s)h(s,t)\,ds - \int_{0}^{1}\int_{0}^{1} \phi(s)h(s,u)\,ds\,du -\int_{0}^{1}\phi(u)\,du\int_{0}^{1}h(t,s)\,ds\\
		& -\int_{0}^{1}\int_{0}^{1} h(s,u)\phi(u)\,ds\,du-\int_{0}^{1} h(s,t)\,ds\int_{0}^{1}\phi(u)\,du
		+2\int_{0}^{1}\phi(u)\,du\int_{0}^{1}\int_{0}^{1}h(s,v)\,ds\,dv\bigg).
	\end{split}
\end{equation}

In particular, we show two special cases on the covariance structure.
\begin{enumerate}
\item $h$ is a diagonal covariance: By setting $h(s,t)=r(t)\delta(t-s)$), we can derive the gradient as: 
\begin{equation}
	\begin{aligned}
		\nabla J(\phi)=  &-2\exp(\phi(t))\int_{t}^{1}q_1(\mu)\dot{q}_2\Big(\int_{0}^{\mu}\exp(\phi(s))ds\Big)\sqrt{\exp(\phi(\mu))}\,d\mu-q_1(t)q_2\Big(\int_{0}^{t}\exp(\phi(s))\,ds\Big)\sqrt{\exp(\phi(t))} \\ 
		&+2\lambda\bigg(r(t)\phi(t)+\int_{0}^{1}\phi(u)\,du\int_{0}^{1}r(s)\,ds-r(t)\int_{0}^{1}\phi(s)\,ds-\int_{0}^{1}\phi(s)r(s)\,ds\bigg). 
	\end{aligned}
\end{equation}

\item  $h$ is an isotropic covariance: By setting $h(s,t)=a\delta(t-s)$, we can derive the gradient as: 
\begin{equation}
	\begin{aligned}
			\nabla J(\phi)=  &-2\exp(\phi(t))\int_{t}^{1}q_1(\mu)\dot{q}_2\Big(\int_{0}^{\mu}\exp(\phi(s))ds\Big)\sqrt{\exp(\phi(\mu))}\,d\mu-q_1(t)q_2\Big(\int_{0}^{t}\exp(\phi(s))\,ds\Big)\sqrt{\exp(\phi(t))} \\ 
		&+2a\lambda\Big(\phi(t)-\int_{0}^{1}\phi(s)\,ds\Big). 
	\end{aligned}
\end{equation}
\end{enumerate}

\end{document}